\documentclass[modern,twocolumn,arydshln]{aastex701}

\usepackage{amsmath}
\usepackage{url}
\usepackage{multirow}
\usepackage{color}
\usepackage{subcaption}
\usepackage{graphicx}
\usepackage{physics}
\usepackage{upgreek}
\usepackage{nicefrac}
\usepackage{euscript}
\usepackage{mathrsfs}

\usepackage{ulem}
\usepackage{soul}

\shorttitle{W50 West} 
\shortauthors{Mac Intyre et al.}
\received{2026 May 1}
\revised{2026 June 3}
\accepted{2026 June 5}
\submitjournal{ApJ}

\newcommand{\unit}[1]{\ensuremath{\,\mathrm{#1}}}
\newcommand{\mysub}[1]{\ensuremath{_{\mathrm{#1}}}}

\newcommand{\ergs}{\ensuremath{~\mathrm{erg\,s^{-1}}}}

\newcommand{\Doppler}{\ensuremath{\mathscr{D}}}

\newcommand\chandra{{\it Chandra}}
\newcommand\xmm{{\it XMM-Newton}}
\newcommand\nustar{{\it NuSTAR \/}}

\begin{document}

\title{The Western Jet of SS~433/W50: Hard X-ray Emission, Spectral Evolution, and Comparison to the Eastern Jet }

\correspondingauthor{Brydyn Mac Intyre}
\email{macintyb@myumanitoba.ca}

\author[0000-0003-0146-3691]{Brydyn Mac Intyre}
\affiliation{University of Manitoba, Department of Physics \& Astronomy, Winnipeg, MB R3T 2N2, Canada}
\email{macintyb@myumanitoba.ca}

\author[0000-0001-6189-7665]{Samar Safi-Harb}
\affiliation{University of Manitoba, 
Department of Physics \& Astronomy,
Winnipeg, MB R3T 2N2, Canada}
\email{Samar.Safi-Harb@umanitoba.ca}

\author[0000-0002-7576-7869]{Dmitry Khangulyan}
\affil{Key Laboratory of Particle Astrophysics, Chinese Academy of Sciences, 100049 Beijing, People's Republic of China}
\affil{Tianfu Cosmic Ray Research Center, 610000 Chengdu, Sichuan, People's Republic of China}
\email{khangulyan@ihep.ac.cn}

\author[0000-0002-0589-765X]{Shuhan Zhang}
\affil{Columbia Astrophysics Laboratory, 550 West 120th Street, New York, NY 10027, USA}
\affil{University of California San Diego, La Jolla, California 92093, USA}
\email{shz091@ucsd.edu}

\author[0000-0002-9709-5389]{Kaya Mori}
\affil{Columbia Astrophysics Laboratory, 550 West 120th Street, New York, NY 10027, USA}
\email{kaya@astro.columbia.edu}

\author[0000-0001-7209-9204]{Naomi Tsuji}
\affil{Institute for Cosmic Ray Research, University of Tokyo, 5-1-5, Kashiwa-no-ha, Kashiwa, Chiba 277-8582, Japan}
\affil{Interdisciplinary Theoretical \& Mathematical Science Center (iTHEMS), RIKEN, 2-1, Hirosawa, Wako, Saitama 351-0198, Japan}
\email{ ntsuji@icrr.u-tokyo.ac.jp}

\author[0000-0003-1157-3915]{Felix Aharonian}
\affil{Tianfu Cosmic Ray Research Center, 610000 Chengdu, Sichuan, People's Republic of China}
\affil{Max-Planck-Institut für Kernphysik, Saupfercheckweg 1, 69117 Heidelberg, Germany}
\affil{Yerevan State University, 1 Alek Manukyan Street,Yerevan 0025, Armenia}
\affil{University of Science and Technology of China, 230026 Hefei, Anhui, People's Republic of China}
\email{felix.aharonian@mpi-hd.mpg.de}

\author[0000-0002-9105-0518]{Laura Olivera-Nieto}
\affil{Anton Pannekoek Institute for Astronomy, University of Amsterdam, Science Park 904, 1098 XH, The Netherlands}
\affil{Gravitation and Astroparticle Physics Amsterdam Institute, University of Amsterdam, Science Park 904, 1098 XH Amsterdam, The Netherlands}
\email{l.oliveranieto@uva.nl}

\begin{abstract}
    The W50 nebula powered by the microquasar SS~433 is a unique laboratory for exploring several fundamental astrophysical phenomena. This study presents observations from \nustar and \xmm, concentrating on the western lobe of W50. Detection of hard non-thermal X-ray emission is reported, extending up to $\sim$30~keV. This emission originates from a compact, knotty area referred to as the ``Head'', located at $\sim$17$^{\prime}$ (equivalent to 26.5 pc at an assumed distance of 5.5~kpc) to the west of SS~433, and characterized by a power-law spectrum with a hard photon index of 1.55$\pm$0.07 (0.5--30~keV). Moving westward from SS~433, the photon index gradually steepens, ultimately reaching a photon index of 2.10$\pm$0.05 in the ``w2'' region centered at $\sim$35$^{\prime}$ or $\sim$56~pc from SS~433. The distinct hard X-ray knots observed serve as clear markers for sites of particle acceleration within the western jet. The synchrotron radiation from the ``Head'' region implies equipartition magnetic field strength $B$~$\sim$~15~$\upmu$G. Notably, these properties (western ``Head'' location, unusually hard spectral index, inferred magnetic field, and spectral evolution away from SS~433) are very similar to what has been observed in the eastern lobe, supporting a symmetric jet-driven origin. Finally, the broadband spectral energy distribution (SED) and X-ray morphology are modeled using semi-analytic jet models, exploring different jet velocity and magnetic field configurations. The results favor a scenario in which in-situ particle acceleration and synchrotron emission dominate, with implications for understanding particle transport, jet dynamics, and W50's role as a Galactic PeVatron.
\end{abstract}

\keywords{ISM: supernova remnants --- ISM: jets and outflows --- ISM: individual objects (W50) --- stars: black holes --- stars: individual (SS~433) --- X-rays: general}

\section{Introduction} \label{sec:intro}
The W50 nebula, catalogued as a Galactic supernova remnant (SNR G39.7--2.0; \cite{1974A&A....32..375V, 1980ApJ...236L..23V}), is associated with the microquasar SS~433, a high-mass X-ray binary system displaying precessing semi-relativistic jets ($v=0.26$c). These jets are believed to emanate from a stellar-mass-sized black hole analog to active galactic nuclei (AGNs; \cite{1984ARA&A..22..507M, 2004ASPRv..12....1F}). Spanning approximately $2^\circ$ (east–west)$\times1^\circ$ (north–south), W50 ranks among the largest known SNRs within the galaxy, extending over 200$\times$100~pc under the assumed distance from Earth $d=5.5$~kpc \citep{2004ApJ...616L.159B}.

\begin{figure*}[htb]
    \centering
    \includegraphics[width=\textwidth]{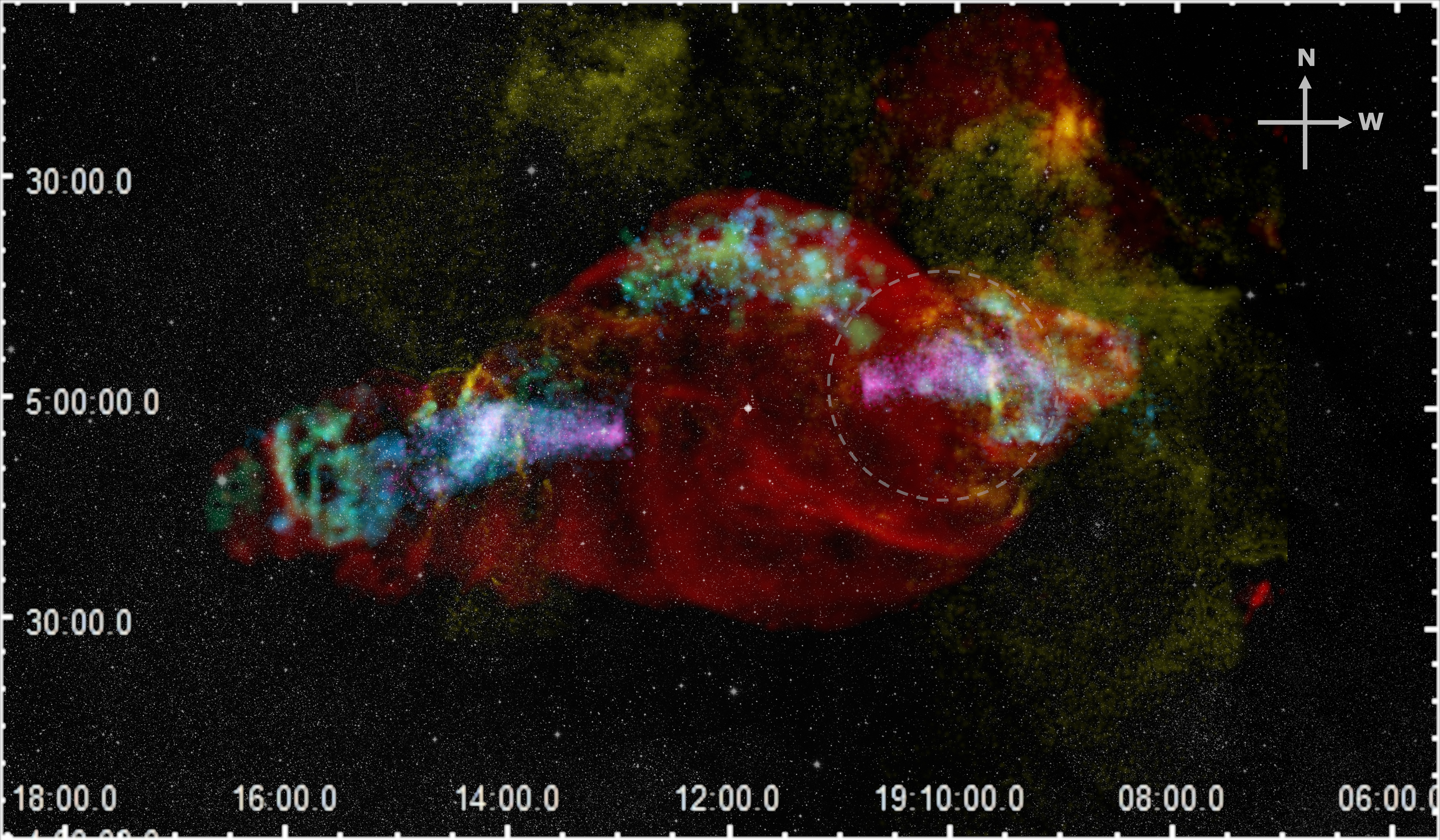}
    \caption{Multi-wavelength image of the W50 nebula. Red: Radio \citep{1998AJ....116.1842D}, Yellow: Optical \citep{2007MNRAS.381..308B}, Green: Soft X-rays (0.5--1~keV), Cyan: Medium energy X-rays (1--2~keV), Magenta: Hard X-ray emission (2--12~keV). 
    The eastern lobe and northern shell shown reflect previously published \xmm\ data \citep{Safi-Harb_2022, 2024ApJ...975L..28C}. 
    The western lobe X-ray image highlights the \xmm\ data presented in this work, revealing a sharp boundary referred to as the ``Head'' at $\sim$17$^{\prime}$ (26.5 pc at an assumed distance of 5.5~kpc) west of SS~433. This region mimics the eastern ``Head'' region counterpart detected 18$^{\prime}$ (29~pc) east of SS~433.
    Sky coordinates are shown in RA and Dec (J2000), and the compass indicates the image orientation (N for North to the top and W for West to the right). The faint dashed circle marks the approximate \xmm\ field of view of the western-lobe observation shown in the top-left panel of Figure~\ref{fig:coord+regs}.}
    \label{fig:mltwvl}
\end{figure*}

Commonly referred to as the ``Manatee" nebula due to its radio characteristics\footnote{https://www.nrao.edu/pr/2013/w50/} \citep{1987AJ.....94.1633E, 1998AJ....116.1842D, farnes2017}, W50's distinct morphology likely results from its interaction with the jets emitted by SS~433. This notion is supported by the alignment of optical filaments and the elongated shape of W50, which corresponds with the precession cone axes of the jets. Numerical simulations have also supported this conclusion \citep{2011MNRAS.414.2838G}. The nebula's ``ears," denoting its eastern and western radio-bright edges, bear resemblance to the lobes observed in AGN and in the bubbles associated with certain ultraluminous X-ray sources. Figure \ref{fig:mltwvl} illustrates the multi-wavelength depiction of this captivating entity.

Various segments of the W50 nebula have been examined within the field of view (FoV) of several X-ray missions: ROSAT, ASCA, and RXTE during the 1990s \citep{1994PASJ...46L.109Y, 1997ApJ...483..868S, 1999ApJ...512..784S}, as well as in the early 2000s with \xmm\ and \chandra\ \citep{ 2007A&A...463..611B,2005AdSpR..35.1062M}. In 2022, motivated by the detection of W50 in TeV $\gamma$-rays, additional joint observations of the eastern lobe with \xmm\ and \nustar resolved and characterized the sites of particle acceleration \citep{Safi-Harb_2022}. Most recently, the Spectrum-Roentgen-Gamma (SRG)/eROSITA mission granted an unprecedented panoramic X-ray view of the entire W50/SS~433 nebula \citep{2026A&A...707A.278S}, demonstrating that the observed large-scale structure supports a scenario in which the hyper-Eddington source SS~433 has reshaped its environment through sustained jet activity. This study further highlights W50/SS~433 as a benchmark system for probing jet-ISM interaction and particle acceleration on parsec scales.

\begin{figure*}[htb]
    \begin{subfigure}[b]{0.475\textwidth}
        \includegraphics[width=\textwidth]{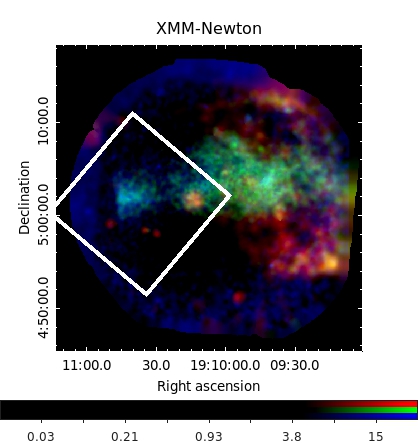}
    \end{subfigure}
    \hfill
    \begin{subfigure}[b]{0.475\textwidth}
        \includegraphics[width=\textwidth]{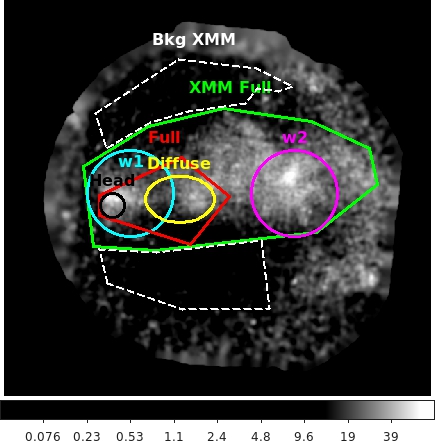}
    \end{subfigure}
    \begin{subfigure}[b]{0.475\textwidth}
        \includegraphics[width=\textwidth]{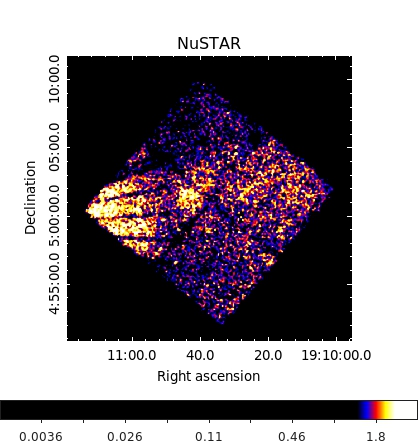}
    \end{subfigure}
    \hfill
    \begin{subfigure}[b]{0.475\textwidth}
        \includegraphics[width=\textwidth]{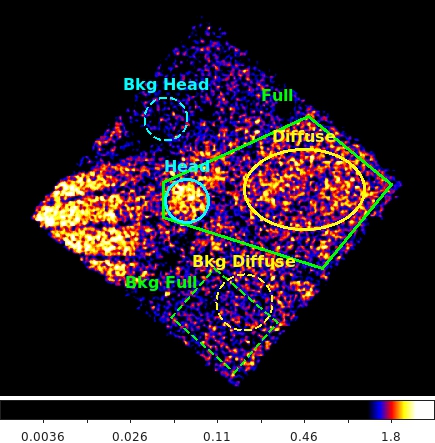}
    \end{subfigure}
    \caption{These images illustrate the FoV, extent of emission, and regions being studied here. [Top Left] A Red-Green-Blue image of W50's western lobe via \xmm\ with the \nustar FoV superimposed in white. Red: 0.5--1.0~keV, Green: 1.0--2.0~keV, Blue: 2.0--12.0~keV. [Top Right] Depiction of all named regions analyzed in this work with \xmm\ data, background extraction area in white. [Bottom Left] \nustar FPMA detector image illustrating its full FoV for comparison with \xmm\ above. [Bottom Right] Region selections for \nustar extraction and analysis are depicted with solid lines while the corresponding background extraction regions use dotted lines of the same color.}
    \label{fig:coord+regs}
\end{figure*}

The X-ray emission, contrary to expectations for an evolved supernova remnant (SNR), displays distinctive characteristics. These include the X-ray lobes' morphology, which fills the gap between SS~433 and the radio ``ears," as well as its alignment within the projected precessing cone of the jets. Additionally, the X-ray spectrum tends to get steeper (or softer, i.e. characterized by a larger photon index) as it moves away from SS~433, along the directions of the precessing jets. Furthermore, while the outermost region of W50, coinciding with the eastern radio ear (identified as region ``e3" in \citep{1997ApJ...483..868S}), is predominantly filled with soft thermal X-rays, the interior sections (represented as regions ``w1'' and ``w2'' in Figure \ref{fig:coord+regs}) exhibit  non-thermal (photon index $\Gamma\sim1.4$--$2.4$) emissions (this work; \cite{1997ApJ...483..868S, 2026A&A...707A.278S}).

The presence of a hard spectrum suggested, in the context of non-thermal synchrotron radiation interpretation,  maximum electron energies of about $\sim$300–450~TeV, thereby pointing to W50/SS~433 as a site of particle acceleration up to PeV energies, or a Galactic PeVatron candidate \citep{1999ApJ...512..784S, 2005A&A...439..635A}. These studies motivated the searches for high-energy $\gamma$-ray emission, using the Major Atmospheric Gamma Imaging Cherenkov telescopes (MAGIC), the High Energy Spectroscopic System (H.E.S.S.), and the Very Energetic Radiation Imaging Telescope Array System (VERITAS) \citep{2018A&A...612A..14M, 2017ICRC...35..713K}, which initially resulted in establishing upper limits on TeV emission from the system.

The PeVatron scenario has recently gained renewed attention due to the discovery of TeV emission by the High Altitude Water Cherenkov (HAWC) Observatory \citep{2018Natur.562...82A}, and more recently by H.E.S.S. \citep{hess_2024}, and the Large High Altitude Air Shower Observatory, LHAASO \citep{2025NSRev..12af496L}. A joint analysis involving Fermi-LAT and HAWC (\cite{2020ApJ...889L...5F} and references therein) identifies common emission sites of GeV-to-TeV $\gamma$-rays within the eastern and western lobes of SS~433. 
H.E.S.S. has in particular helped resolve the peaks for TeV emission found to closely align with the ``e1--e2'' and ``w1--w2'' regions. 
LHAASO can not resolve the emission of ultra-high energies, but showed that the $\gamma$-ray emission cannot be described by a single leptonic component. While the eastern and western lobes dominate below 100~TeV, a separate extended component detected at energies above 100 TeV coincides with a nearby cloud suggesting a hadronic component.

The X-ray band is crucial to resolve the sites of particle acceleration. While the origin and properties of the hard X-ray emission in the inner eastern lobe have now been constrained \citep{Safi-Harb_2022, 2025ApJ...993L..24T}, the spatial extent and spectrum of the hard X-ray emission across the western jet and lobe remain poorly understood. This work aims at zooming in on the western lobe using \xmm\ and \nustar\ (Sections~\ref{sec:Obs}, \ref{sec:Analysis}), comparing it to the eastern lobe (Section~\ref{sec:Analysis-XSRSvC}), and performing a modeling study (Section~\ref{sec:SED}, \ref{sec:Mod}) to shed light on the mechanism responsible for the X-ray emission and particle acceleration. A discussion of the results is presented in Section~\ref{sec:Conc}.

\section{Observations} \label{sec:Obs}
Following the study of the eastern lobe of W50 by \cite{Safi-Harb_2022}, the western lobe was targeted for follow-up observations with both \xmm\ and \nustar (see Table~\ref{tab:obs}). This joint observation provides complementary capabilities: \xmm\ is capable of constraining the column density in the soft X-ray band,  while \nustar data extends the spectral coverage to the hard X-ray band beyond 10~keV, thus allowing for broadband spectroscopy.

\begin{table*}[htb]
    \caption{} \label{tab:obs} 
    \centering
    \begin{tabular}{lcccc}
        \multicolumn{5}{c}{Log of the X-ray observations presented in this study.} \\ \hline
        Satellite & Date & ObsID & Pointing & Total Exposure \\
        &&&RA, Dec (J2000) & (ks)  \\ \hline
        \nustar  & 2021-07-14 & 40701002002  & 19 10 39.4, +05 01 30 & 97.6        \\
        \xmm\ & 2022-04-05 & 0890800101 & 19 10 08.0, +05 01 42 & 60.8        \\ 
        \xmm\ & 2020-03-24 & 0840490101 & 19 13 13.4, +04 57 30 & 69.1        \\ 
        \xmm\ & 2004-09-30 & 0075140401 & 19 14 12.0, +04 55 47 & 32.5        \\ \hline
    \end{tabular}
\end{table*}

\subsection{NuSTAR} \label{sec:Obs-Nu}
\nustar observed the western lobe of W50 targeting the ``w1'' region with a 69 ks exposure on 2021 July 14. \nustar is comprised of two co-aligned hard X-ray telescopes with corresponding focal plane detector modules FPMA and FPMB. Each module has a $13\farcm2 \times 13\farcm2$ FoV, with an angular resolution of $18\farcs$ FWHM ($58\farcs$ HPD). The detectors are sensitive to X-rays in the 3--79~keV range and provide a FWHM energy resolution of 0.4~keV below 50~keV. \nustar data processing and analysis were conducted using {\tt FTOOLS} (HEAsoft 6.28) with \nustar Calibration Database (CALDB) files of 2022 May 10. Background contamination was first assessed in the \nustar data dominated by two components --- (1) ghost-ray background photons from SS~433 and (2) stray-light photons from nearby bright X-ray sources and the CXB. For the former, ghost-ray background simulations with {\tt NuSIM} were run, a ray-tracing Monte-Carlo simulator specifically developed for \nustar. For the latter, {\tt Nuskybgd} \citep{Wik2014} was applied, an IDL module that models stray-light and internal detector backgrounds, and {\tt {nustar\_stray\_light}}, which calculates stray-light background patterns on the detector plane from bright X-ray sources within $\sim5^\circ$. With these software tools, the different background components were fully investigated to perform background subtraction for \nustar data. Detailed \nustar background analysis is presented in \ref{sec:nustarbkg}. The background-subtracted images obtained from {\tt Nuskybgd} were used to generate images and exposure maps using {\tt nuexpomap} and {\tt FTOOLS} extractor.

\subsection{XMM-Newton} \label{sec:Obs-XMM}
With its larger FoV, \xmm\ covered the western lobe of W50 on 2022 April 5 for 60.1~ks of exposure time. On board \xmm, the European Photon Imaging Camera (EPIC) comprises three detectors: the EPIC pn detector \citep{2001A&A...365L..18S}, and EPIC MOS1 and MOS2 \citep{2001A&A...365L..27T}. At 1.5~keV, the on-axis point-spread function exhibits an FWHM of approximately $12\farcs5$ and $4\farcs3$ for the pn and MOS detectors, respectively. These EPIC detectors are capable of detecting X-rays in the range of 0.15~to~12~keV with a moderate energy resolution of $E/\Delta E$(pn) $\sim$~20--50. The data underwent reduction and analysis using the most up-to-date calibration files and the Standard Analysis Software (SAS) v.19.1.0. Detailed descriptions of the source and background region selections can be found in Section \ref{sec:AnalysisImg}, and they were extracted using {\tt{evselect}}. To account for extended emission, the SAS commands {\tt{rmfgen}} and {\tt{arfgen}} were utilized to generate the response matrix file (rmf) and ancillary response file (arf). After filtering out background flares, the effective exposure times for the MOS1, MOS2, and pn data were 45.5~ks, 45.6~ks, and 38.3~ks, respectively.
The final two \xmm\ observations from 2020 and 2004 respectively cover the eastern X-ray jet \citep{Safi-Harb_2022}. These observations were both analyzed with the same methods as the new western observation to maintain consistency. Although not used for spectroscopy in this work, two additional observations: 0882560101 and 0882560201, as discussed in \citep{2024ApJ...975L..28C}, were additionally included in the multiwavelength image creation for the northern shell as seen in Figure~\ref{fig:mltwvl}.

\section{Analysis Results} \label{sec:Analysis}

\subsection{Imaging Analysis}\label{sec:AnalysisImg}
Figure~\ref{fig:coord+regs} shows the \xmm\ and \nustar images of the western lobe of W50 and details the regions selected to examine areas of interest within. The top-left image is a red-green-blue representation of the \xmm\ X-ray data where red corresponds to the 0.5--1.0~keV range, green to 1.0--2.0~keV, and blue 2.0--12.0~keV. This energy separation clearly shows the innermost region is hardest in nature, while it appears to soften westward along the jet axis until the furthest edge of the observation where the diffusive low energy emission is most noticeable. The white box in this image represents the FoV of \nustar's observation, as depicted in the bottom-left image, centered on the head of the western jet.

As \nustar must contend with a smaller FoV, and much higher levels of contamination as discussed later in Section~\ref{sec:Analysis-Spec}, these constraints are considered with the image analysis results to determine areas of interest for further study. From the \nustar FoV, three areas are determined for joint \xmm\ and \nustar broadband analysis; being the ``Head'', ``Diffuse'', and ``Full'' regions depicted in the bottom-right image of Figure~\ref{fig:coord+regs}. As discussed, these regions cover the hardest emission from the start of the jet, as well as the mid-range hardness shown in green in Figure~\ref{fig:coord+regs}, and the full jet as captured by \nustar. As \nustar does not cover the full extent of the \xmm\ observation, additional regions were selected for \xmm\ analysis only. These include the ``w1'' and ``w2'' regions as defined in previous literature \citep{1997ApJ...483..868S}, as well as the ``XMM Full" region which encompasses the full hard emission. The spectral analysis of these spatially resolved regions is detailed in Section~\ref{sec:Analysis-SRS}.

\subsubsection{Spatial Profile with NuSTAR and XMM-Newton} \label{sec:Analysis-SPNX}
The X-ray spatial profiles were investigated in different energy bands using both \nustar and \xmm\ data in order to study how the X-ray morphology and hardness distribution vary along the western jet. For this study, a $8\farcm19 \times 3\farcm07$ region near the ``Head'' from the exposure-corrected, background-subtracted image was selected with combined FPMA and FPMB data. The region was sliced with linearly-spaced bins perpendicular to the jet direction, summed the image values in each slice, and plotted X-ray flux distribution along the jet.

\begin{figure}[t!]
    \centering
    \includegraphics[width=0.45\textwidth]{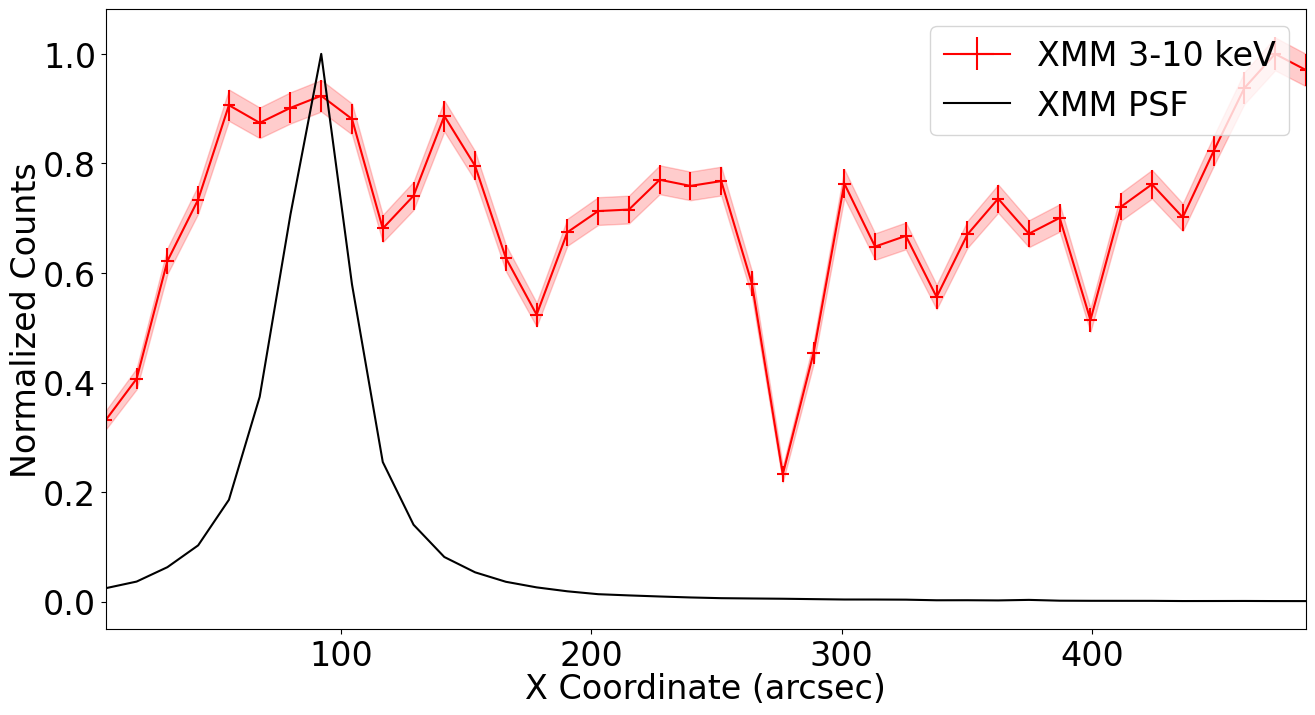}
    \caption{Spatial profile along the jet for the 3--10~keV \xmm\ image compared to 6~keV \xmm\ 9\arcmin\ off-axis PSF. This illustrates the extended emission relative to the PSF. Binning was set to 5 pixels. }
    \label{fig:xmmpsf3to10}
\end{figure}

\begin{figure}[t!]
    \centering
    \includegraphics[width=0.45\textwidth]{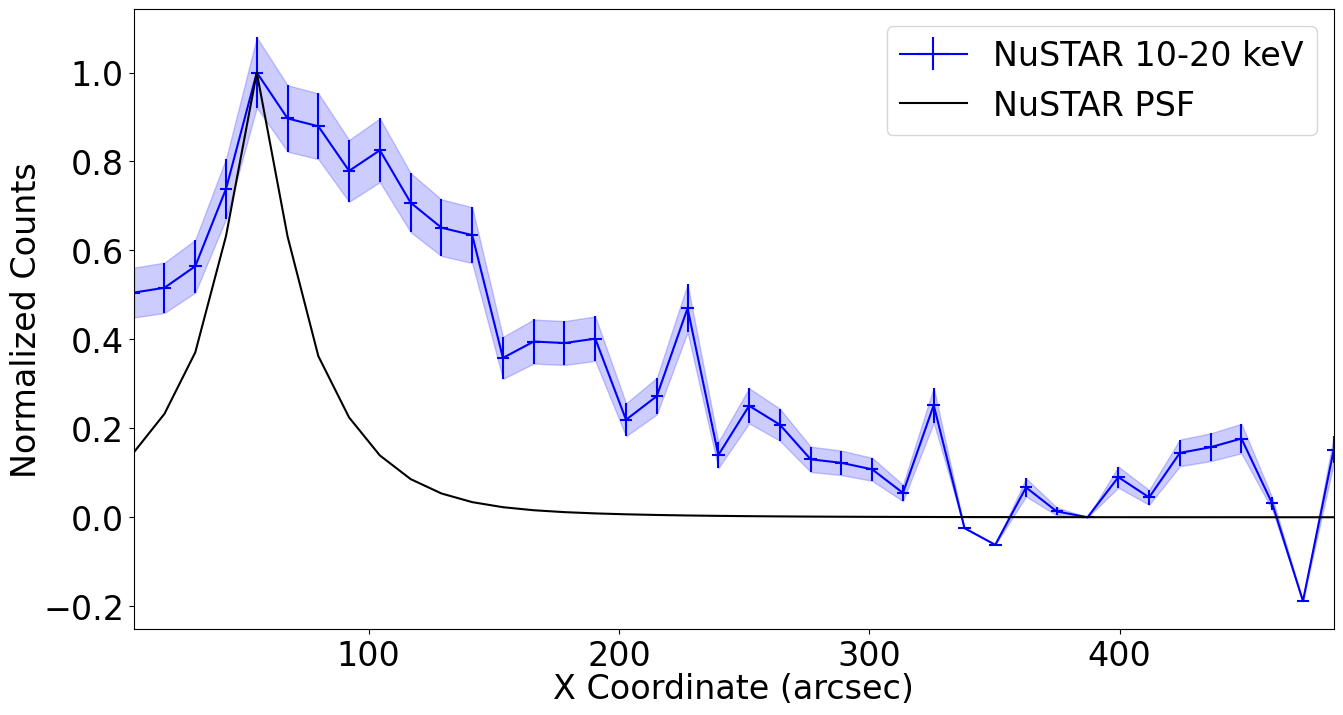}
    \caption{Spatial profile along the jet for the 10--20~keV \nustar image compared to 12--20~keV \nustar effective PSF. While the difference is less pronounced than in the \xmm\ case, the extent relative to the PSF is still evident. Binning was set to 5 pixels.}
    \label{fig:nupsf10to20}
\end{figure}

In order to study the morphology of the X-ray emission in both soft (3--10 keV) and hard (10--20 keV) bands, linear profiles were normalized and compared them to the point spread function (PSF) of the corresponding telescopes. Figure~\ref{fig:xmmpsf3to10} shows the normalized 3--10 keV profile from \xmm\ MOS1 image in comparison with the MOS1 PSF at $E=6$~keV and 9\arcmin\ off-axis angle, which was centered at the $x$ coordinate with the highest normalized counts in the profile. The MOS1 image is initially binned to 2\farcs45 per pixel, and the binning of the profile was set to 5 pixels in order to reduce statistical noise while retaining substantial structural details. It is noted that the dips in the soft band profile are attributed to MOS1 detector gaps. For the hard band profile, in Figure~\ref{fig:nupsf10to20}, the 10--20 keV \nustar image is presented in comparison with the 12--20~keV \nustar effective PSF. Background subtraction was applied to the \nustar FPMA and FPMB images using {\tt Nuskybgd} with the procedure described in \ref{sec:nustarbkg}. Then the FPMA and FPMB images were combined and exposure-corrected with vignetting effects using the procedure described in \cite{2014ApJ...789...72N}. The resulting \nustar image with a pixel size of 2\farcs46 was used to make the linear profile with binning set to 5 pixels. As shown in Figure~\ref{fig:xmmpsf3to10} and Figure~\ref{fig:nupsf10to20}, emission from the ``Head'' region is extended compared to the \nustar PSF. 

\begin{figure}[t!]
    \centering
    \includegraphics[width=0.45\textwidth]{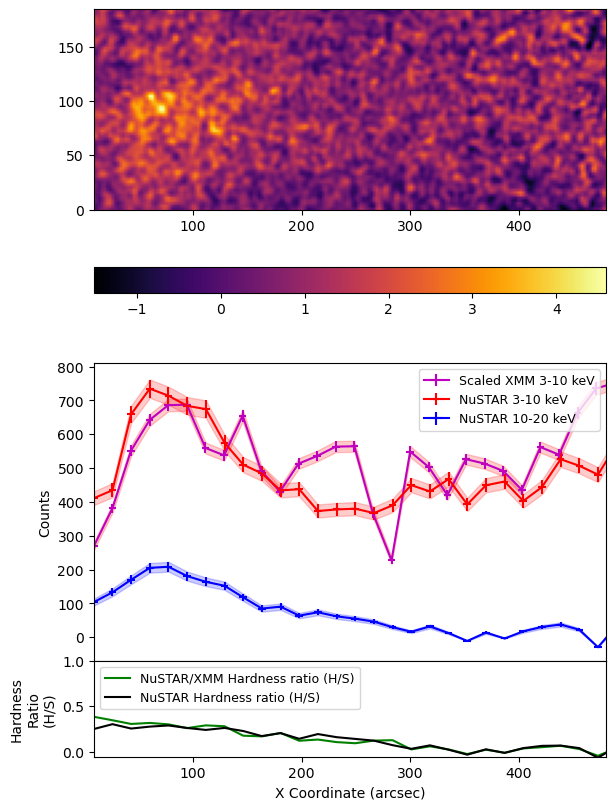}
    \caption{\nustar spatial profiles (middle) around the ``Head'' region (top) and the hardness ratio (bottom). Binning was set to 8 pixels.}
    \label{fig:nuhardness}
\end{figure}

In addition, the hardness ratio profile was plotted by dividing the hard band flux values by those in the soft band. As shown in Figure~\ref{fig:nuhardness}, the hard X-ray emission in 10--20 keV decreases away from the core of ``w1'' faster than the soft X-ray emission. This is expected from the synchrotron origin of the X-ray emission since the cooling time is faster for more energetic particles emitting harder X-rays. 

\subsection{Spectral Analysis} \label{sec:Analysis-Spec}
All potential sources of contamination were examined to ensure that the extracted spectra represent the intrinsic source emission. The analysis began with a comprehensive review of the various background components and strategies for treating them. Due to the complicated nature of W50's environment with extensive extended emission, nearness to the Galactic plane, and to its bright engine SS~433, local background selections were identified as better able to represent this complex structure after modeling and review. For \xmm, 11 different background regions covering the entirety of the observation's FoV were examined before settling on the background shown in white on the top-right image of Figure~\ref{fig:coord+regs}. Additionally, when fitting the ``Head'', ``Diffuse'', and ``Full'' regions, three identical background regions as depicted in the bottom-right image of Figure~\ref{fig:coord+regs} were also extracted from \xmm\ for an additional consistency check between the observations.

\nustar must additionally contend with stray light and ghost ray contamination. While previous extractions from the eastern lobe made use of extensive background simulations \citep{Safi-Harb_2022} which were also completed for this work, they did not accurately represent the background needed for the western lobe. After all options were examined, the optimal solution was again determined to be local background selections. These backgrounds as much as possible follow radial symmetry with their associated region's distance from SS~433, while being source-free and staying as close as possible to the extraction regions.

\begin{figure}[htb]
    \centering
    \begin{subfigure}{\columnwidth}
        \centering
        \includegraphics[width=\columnwidth]{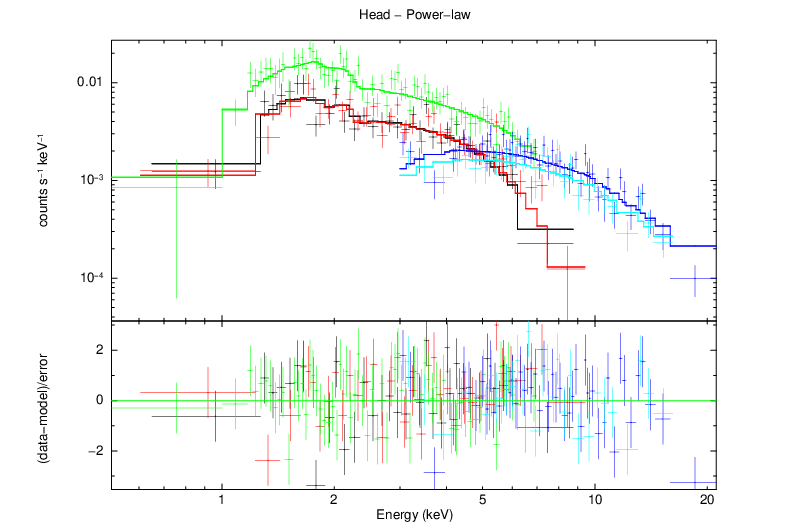}
    \end{subfigure}
    \newline 
    \begin{subfigure}{\columnwidth}
        \centering
        \includegraphics[width=\columnwidth]{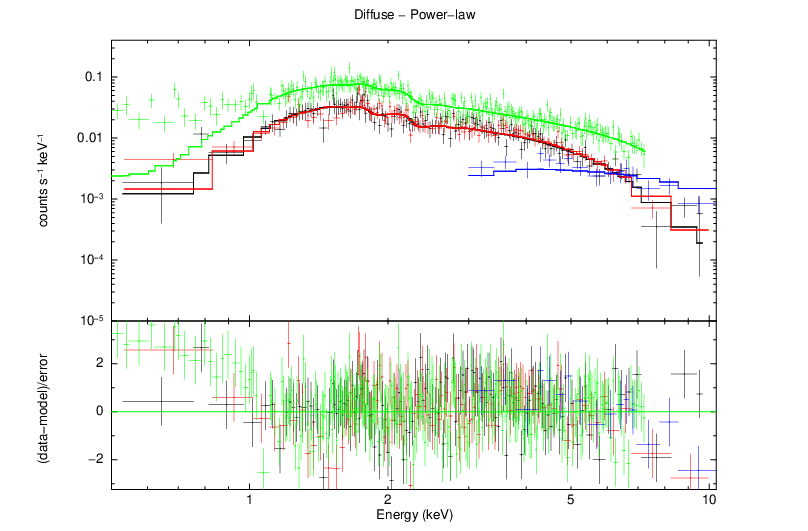}
    \end{subfigure}
    \newline 
    \begin{subfigure}{\columnwidth}
        \centering
        \includegraphics[width=\columnwidth]{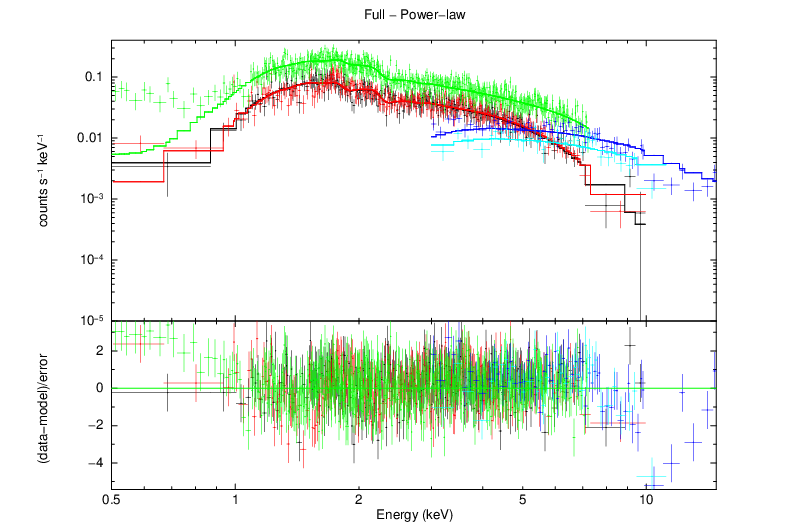}
    \end{subfigure}
    \caption{ The joint \xmm\ and \nustar spectra for the ``Head” (top), ``Diffuse” (middle), and ``Full” (bottom) \nustar regions, fitted with the absorbed power-law model as described in the text. Black, red, and green spectra correspond to MOS1, MOS2, and pn, respectively; blue and cyan correspond to the \nustar FPMA and FPMB spectra. Note that the ``Diffuse'' fit omitted poor data from FPMB. The regions are shown in Figure \ref{fig:coord+regs}, and the spectral parameters for these regions are summarized in Table \ref{tab:data}.}
    \label{fig:spectra}
\end{figure}

\subsubsection{NuSTAR and XMM-Newton Spatially Resolved Spectroscopy} \label{sec:Analysis-SRS}
Based on previous X-ray observations of the W50 system (see Section~\ref{sec:intro}), various models were explored, including non-thermal (power-law and broken power-law) and thermal models, to perform spectral fits. The X-ray spectral fitting package XSPEC v12.11.1 \citep{1996ASPC..101...17A}\footnote{https://heasarc.gsfc.nasa.gov/xanadu/xspec/} was employed for the spectral analysis, incorporating the {\tt{tbabs}} model to account for photoelectric absorption by the interstellar medium with the abundances from \cite{2000ApJ...542..914W}. This choice of abundances generally led to higher column densities, $N_H$, compared to early (pre-2020) studies that used different abundance models, but is consistent with that used for the  joint \xmm\ and \nustar study of the eastern lobe \citep{Safi-Harb_2022}.

The \xmm\ and \nustar data were separately fit in the energy ranges of 0.5--10~keV and 3--30~keV, respectively. For \xmm , MOS/pn data were binned with a minimum of 30 counts per bin. As for \nustar, the background-subtracted spectra were grouped to achieve at least a 3$\sigma$ detection level per data bin. Given \nustar's insensitivity to the soft X-ray band, the column density was fixed to the value obtained from \xmm . To account for potential calibration differences between instruments, a constant factor was employed in all joint fits. The uncertainties reported are at the 90\% confidence level (CL), and the fitting procedure used the chi-square ($\chi^2$) statistic, unless stated otherwise.

In Figure~\ref{fig:coord+regs}, several regions are highlighted and elaborated on in Section~\ref{sec:AnalysisImg}. For each of these regions, separate and combined fits were conducted for the \xmm\ spectra in the 0.5--10~keV range and the \nustar spectra in the 3--30~keV range. The fitting was performed using an absorbed power-law model ({\tt{const}}{$\ast$}{\tt{tbabs}}{$\ast$}{\tt{power}} in XSPEC). In the case of individual \nustar fits, a fixed column density was maintained, matching the value obtained from \xmm. The spectral fits from these analyses are presented comprehensively in Table~\ref*{tab:data}.

\begin{table*}[htb]
\caption{} \label{tab:data} \vspace{-0.25cm}
\centering
\begin{tabular}{llccc}
\multicolumn{5}{c}{XSPEC Fits Using {\tt{const}}{$\ast$}{\tt{tbabs}}{$\ast$}{\tt{power}} with the {\tt{cFlux}} Model Added to Acquire Flux Values. }
\\ \hline
                                       & Parameter                                                 & \xmm\    & \nustar  & Joint$^{c}$  \\ 
                                       &                                                  & 0.5--10 keV   & 3--30 keV  & 0.5--30 keV  \\ \hline
Head    & Photon Index $\Gamma$                               & $1.5\pm0.14$  & $1.7\pm0.11$  & $1.55\pm0.07$ \\
$N_H = (1.9\pm0.3)\times10^{22}$~cm$^{-2}$                                       & $F\left[\times10^{-12}\right]$ (abs.)$^{a,c}$             & $0.62\pm0.02$ & $1.01\pm0.05$ & $1.19\pm0.03$ \\
                                       & $F\left[\times10^{-12}\right]$ (unabs.)$^{b,c}$           & $0.88\pm0.03$ & $1.03\pm0.05$ & $1.46\pm0.03$ \\
                                       & $\chi^2_\nu$ (DoF)                                        & $1.20~(160)$  & $1.21~(81)$   & $1.22~(243)$ \\ \hline
Diffuse    & Photon Index $\Gamma$                           & $1.66\pm0.07$ & $2.4\pm0.4$   & $1.67\pm0.04$ \\
$N_H = (1.5\pm0.12)\times10^{22}$~cm$^{-2}$                                       & $F\left[\times10^{-12}\right]$ (abs.)                     & $1.45\pm0.03$ & $1.5\pm0.18$  & $1.95\pm0.07$ \\
                                       & $F\left[\times10^{-12}\right]$ (unabs.)                   & $2.12\pm0.04$ & $1.57\pm0.19$ & $2.68\pm0.07$ \\
                                       & $\chi^2_\nu$ (DoF)                                        & $1.27~(650)$  & $0.50~(14)$   & $1.27~(666)$ \\ \hline
Full    & Photon Index $\Gamma$                     & $1.68\pm0.05$ & $2.2\pm0.10$  & $1.72\pm0.03$ \\
$N_H = (1.61^{+0.09}_{-0.08})\times10^{22}$~cm$^{-2}$                                       & $F\left[\times10^{-12}\right]$ (abs.)                     & $4.02\pm0.05$ & $8.2\pm0.3$   & $7.3\pm0.15$ \\
                                       & $F\left[\times10^{-12}\right]$ (unabs.)                   & $6.06\pm0.07$ & $8.4\pm0.3$   & $9.1\pm0.15$ \\
                                       & $\chi^2_\nu$ (DoF)                                        & $1.13~(1482)$ & $1.44~(96)$   & $1.19~(1580)$ \\ \hline
w1    & Photon Index $\Gamma$                      & $1.71\pm0.07$ & ... & ... \\
$N_H = (1.85^{+0.12}_{-0.11}) \times10^{22}$~cm$^{-2}$                                       & $F\left[\times10^{-12}\right]$ (abs.)                     & $3.23\pm0.05$ & ... & ... \\
                                       & $F\left[\times10^{-12}\right]$ (unabs.)                   & $5.06\pm0.08$ & ... & ... \\
                                       & $\chi^2_\nu$ (DoF)                                        & $0.95~(1184)$ & ... & ... \\ \hline
w2    & Photon Index $\Gamma$                               & $2.10\pm0.05$ & ... & ... \\
$N_H = (0.80\pm0.04)\times10^{22}$~cm$^{-2}$                                       & $F\left[\times10^{-12}\right]$ (abs.)                     & $2.68\pm0.04$ & ... & ... \\
                                       & $F\left[\times10^{-12}\right]$ (unabs.)                   & $4.28\pm0.06$ & ... & ... \\
                                       & $\chi^2_\nu$ (DoF)                                        & $1.09~(1174)$ & ... & ... \\ \hline
XMM Full    & Photon Index $\Gamma$                         & $1.67\pm0.03$ & ... & ... \\
$N_H = (0.92\pm0.04)\times10^{22}$~cm$^{-2}$                                       & $F\left[\times10^{-12}\right]$ (abs.)                     & $15.4\pm0.13$ & ... & ... \\
                                       & $F\left[\times10^{-12}\right]$ (unabs.)                   & $21.1\pm0.17$ & ... & ... \\
                                       & $\chi^2_\nu$ (DoF)                                        & $1.10~(3277)$ & ... & ... \\ \hline
\end{tabular}
\caption*{\footnotesize{\textbf{Notes.} Fig.~\ref{fig:coord+regs} shows the selected regions. Column densities $N_H$ are frozen to their best fit value using the \xmm\ data. Quoted uncertainties are 90\% C.L.\\
$^{a}$ Absorbed flux in erg cm$^{-2}$ s$^{-1}$.\\
$^{b}$ Unabsorbed flux in erg cm$^{-2}$ s$^{-1}$.\\
$^{c}$ For the joint 0.5--30~keV flux, the combined 0.5--10~keV (\xmm) and 10--30~keV (\nustar) fluxes with the model parameters frozen to their best values from the joint fit are listed.}}
\end{table*}

The analysis reveals significant variations in the column density, $N_H$, across the inner western lobe. The ``Head'' region exhibits $N_H = (1.9\pm0.3) \times10^{22}$~cm$^{-2}$, while the ``Diffuse'' region shows $N_H = (1.5\pm0.12) \times10^{22}$~cm$^{-2}$, and the ``Full'' region has $N_H = (1.61^{+0.09}_{-0.08})\times10^{22}$~cm$^{-2}$. Further west, the ``w2'' region reveals the continuation of the decreasing trend showing $N_H = (0.80\pm0.04)\times10^{22}$~cm$^{-2}$. The spectral parameters provided in Table~\ref{tab:data} correspond to $N_H$ fixed at its best-fit value derived from the \xmm\ fits.

A power-law model with photon indices of $1.55\pm0.07$, $1.67\pm0.04$, and $1.72\pm0.03$ (0.5–30~keV) adequately describes the emission in the ``Head," ``Diffuse," and ``Full'' regions, respectively. Figure~\ref{fig:spectra} illustrates the corresponding spectra and fitted models. When $N_H$ is allowed to vary, the error bars slightly increase as expected, but the overall results remain consistent within the uncertainties. The fits with varying $N_H$ yield photon indices of $1.62^{+0.09}_{-0.10}$, $1.69\pm0.07$, and $1.79\pm0.05$ (0.5–30~keV) for the ``Head," ``Diffuse," and ``Full'' regions, respectively. Considering the spectral outcomes for the ``w2'' region (Table~\ref{tab:data}), there appears to be a gradual spectral steepening (i.e. an increase in the photon index) moving westward from SS~433, with the ``Head'' region exhibiting the hardest emission and marking the onset of X-ray emission in the western lobe. 

The index for the ``Head'' region remains consistent between the two instruments (at $\Gamma\sim$1.6), while for the other regions, \nustar shows a slightly steeper index than \xmm. In the larger-scale region ``Full," which covers the emission across the \nustar field, the discrepancy primarily arises from the mixture of multiple components with different spectral indices, as demonstrated by the spatially resolved spectroscopy. For the fainter ``Diffuse'' and ``Full'' regions, a broken power-law model was employed to examine any steepening in the \nustar hard band above 10~keV. However, these fits did not yield conclusive results, as small variations in starting conditions significantly affected the break energy and photon index on each side. This behavior is likely due to mimicking the spectral evolution (steepening) observed across the wide ``Full'' region, extending from the ``Head'' toward the ``w2'' region, where the index steepens to $\sim$2.1.

Additional checks were carried out to determine whether cut-off power-law models would instead be preferred for fitting the spectra. The XSPEC model {\tt{cutoffpl}} resulted in cutoff energies falling much past the fitting energy range and did not improve the residuals from power-law fits. The more physical model of \cite{2007A&A...465..695Z} was also tested with similar results and did not find physical or statistical fits, likely due to the photon index being frozen to 2 in its derivation.

Finally, different thermal models were tested to explore alternative possibilities. Although some of these models yielded statistically acceptable fits, none demonstrated an improvement over the power-law fits. The presence of unrealistically high temperatures, absence of emission lines, and the lack of statistical improvement in comparison to the power-law fits all support the preference for the power-law model over thermal models. This conclusion aligns with previous investigations of the western lobe using joint \textit{ROSAT} and \textit{ASCA} data, as well as \textit{Chandra} data (e.g., \cite{1997ApJ...483..868S, 2005AdSpR..35.1062M, 2022PASJ...74.1143K} and references therein).

\subsubsection{XMM-Newton Counterparts to H.E.S.S. Analysis Regions} \label{sec:XMMHESS}
\begin{table*}[t]
    \caption{} \label{tab:xmm-hess} 
    \centering
    \begin{tabular}{lccc}
        \multicolumn{4}{c}{Fit Results of H.E.S.S.-matched Regions with \xmm\ Spectra.} \\ \hline
        Parameter & west\_low\_energy & west\_med\_energy & west\_high\_energy  \\ \hline
        $N_H$ (Frozen)  & $0.92 $& $0.92$  &  $0.92$       \\
        Photon Index $\Gamma$ & $2.03\pm0.02$ & $1.59\pm0.02$ & $1.38\pm0.04$        \\ 
        $F\left[\times10^{-12}~\mathrm{erg}~\mathrm{cm}^{-2}~\mathrm{s}^{-1}\right]$ (abs.)  & $4.6\pm0.13$ & $5.6\pm0.14$ & $3.8_{-0.16}^{+0.17}$         \\ 
        $F\left[\times10^{-12}~\mathrm{erg}~\mathrm{cm}^{-2}~\mathrm{s}^{-1}\right]$ (unabs.) & $7.3\pm0.18$ & $7.4\pm0.16$ & $4.7\pm0.17$ \\
        $\chi^2_\nu~(DoF)$ & 1.061 (1396) & 1.098 (1750) & 1.262 (756) \\ \hline
        $N_H$ (Thawed)  & $0.95\pm0.05$ & $1.19\pm0.06$  &  $2.39_{-0.13}^{+0.14}$       \\
        Photon Index $\Gamma$ & $2.06\pm0.05$ & $1.75\pm0.04$ & $1.63\pm0.08$        \\ 
        $F\left[\times10^{-12}~\mathrm{erg}~\mathrm{cm}^{-2}~\mathrm{s}^{-1}\right]$ (abs.)  & $4.6\pm0.13$ & $5.3\pm0.14$ & $3.5_{-0.15}^{+0.18}$         \\ 
        $F\left[\times10^{-12}~\mathrm{erg}~\mathrm{cm}^{-2}~\mathrm{s}^{-1}\right]$ (unabs.) & $7.4\pm0.18$ & $7.9\pm0.17$ & $5.1\pm0.18$ \\
        $\chi^2_\nu~(DoF)$ & 1.061 (1395) & 1.061 (1749) & 1.210 (755) \\ \hline
    \end{tabular}
\end{table*}

\begin{figure}[htb]
    \centering
    \includegraphics[width=\columnwidth]{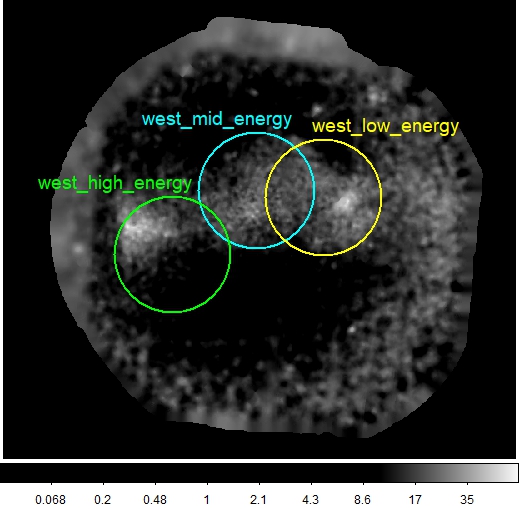}
    \caption{The three regions as defined in the auxiliary documents for the H.E.S.S. study \citep{hess_2024}, displayed on the \xmm\ image of the western lobe.}
    \label{fig:xmm-hess}
\end{figure}

As described in the Auxiliary Information for the H.E.S.S. study \citep{hess_2024}\footnote{\url{https://www.mpi-hd.mpg.de/hfm/HESS/pages/publications/auxiliary/2023_SS433/}}, X-ray fluxes were extracted from selected regions along the western jet shown in Figure~\ref{fig:xmm-hess}. To maintain consistency and to enable direct modeling and SED fitting later in Sections \ref{sec:Mod} and \ref{sec:SED}, spectra from the same regions were extracted from the \xmm\ data and analyzed with the same methods as the other regions in this work. Table~\ref{tab:xmm-hess} details the results for the power-law fits used for these H.E.S.S.-matched regions including the photon indices, fluxes, and reduced chi-square values. In Section~\ref{sec:Disc}, the correlation between the X-ray and TeV studies is discussed.

\section{XMM-Newton Spatially Resolved Spectroscopy via Contbin} \label{sec:Analysis-XSRSvC}
\subsection{Western Contbin Analysis}
\begin{figure*}[htb]
    \centering
    \begin{subfigure}[t]{0.475\textwidth}
        \centering
        \includegraphics[width=\textwidth]{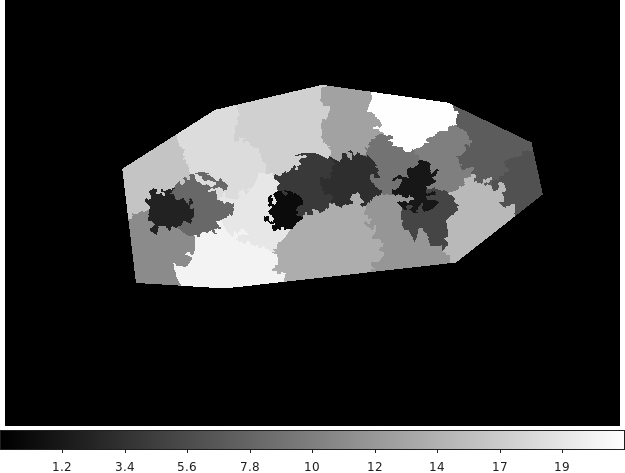}
        \caption{Contbin binmap regions.} \label{fig:contbinA}
    \end{subfigure}
    \hfill
    \begin{subfigure}[t]{0.475\textwidth}
        \centering
        \includegraphics[width=\textwidth]{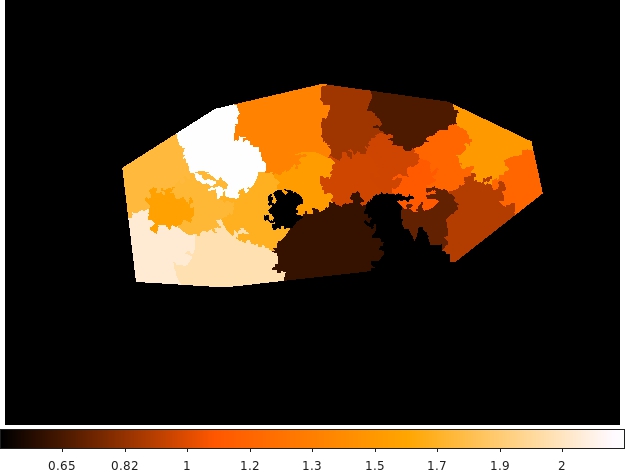}
        \caption{Fit $N_H$ values.} \label{fig:contbinB}
    \end{subfigure}
    \begin{subfigure}[t]{0.475\textwidth}
        \centering
        \includegraphics[width=\textwidth]{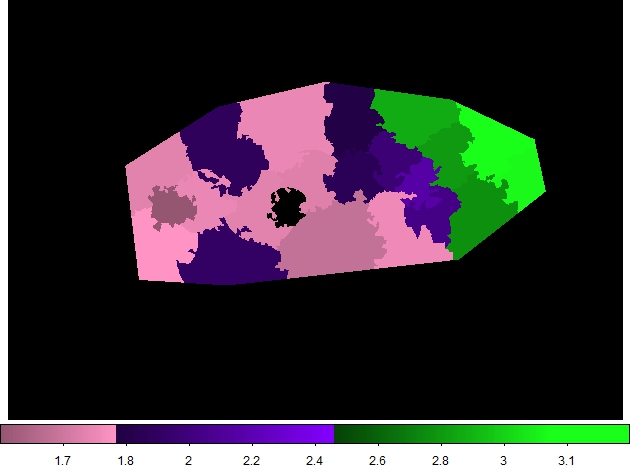}
        \caption{$\Gamma$ when $N_H$ is allowed to vary.} \label{fig:contbinC}
    \end{subfigure}
    \hfill
    \begin{subfigure}[t]{0.475\textwidth}
        \centering
        \includegraphics[width=\textwidth]{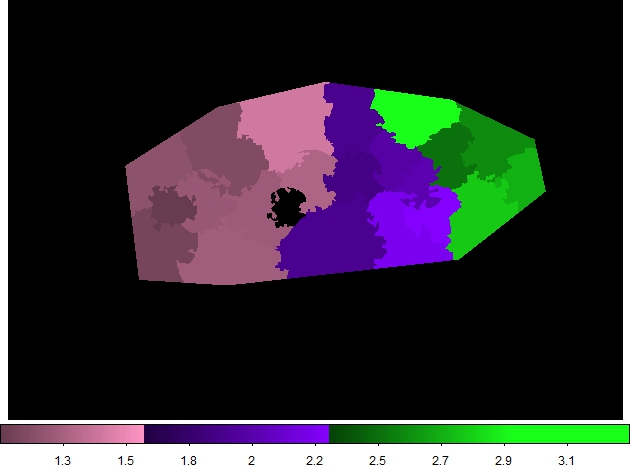}
        \caption{$\Gamma$ with $N_H$ frozen to an average value of $0.92\times10^{22}$cm$^{-2}$.} \label{fig:contbinD}
    \end{subfigure}
    \begin{subfigure}[t]{0.475\textwidth}
        \centering
        \includegraphics[width=\textwidth]{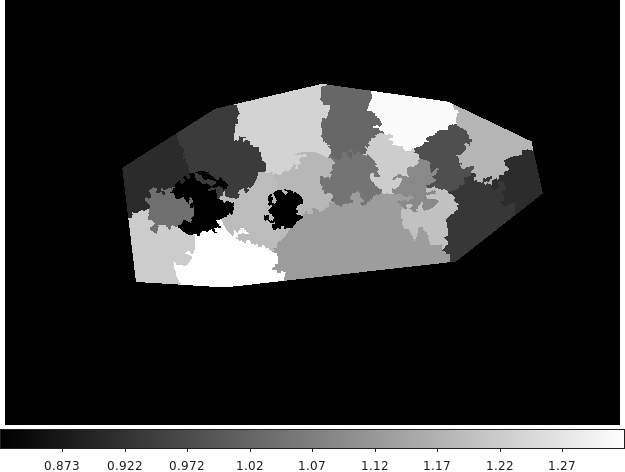}
        \caption{$\chi^2_\nu$ when $N_H$ is thawed.} \label{fig:contbinE}
    \end{subfigure}
    \hfill
    \begin{subfigure}[t]{0.475\textwidth}
        \centering
        \includegraphics[width=\textwidth]{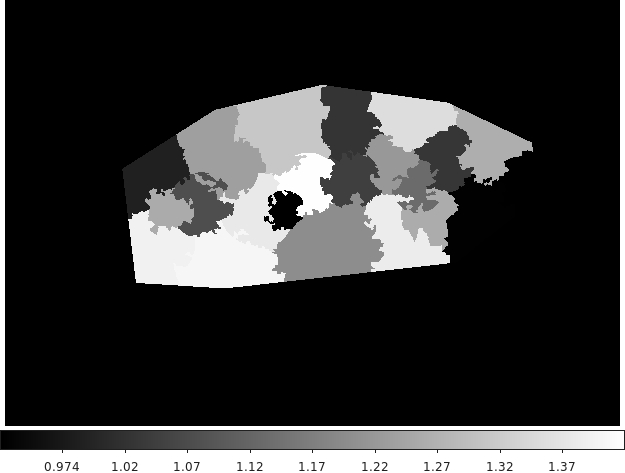}
        \caption{$\chi^2_\nu$ when $N_H$ is frozen.} \label{fig:contbinF}
    \end{subfigure}

    \caption{{Western lobe \tt{Contbin}} regions and results of fitting with a power-law model. Note that region ``0" contains a bright point-source and has been excluded from the fits in Figures~\ref{fig:contbinB}-\ref{fig:contbinF}. [Top] Images depict the photon index across all regions within ``XMM Full" as selected by {\tt{Contbin}}. Values are listed along the color-bar at the bottom and colored red, green, and blue to illustrate the spectral steepening along the jet. [Middle] The reduced chi-square values of the corresponding fits illustrate acceptable fits across the entire jet. [Bottom Left] Column density values are shown to vary significantly across the jet when allowed to vary during fitting. [Bottom Right] Illustration of the regions selected by {\tt{Contbin}} with integer numbering from dark-to-light.} \label{fig:contbin}
\end{figure*}

In order to further examine the spatial properties of W50's western lobe, the adaptively smoothed binning program, {\tt{Contbin}}, was used \citep{2006MNRAS.371..829S}. This software enables the automation of spatial region selection for X-ray spectroscopy by following contours on a smoothed and masked image of the object. This approach allows the entire lobe to be studied, and divided into regions that follow its morphology rather than imposing arbitrary geometric boundaries. The selected regions are intended to contain comparable numbers of counts and are based on surface brightness, given the broadly observed correlation between spectral properties and surface-brightness variations. The results can be seen in Figure~\ref{fig:contbinA}, which displays the 21 binmap regions output from the software. {\tt{Contbin}} has been used to separate the entire ``XMM Full" region into smaller resolved areas while maintaining the detail and morphology within. This thus allows previously discussed properties of the jet structure, such as the variation of $N_H$ and Photon Index $\Gamma$, to be examined continuously across the region.

First, Figure~\ref{fig:contbinB} demonstrates the variation in $N_H$ as discussed in Section~\ref{sec:Analysis-SRS}. All regions were fit with the same ({\tt{const}}{$\ast$}{\tt{tbabs}}{$\ast$}{\tt{power}}) model as used previously, showing the column density clearly decreases westward along the jet and in the regions north and south of the main emission regions. Figure~\ref{fig:contbinC}~and~\ref{fig:contbinD} illustrate the photon index with $N_H$ thawed, and frozen to an average value of $0.92\times10^{22}$~cm$^{-2}$, respectively. Both images show the same spectral steepening of the jet westward, with the triangular head being hardest (note the only two regions shown in green of \ref{fig:contbinC} are two of the three regions that contain the least emission from the region). $\chi^2_\nu$ plots of Figure~\ref{fig:contbinE}~and~\ref{fig:contbinF} are included to show the relatively good fits across all regions, as well as the comparatively small change when freezing the column density parameter.

Thus, the {\tt{Contbin}} region selection used to fit the entire ``XMM Full" region for fine spatially resolved spectroscopy finds agreement with the conclusions from the discrete regions above. The column density is found to vary significantly across the region, decreasing in general from east-to-west. The steepening of the photon index along the jet as previously noted is exemplified by the continuous region selection, also showing the pattern holds regardless of the column densities being frozen or thawed. With deep enough observation and high enough counts, {\tt{Contbin}} is shown to be a powerful tool for fine spatial region selection and its ability to provide a continuous view of the system.

\subsection{Eastern Contbin Comparison}
To enable a direct comparison between the two jet lobes of W50, analogous X-ray {\tt{Contbin}} analysis was performed for both the eastern and western jets. In both cases, spatial binning was restricted to the hard X-ray jet regions, excluding the surrounding thermal ``ear" structures  (referred to as ``e3" in the east and the ``shell" in the north).

\begin{figure*}[htb]
    \centering
    \begin{subfigure}[t]{0.475\textwidth}
        \centering
        \includegraphics[width=\textwidth]{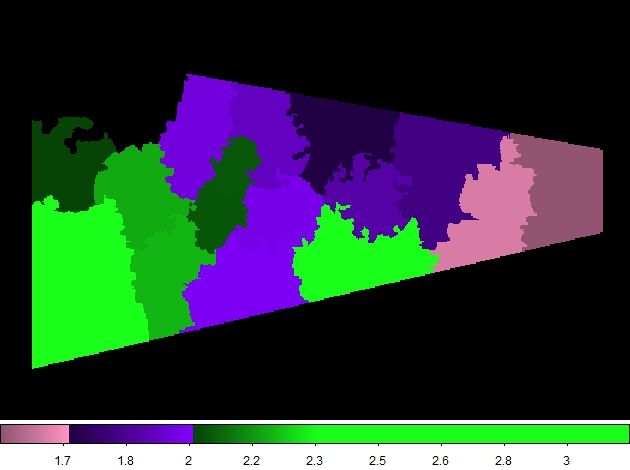}
        \caption{$\Gamma$ when $N_H$ is allowed to vary.} \label{fig:EcontbinA}
    \end{subfigure}
    \hfill
    \begin{subfigure}[t]{0.475\textwidth}
        \centering
        \includegraphics[width=\textwidth]{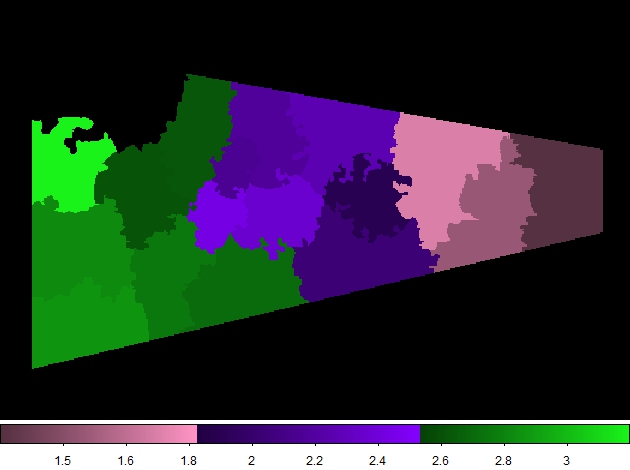}
        \caption{$\Gamma$ with $N_H$ frozen to an average value of $1.03\times10^{22}$cm$^{-2}$.} \label{fig:EcontbinB}
    \end{subfigure}
    \begin{subfigure}[t]{0.475\textwidth}
        \centering
        \includegraphics[width=\textwidth]{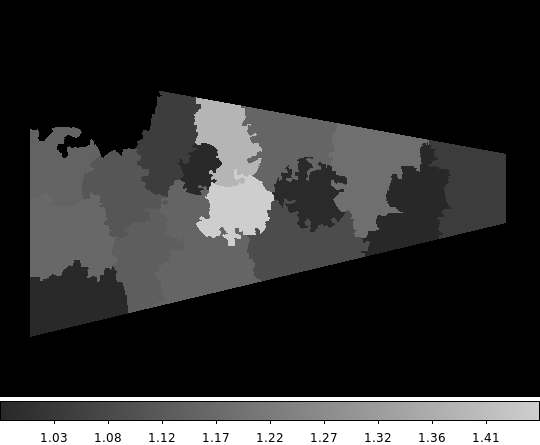}
        \caption{$\chi^2_\nu$ when $N_H$ is thawed.} \label{fig:EcontbinC}
    \end{subfigure}
    \hfill
    \begin{subfigure}[t]{0.475\textwidth}
        \centering
        \includegraphics[width=\textwidth]{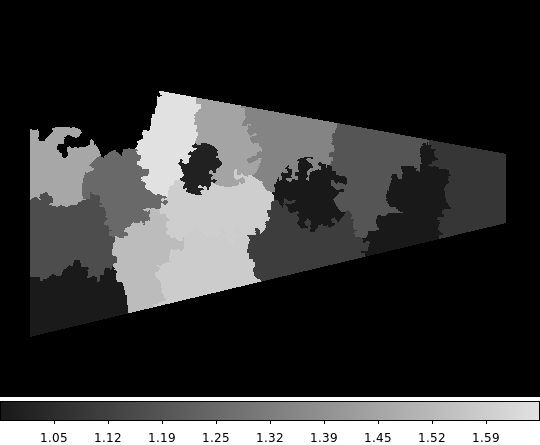}
        \caption{$\chi^2_\nu$ when $N_H$ is frozen.} \label{fig:EcontbinD}
    \end{subfigure}
    \begin{subfigure}[t]{0.475\textwidth}
        \centering
        \includegraphics[width=\textwidth]{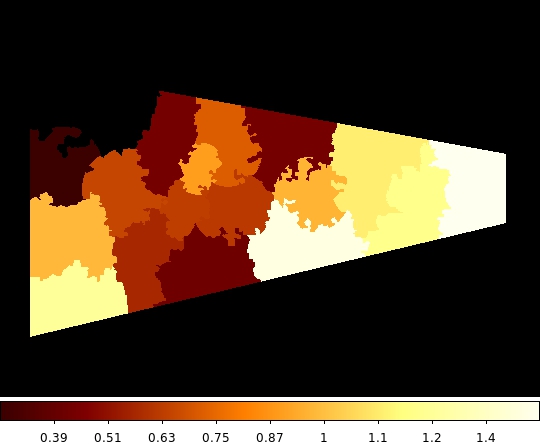}
        \caption{Fit $N_H$ values.} \label{fig:EcontbinE}
    \end{subfigure}
    \hfill
    \begin{subfigure}[t]{0.475\textwidth}
        \centering
        \includegraphics[width=\textwidth]{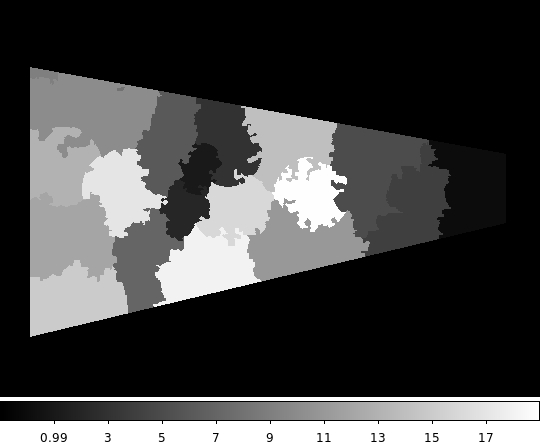}
        \caption{Contbin binmap regions.} \label{fig:EcontbinF}
    \end{subfigure}
    \caption{{Eastern lobe \tt{Contbin}} regions and results of fitting with a power-law model. Note that region ``0" contains a bright point-source and has been excluded from the fits in Figure~\ref{fig:EcontbinA}-\ref{fig:EcontbinE}. [Top] Images depict the photon index across all regions as selected by {\tt{Contbin}}. Values are listed along the color-bar at the bottom and colored red, green, and blue to illustrate the spectral steepening along the jet. [Middle] The reduced chi-square values of the corresponding fits illustrate acceptable fits across the entire jet. [Bottom Left] Column density values are shown to vary significantly across the jet when allowed to vary during fitting. [Bottom Right] Illustration of the regions selected by {\tt{Contbin}} within the cone region masking for the extent of \xmm\ X-ray jet emission, with integer numbering from dark-to-light.} \label{fig:contbinEast}
\end{figure*}

\begin{figure*}[htb]
    \begin{subfigure}[t]{0.475\textwidth}
        \includegraphics[width=\textwidth]{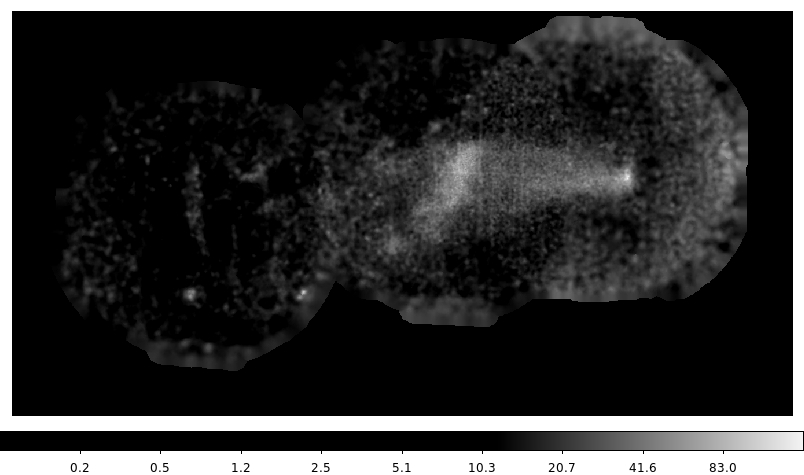}
        \label{fig:East3obs}
    \end{subfigure}
    \hfill
    \begin{subfigure}[t]{0.475\textwidth}
        \includegraphics[width=\textwidth]{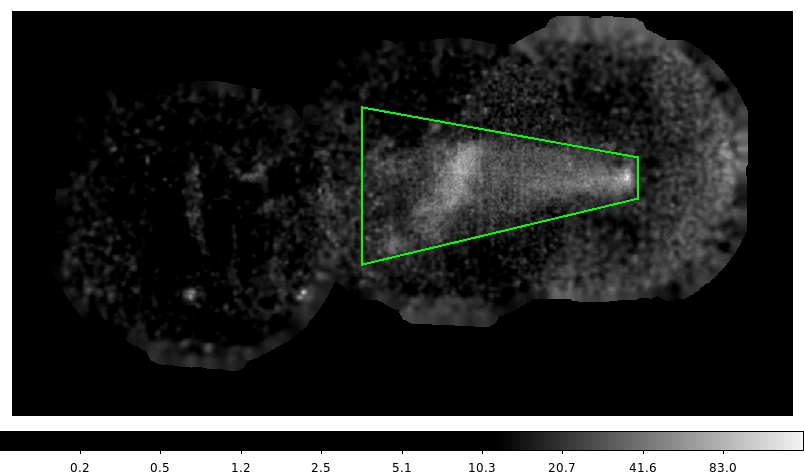}
        \label{fig:East3obsCone}
    \end{subfigure}
    \caption{[Left] depicts the three \xmm\  observations used for the Contbin analysis of the Eastern jet over 0.5--10.0~keV. (0840490101, 0075140401, 0075140501). [Right] The region selected to use as a mask for Contbin analysis. This cone was selected to include the X-ray jet, and exclude any other fainter contamination or sources, such as the thermal emission in the e3, ear.} \label{fig:contbinEastMask}
\end{figure*}

In the eastern lobe seen in Figure~\ref{fig:contbinEast}, a cone-shaped region was selected due to the symmetric appearance of the eastern jet as shown in Figure~\ref{fig:contbinEastMask}, while a polygonal region was used for the western lobe to match its more irregular morphology. In both regions, the {\tt{Contbin}} maps were generated to trace jet-structured features at comparable signal-to-noise thresholds.

Both lobes show a gradual steepening (increase) of the photon index ($\Gamma$) as one moves away from SS~433. In both east and west, the jet ``Head'' regions exhibit hard photon indices ($\Gamma \approx 1.5$), which steadily increase to $\Gamma > 2.0$ at the jet termination. This trend supports the interpretation that both jets undergo similar reacceleration and cooling processes despite differences in local environment and morphology.
Additionally, the hydrogen column density distribution is qualitatively consistent between lobes: higher column densities are observed near SS~433, decreasing outward. As suggested for the eastern lobe study, this could be evidence of jet entrainment, which would contribute to mixing and turbulence impacting the shock acceleration process, e.g., through internal shocks \citep{2013A&A...558A..19W, Safi-Harb_2022}. Furthermore, this trend in both the eastern and western lobes is consistent with the large-scale gaseous environment study with both CO and HI revealing concentrations of molecular clouds or clumps aligned with the jet precession geometry on large scales and supporting evidence of interaction with nearby molecular gas \citep{2018ApJ...863..103S, 2023PASJ...75..338S}. Finally, the  slightly overall higher value of the column density in the western lobe can be attributed to its proximity to the Galactic Plane and an overall denser environment than in the east.

Together, these results reinforce the jet-driven nature of W50's high-energy emission and large-scale morphology further impacted by interaction with the interstellar medium. While some asymmetries exist in morphology (see Fig.~\ref{fig:mltwvl}), the underlying spectral behavior and evolution are largely symmetric.

These spatially resolved constraints on photon index, column density, and morphology in both lobes provide a natural basis for testing simplified jet models. The following Section~\ref{sec:Mod} introduces a semi-analytic framework designed to reproduce the broadband emission and surface-brightness profiles along the western jet, which are subsequently used to model the spectral energy distribution (SED) of the system (Section~\ref{sec:SED}).

\section{Jet Models and Their Implications on the Spectral-Energy-Distribution and X-ray Morphology}  \label{sec:SED}

X-ray observations of sources that feature sub-milli-Gauss magnetic fields probe the distribution of multi-TeV electrons. While these electrons generate TeV $\gamma$-rays through the inverse Compton (IC) channel, the superior angular resolution of X-ray instruments allows the internal structure of these sources, particularly the acceleration site, to be resolved and studied in great detail. 

The structure of $\gamma$-ray sources is governed by the location of the acceleration sites, particle transport, and the efficiency of the radiation channels. The interplay and balance of these processes define the spectral and morphological properties of the emission. In the case of SS~433, the consensus scenario interprets the non-thermal X-ray lobes as synchrotron emission from efficiently accelerated electrons in the innermost regions of the lobes, followed by particle advection away from the acceleration sites and cooling through synchrotron, IC, and adiabatic processes. In \ref{sec:Mod}, a simplified semi-analytic approach is introduced to pursue a detailed modeling study of the non-thermal emission from the jet.

This treatment is closely related to the general framework of \citet{2009ApJ...703..662R}, who showed that synchrotron-loss breaks in inhomogeneous outflows depend on the spatial evolution of flow-tube radius, magnetic field strength, density, and velocity, rather than following only the canonical homogeneous-source steepening. The present calculation applies this physical point to the specific geometry of SS~433/W50: the observed X-ray steepening is not treated as a purely radiative cooling diagnostic, but as the combined result of synchrotron cooling, advection, magnetic field evolution, and source dynamical and geometric properties.

While there is little doubt about this general scenario, the essential characteristics of the jets in SS~433 remain unconstrained in the region where the observed X-ray emission is generated. In the first place the uncertainties concern the spatial dependence of the jet speed, evolution of the magnetic field, and possible in-situ acceleration processes operating in the jet flow downstream from the main accelerator.   

The velocity relevant for the large-scale X-ray-emitting flow is not directly measured. Spectroscopic measurements constrain the inner baryonic jets of SS~433 to mildly relativistic speeds, with recent \textit{XRISM} and simultaneous optical observations finding jet speeds fluctuating around \(0.26\pm0.01c\) and \(0.30\pm0.01c\) in different epochs \citep{Sakai2026}. These measurements apply close to the binary and do not directly determine the advective speed of the particle-carrying flow tens of parsecs from SS~433. Conversely, the first \chandra\ proper motion search of the eastern and western X-ray knots found no statistically significant motion over a baseline of approximately 20 years, placing upper limits of \(<0.019\)--\(0.033c\) on the apparent knot speeds and favoring a standing recollimation-shock interpretation \citep{2025ApJ...993L..24T}.

More generally, jet systems can sustain fast internal flow while the jet head advances much more slowly, since the large-scale propagation is governed by momentum balance with the ambient medium and the evolution of the shocked cocoon \citep{1992ApJ...392..458C}. This provides a natural basis for expecting sub-relativistic speeds on large scales even when the inner jet remains mildly relativistic.
The velocity \(v(z)\) used below should therefore be interpreted as a phenomenological transport speed of the emitting plasma, not as the observed proper motion of the X-ray knots themselves.
Since some illustrative models require mildly relativistic transport speeds, Doppler boosting is included in the synthetic surface-brightness calculation described in Appendix~\ref{sec:Mod}; however, it is not invoked as the primary origin of the hard X-ray morphology, which is instead controlled mainly by particle transport, magnetic field evolution, and local acceleration.

There are two very general scenarios, which one can link to large-scale outflows: jets with constant speed (e.g., unperturbed supersonic outflows) and jets with constant density (typical for subsonic, weakly magnetized flows). Real outflows might deviate significantly from these models, especially when one considers extended regions or flows in which magnetic fields play an important dynamic role. However, these two limiting cases can serve as illustrative toy models.

Another characteristic critical for the non-thermal emission is the strength of the magnetic field and its evolution. As model cases one can adopt two different orientations of the magnetic field: toroidal and poloidal, which imply different radial evolution of the field strength. Although both field orientations are explored as limiting cases, the interpretation favored below is a hybrid case, as discussed further in Section~\ref{sec:bejm}. Radio polarimetry can provide projected magnetic field information, but it does not by itself uniquely separate toroidal and poloidal components in the W50 lobes because projection, Faraday effects, and shell/lobe superposition complicate the interpretation. Synthetic spectra and X-ray morphology for four combinations (Models A, B, C, and D, see \ref{sec:Mod} for details) of the jet speed and magnetic field orientation are presented  in \ref{sec:Mod}. The obtained results can be qualitatively understood based on order-of-magnitude estimates presented below. 
 
The jet speed and its spatial dependence determine the age, \(T\), of the  particle population producing the detected emission. For dominant advection transport, the age is the advection time between the acceleration site at \(Z\mysub{min}\approx26.6\)~pc and the extension of the region of interest, \(Z\mysub{max}\) (for the case of SS~433 one has \(Z\mysub{max}\lesssim3Z\mysub{min}\)). For a constant speed jet, which propagates with \(v_0\),  one obtains
\begin{equation}
  \begin{split}
    T &= \frac{Z\mysub{max}-Z\mysub{min}}{v_0}\\
      &\approx 1.7 \qty(\frac{v_0}{0.1c})^{-1}\unit{kyr}
  \end{split}
\end{equation}
and for a constant density case, i.e., \(v(z) = v_0 \qty(Z\mysub{min}/Z)^2\), the advection time is longer:  
\begin{equation}
  \begin{split}
    T &= \frac{Z\mysub{max}^3-Z\mysub{min}^3}{3Z\mysub{min}^2v_0}\\
      &\approx 7 \qty(\frac{v_0}{0.1c})^{-1}\unit{kyr}\,.
  \end{split}
\end{equation}
The difference in these timescales is noticeable, and can influence the position of the cooling break and thus the
spectral shape. However, the magnetic field strength and its evolution may have even a stronger impact on the spectral
shape. If the magnetic field is poloidal, then its evolution does not depend on the jet speed, \(B_z\propto Z^{-2}\). The electron cooling break is then approximately defined by \(t\mysub{syn}(E,B_0) = Z\mysub{min}/\zeta v_0\), where \(E\) is the electron energy and
\(\zeta\approx 1\) and \(2.5\) for a constant-speed and constant-density jets, respectively. For \(Z\mysub{min}=26.5\)~pc
one obtains:
\begin{equation}
E\mysub{br}\approx 140 \qty(\frac{B_0}{10~\upmu\unit{G}})^{-2}\qty(\frac{v_0}{0.1\zeta c})\unit{TeV}\,,
\end{equation}
and the corresponding photon-energy break in the synchrotron component should appear at
\begin{equation}
  \hbar\omega\mysub{br}\approx 12 \qty(\frac{B_0}{10~\upmu\unit{G}})^{-3}\qty(\frac{v_0}{0.1\zeta c})^2\unit{keV}\,.
\end{equation}
If the dominant magnetic field is toroidal, then its evolution is strongly affected by the speed profile. For the constant-speed case, the toroidal field decreases as \(B \propto Z^{-1}\), and the position of the cooling break should be similar to the case of a poloidal magnetic field. If the density is constant, then the magnetic field increases with distance, \(B\propto Z\) (approximately to the distance where the strength reaches the equipartition to the plasma pressure value). If the initial jet magnetization is small, \(\sigma\lesssim 0.1\), then the linear increase of the magnetic field occurs in the entire region of interest. Thus, the position of the cooling break is approximately defined as \(t\mysub{syn}(E,B_0) = Z\mysub{min}/ v_0\qty(Z\mysub{max}/Z\mysub{min})^5\) and one obtains
\begin{equation}
E\mysub{br}\approx 0.7 \qty(\frac{B_0}{10~\upmu\unit{G}})^{-2}\qty(\frac{v_0}{0.1 c})\unit{TeV}\,,
\end{equation}
and the corresponding spectral break in the synchrotron component should appear at
\begin{equation}
  \hbar\omega\mysub{br}\approx 0.3 \qty(\frac{B_0}{10~\upmu\unit{G}})^{-3}\qty(\frac{v_0}{0.1 c})^2\unit{eV}\,.
\end{equation}

For the photon fields expected in the jets of SS~433 \citep[for details see, e.g.,][]{hess_2024}, the interaction of TeV electrons proceeds predominantly in the Thomson regime with Cosmic Microwave Background Radiation (CMBR) and Far-infrared radiation (FIR) photon fields, while for the Near-Infrared  regime, the Klein-Nishina cutoff might be important, especially for \(100\)~TeV electrons. Also, for hard electron distributions the role of lower temperature photon fields is higher \citep{1997MNRAS.291..162A}, thus the ratio of the synchrotron and IC components can be estimated as \({L\mysub{syn}}/{L\mysub{IC}}\approx 10 \qty(\frac{B_0}{10\unit{\upmu G}})^2\). 
However, this relation ignores the evolution of the magnetic field and the impact of the adiabatic losses. In constant speed jets, these two factors approximately cancel each other, thus one obtains:
\begin{equation}
  \frac{L\mysub{syn}}{L\mysub{IC}}\approx 10 \qty(\frac{B_0}{10\unit{\upmu G}})^2\,.
\end{equation}
If the jet density is constant (i.e., no strong adiabatic cooling) and the magnetic field is poloidal (i.e. it decreases with distance), then this ratio is reduced by a factor of \(\sim10(Z\mysub{min}/Z\mysub{max})^3\approx0.5\) due to a longer propagation time through the jet and weakening of the magnetic field. In a constant density jet with toroidal magnetic field, the situation changes dramatically because of increasing strength of the magnetic field:   
\begin{equation}
  \frac{L\mysub{syn}}{L\mysub{ic}}\approx 2.5 \qty(\frac{B_0}{10\unit{\upmu G}})^2\qty(\frac{Z\mysub{max}}{Z\mysub{min}})^2\,,
\end{equation}
where Thomson scattering is also accounted for on the NIR photon field, provided that the cooling break appears in the sub-TeV domain.

The above relations determine the basic properties of the synchrotron and IC components and allow one to infer the key jet parameters needed to reproduce the broadband spectral detected from the jet. It might be convenient to define the strength of the magnetic field via a phenomenological magnetization parameter:
\begin{equation}
  \frac{B_0^2v_0R\mysub{min}^2}{4}=\sigma L\mysub{j}\,,
\end{equation}
Thus one obtains
\begin{equation}
  B_0\approx70\qty(\frac{0.1 c\sigma}{v_0})^{1/2}\unit{\upmu G}\,,
\end{equation}
where \(L\mysub{j}\) was set to \(10^{39}\unit{erg\,s^{-1}}\) and \(R\mysub{min}\approx5.6\)~pc.

\begin{figure*}[htb]
    \begin{subfigure}[t]{0.475\textwidth}
        \includegraphics[width=\textwidth]{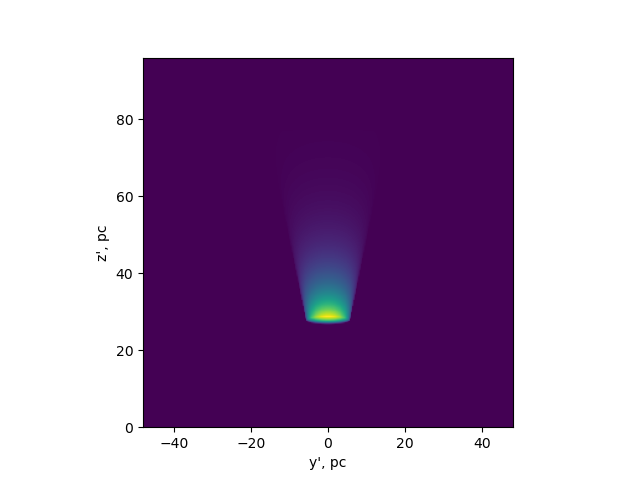}
        \caption{Model A: a constant speed jet carrying toroidal magnetic field.} 
    \end{subfigure}
    \hfill
    \begin{subfigure}[t]{0.475\textwidth}
        \includegraphics[width=\textwidth]{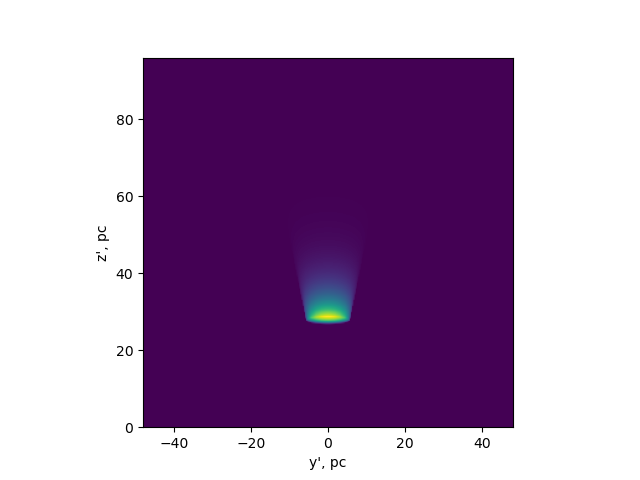}
        \caption{Model B: a constant speed jet carrying poloidal magnetic field.} 
    \end{subfigure}
    \begin{subfigure}[t]{0.475\textwidth}
        \includegraphics[width=\textwidth]{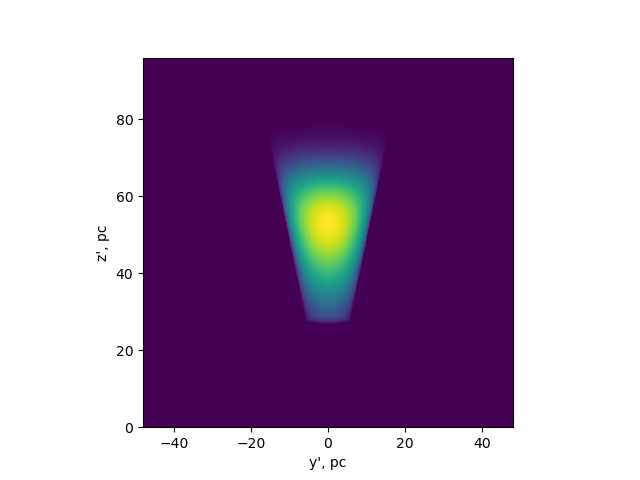}
        \caption{Model C: a constant density jet carrying toroidal magnetic field.}
      \end{subfigure}
    \hfill
    \begin{subfigure}[t]{0.475\textwidth}
        \includegraphics[width=\textwidth]{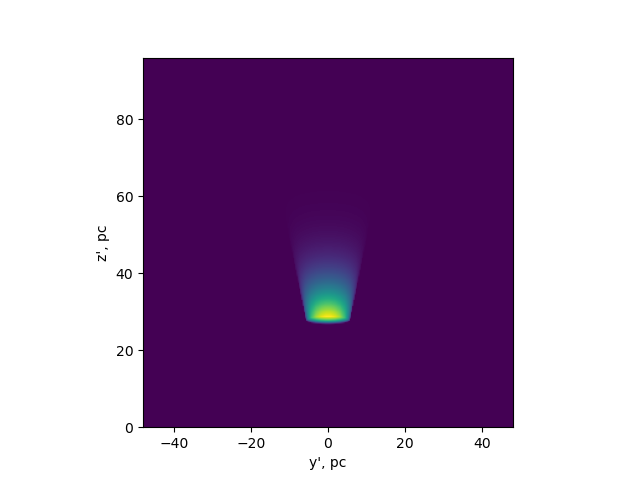}
        \caption{Model D: a constant density jet carrying poloidal magnetic field.}
    \end{subfigure}
    \begin{subfigure}[t]{0.475\textwidth}
        \includegraphics[width=\textwidth]{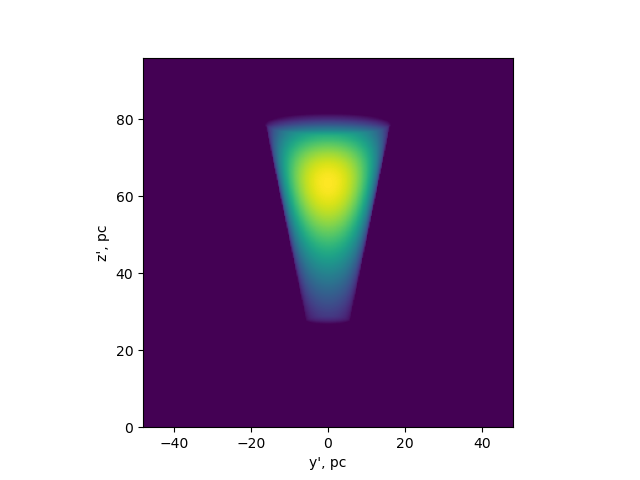}
        \caption{Model C hard: a jet similar to Model C with a harder adopted injection spectrum (\(\alpha_e=2\rightarrow 1.6\)).}
      \end{subfigure}
    \hfill
    \begin{subfigure}[t]{0.475\textwidth}
        \includegraphics[width=\textwidth]{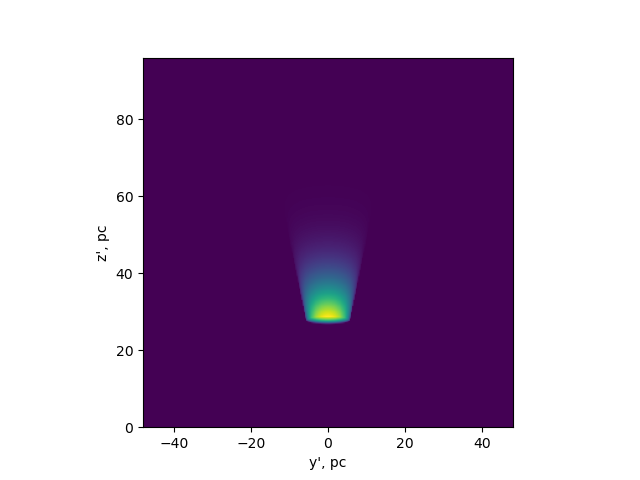}
        \caption{Model D hard: a jet similar to Model D with a harder adopted injection spectrum (\(\alpha_e=2\rightarrow 1.6\)).} 
    \end{subfigure}
    \caption{Synchrotron surface brightness at \(3\)~keV.} \label{fig:sb_west}
\end{figure*}

The above consideration demonstrates that it is possible to achieve at least a qualitative agreement between the
broadband Spectral-Energy-Distribution (SED) and observational data obtained in X- and $\gamma$-ray bands for different models for the jet. These models
however imply very different X-ray morphology as shown in Fig.~\ref{fig:sb_west}. The demonstrated cases correspond to four models introduced in \ref{sec:Mod}. These calculations emphasize the importance of
morphological studies of Galactic jet sources, and X-ray observations allow obtaining essential information about the
structure of such jets. 

The surface brightness maps shown in Fig.~\ref{fig:sb_west} demonstrate the strong impact of the
magnetic field strength evolution on the synchrotron emissivity: in Models A (i.e., constant jet speed toroidal magnetic field), B (i.e., constant jet speed poloidal magnetic field), and D (i.e., constant density speed poloidal magnetic field), the magnetic field strength rapidly
decreasing with distance, which causes the dimming of the emission, clearly seen in the figure. In the case of Model C (i.e., constant jet density toroidal magnetic field),
the magnetic field strength increases with distance, causing brightening of the jet up to the distance of \(30\)~pc from
the accelerating shock. At larger distances the effects related to the particle cooling start to dominate, causing the
jet dimming.

\begin{figure*}[htb]
    \begin{subfigure}[t]{0.475\textwidth}
        \includegraphics[width=\textwidth]{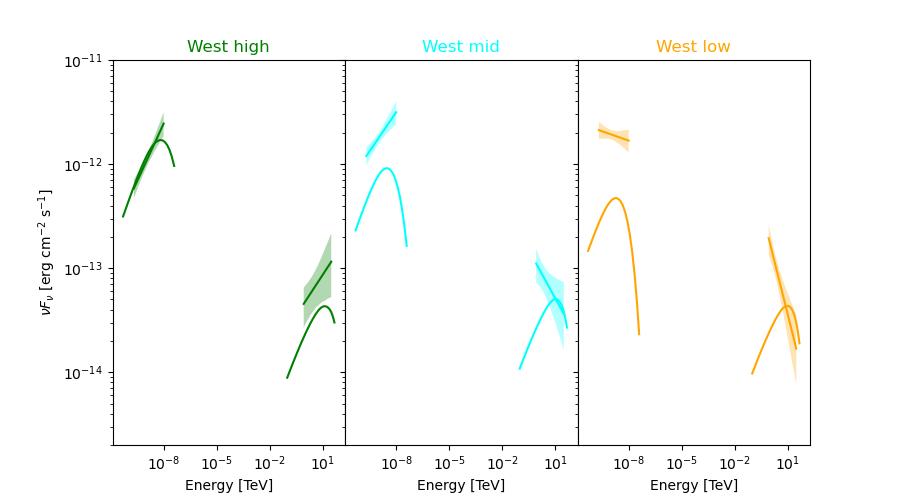}
        \caption{Model A: a constant speed jet carrying toroidal magnetic field.}
    \end{subfigure}
    \hfill
    \begin{subfigure}[t]{0.475\textwidth}
        \includegraphics[width=\textwidth]{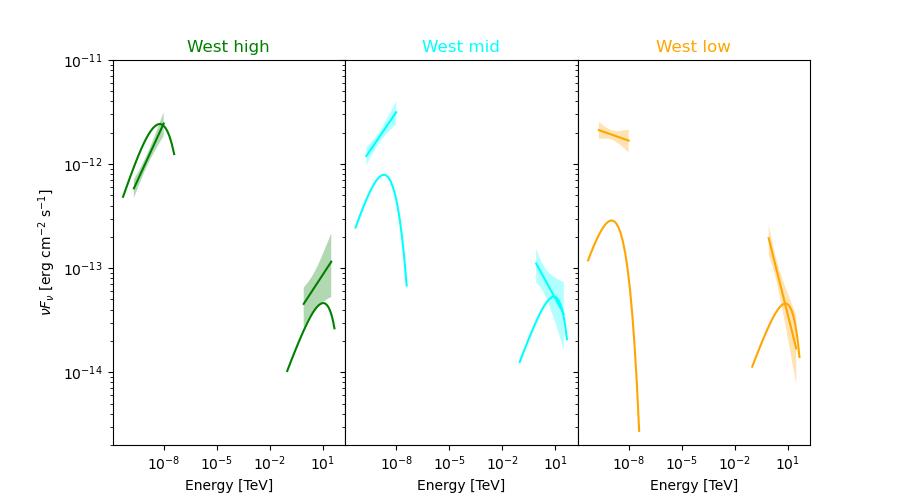}
        \caption{Model B: a constant speed jet carrying poloidal magnetic field.}
    \end{subfigure}
    \begin{subfigure}[t]{0.475\textwidth}
        \includegraphics[width=\textwidth]{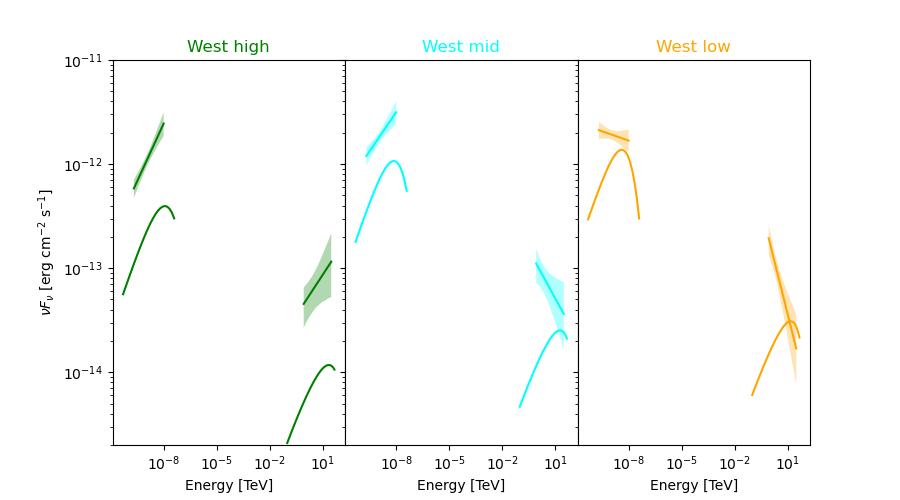}
        \caption{Model C: a constant density jet carrying toroidal magnetic field.}
    \end{subfigure}
    \hfill
    \begin{subfigure}[t]{0.475\textwidth}
        \includegraphics[width=\textwidth]{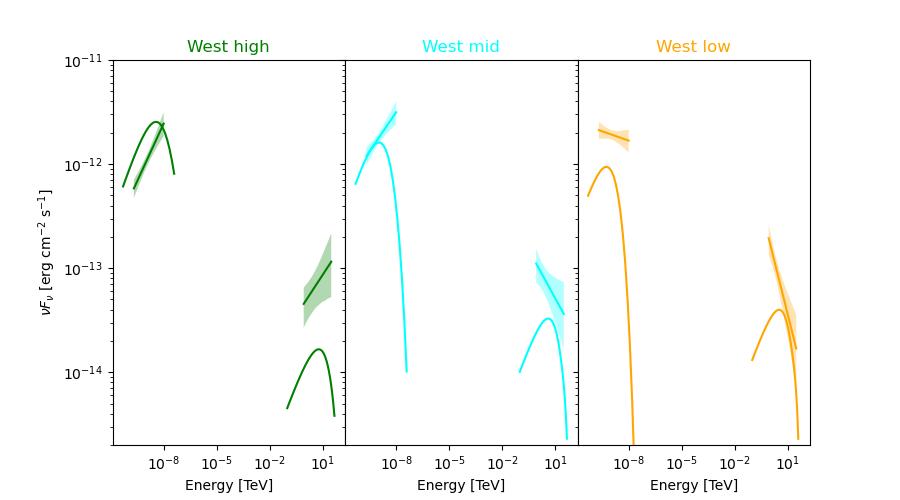}
        \caption{Model D: a constant density jet carrying poloidal magnetic field.}
    \end{subfigure}
    \begin{subfigure}[t]{0.475\textwidth}
        \includegraphics[width=\textwidth]{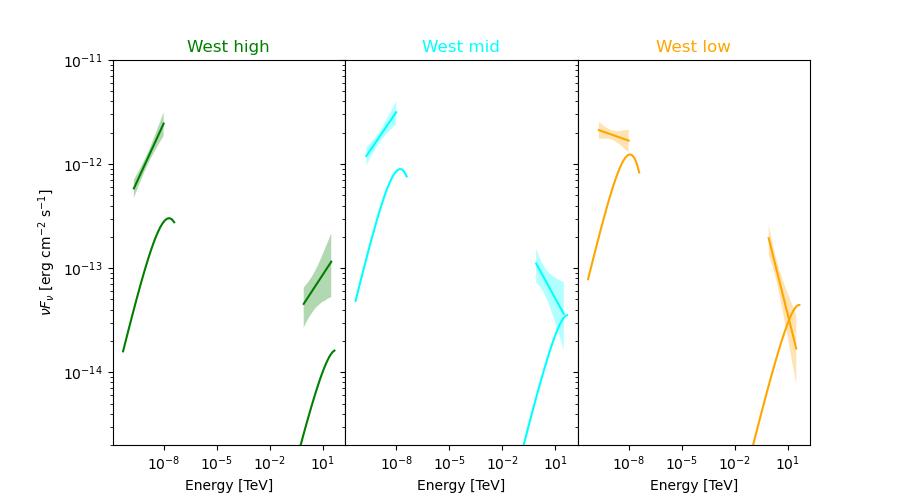}
        \caption{Model C hard: a jet similar to Model C with a harder adopted injection spectrum (\(\alpha_e=2\rightarrow 1.6\)).} 
    \end{subfigure}
    \hfill
    \begin{subfigure}[t]{0.475\textwidth}
        \includegraphics[width=\textwidth]{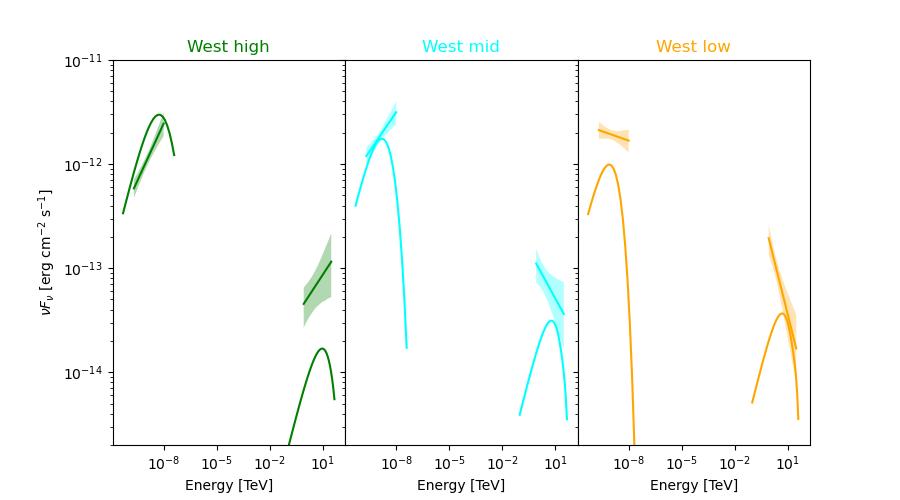}
        \caption{Model D hard: a jet similar to Model D with a harder adopted injection spectrum (\(\alpha_e=2\rightarrow 1.6\)).} 
    \end{subfigure}
    \caption{Comparison of narrow region SED for three regions defined in the western jet of SS~433 (left panel ``west high'', middle panel ``west mid'', and right panel ``west low'', see in Figure~\ref{fig:xmm-hess}).} \label{fig:regional_sed_west}
\end{figure*}

To compare the morphology predicted by models to the observational data,  the synchrotron and IC spectra are computed from three small regions in the western jet for which TeV spectra were extracted (see in Fig.~\ref{fig:xmm-hess}). The results of the calculations are shown in Fig.~\ref{fig:regional_sed_west}.  It can be seen that models with constant speed (``A'' and ``B'') underestimate the synchrotron emission at large distances from the acceleration site. This is caused by a combination of factors, such as adiabatic cooling, decrease of magnetic field strength and plasma density. Among these factors, the decrease of magnetic field strength does not influence the IC component, thus it should be the critical factor that changes the balance of the X-ray and TeV components.

The models with constant density (``C'' and ``D'') provide a better approximation to the jet's multiwavelength morphology. However, given the simplified framework adopted here, a close match is not expected, since additional effects such as enhancement of the magnetic field strength or particle reacceleration may occur in the jet. Recall that IXPE results \citep{2024ApJ...961L..12K} favor dominant poloidal magnetic field, which correspond to the models ``D'' and ``D hard'' (in which the jet speed and orientation of the magnetic field are the same as in model D, but the injection spectrum is assumed to be \(1.6\)) in these figures. These models also rather successfully reproduce the narrow-region SEDs.

\begin{figure*}[htb]
    \begin{subfigure}[t]{0.475\textwidth}
        \includegraphics[width=\textwidth]{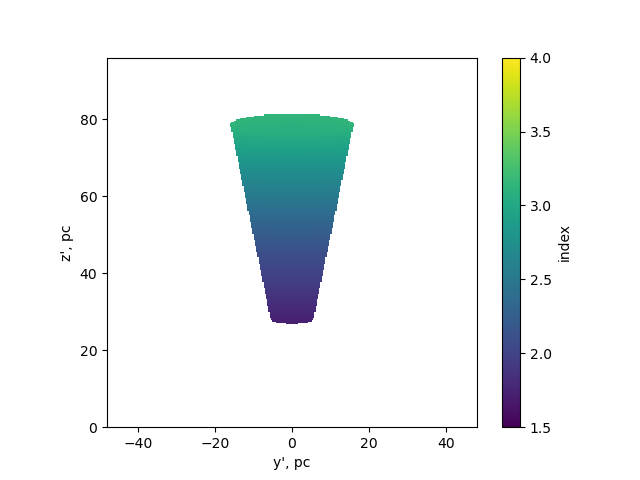}
        \caption{Model A: a constant speed jet carrying toroidal magnetic field.} 
    \end{subfigure}
    \hfill
    \begin{subfigure}[t]{0.475\textwidth}
        \includegraphics[width=\textwidth]{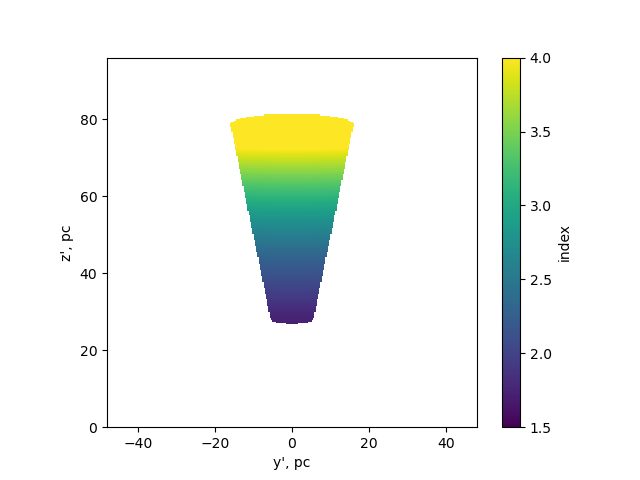}
        \caption{Model B: a constant speed jet carrying poloidal magnetic field.} 

    \end{subfigure}
    \begin{subfigure}[t]{0.475\textwidth}
        \includegraphics[width=\textwidth]{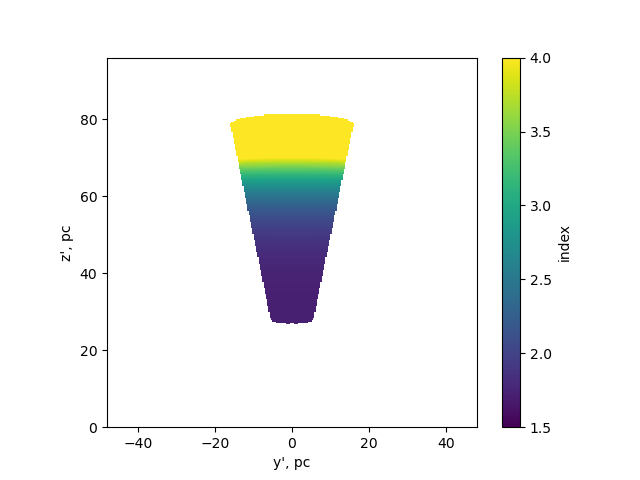}
        \caption{Model C: a constant density jet carrying toroidal magnetic field.}
        
    \end{subfigure}
    \hfill
    \begin{subfigure}[t]{0.475\textwidth}
        \includegraphics[width=\textwidth]{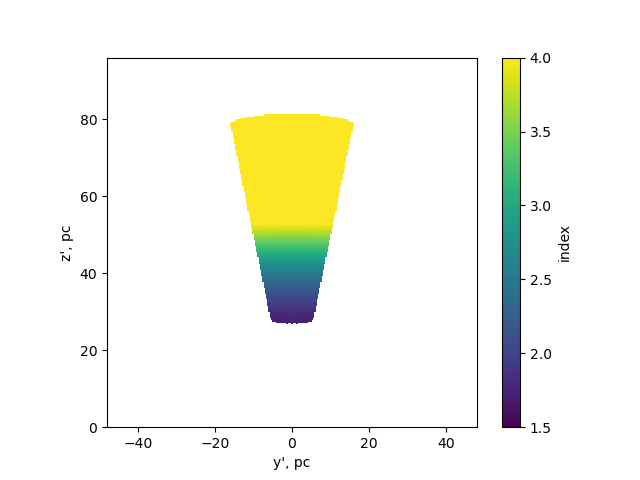}
        \caption{Model D: a constant density jet carrying poloidal magnetic field.}
    \end{subfigure}
    \begin{subfigure}[t]{0.475\textwidth}
        \includegraphics[width=\textwidth]{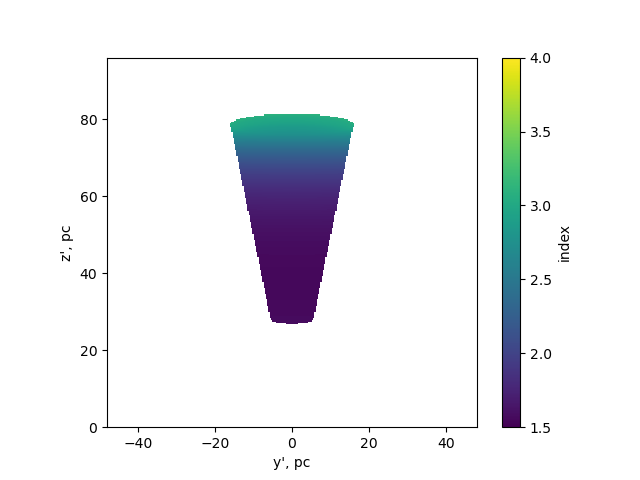}
        \caption{Model C hard: a jet similar to Model C with a harder adopted injection spectrum (\(\alpha_e=2\rightarrow 1.6\)).} 
      \end{subfigure}
    \hfill
    \begin{subfigure}[t]{0.475\textwidth}
        \includegraphics[width=\textwidth]{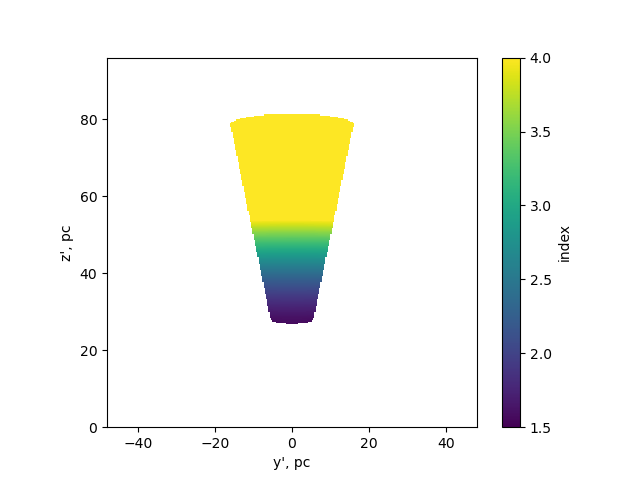}
        \caption{Model D hard: a jet similar to Model D with a harder adopted injection spectrum (\(\alpha_e=2\rightarrow 1.6\)).} 
    \end{subfigure}
    \caption{Spectral index of the synchrotron emission computed in a narrow range of photon energies around \(3\)~keV.} \label{fig:ss_west}
\end{figure*}

Finally, in Fig.~\ref{fig:ss_west} the synthetic maps for the photon index of the synchrotron emission at
photon index, between \(1.5\) and \(2.5\), which is consistent with dominant adiabatic cooling (which does not influence
the spectral shape), adopted fast jet speed (for slower jets the spectral steepening is more pronounced). Models with constant
density demonstrate a significantly stronger steepening of the X-ray spectrum, which is expected due to slower particle
transport. Comparison of Models C and D demonstrate that for the adopted parameters, the effects related to evolution of
magnetic field dominate over the effects related to the particle cooling.

\section{Discussion} \label{sec:Disc}

\subsection{Western Lobe Morphology and the Hard X-ray Head}

In Figure~\ref{fig:mltwvl}, the multi-wavelength perspective of W50-west is presented. 
As for the eastern lobe, the interior hard X-ray lobe is positioned within a region of diminished radio emission, creating a radio ``hole''. Additionally, the optical filaments partially coincide with the projection of the radio shell along the SS~433 jet axis. This combined view, coupled with a spatially resolved spectroscopic analysis, lends support to the notion that the ``Head'' region serves as an acceleration zone, marking the initiation of X-ray emission within the western lobe. The thermalization process of the jet seems to commence with its interaction with the SNR, notably in the region marked as ``w2'', just beyond the intersection of the extrapolated radio shell and the jet axis. Although not observed here, if continuation is assumed as in the eastern lobe, the thermalization process progresses to the point of the jet's termination shock with the interstellar medium, forming the western radio ``ear''.

Within this study, the comprehensive X-ray emissions of W50-west have been mapped, spanning from SS~433 to just shy of the western radio ``ear''. Through this investigation, it is seen that the most robust non-thermal X-ray emission originates in the innermost portion of the western lobe. Its peak is specifically discerned in the ``Head'' region, situated approximately $16\farcm6$ west of SS~433 (equivalent to roughly 26.5 pc at a distance of $d=5.5$~kpc from Earth). This emission showcases an unusually hard spectral index. With a photon index $\Gamma$ of approximately 1.5, it implies a particle index $p$ of around 2 $(\frac{dN}{dE}\sim$E$^{-p}; p=2\alpha+1 $, where $\alpha=\Gamma-1)$. As a result, an $E^{-2}$ distribution of electrons becomes evident. This index, very similar to that of the eastern counterpart,  actually aligns more closely with the indices observed in pulsar wind nebulae (PWNe) fuelled by the relativistic winds of neutron stars \citep{2003ApJ...591..361G, 2008AIPC..983..171K}, and some extra-galactic jets \citep{2003ASPC..300..151H}. It is noteworthy that X-ray knots within numerous active galactic nuclei (AGN) jets also exhibit photon indices $\Gamma$ around 1.1--1.6 \citep{2004ApJ...608..698S}.

\subsection{Particle Acceleration and Synchrotron Cooling}

Furthermore, it is essential to account for the cooling of such energetic electrons. The appearance of the cooling break or spectra emission generated in the fast cooling regime depend strongly on the realized jet model and magnetic field strength (see in Section~\ref{sec:SED}). If the jet plasma moves with a constant speed, of the order of the speed of light, then for \(\sim10\unit{\upmu G}\) magnetic field the cooling break should appear above \(1\unit{keV}\) energies. For a smaller jet speed, constant-density jet models, or stronger magnetic fields to manifest an $E^{-2}$-type spectrum, an extraordinarily hard differential injection spectrum—harder than $E^{-1}$—would be required. For constant-speed jet models, adiabatic losses could potentially overshadow synchrotron losses (see in Section~\ref{sec:SED}), similar to the mechanism suggested for explaining the hard spectra of blazars \citep{2011ApJ...740...64L}. Interestingly, the index undergoes a steepening to $E^{-3}$ (for $\Gamma$ around 2) in the more distant regions as the jet undergoes thermalization. This leads to the possibility that particles are accelerated within the ``Head'' region following an $E^{-2}$ spectrum. While within the accelerator, the spectrum remains unaltered (due to adiabatic losses, decay of the magnetic field, or fast advection), electrons, upon exiting the source, encounter synchrotron losses. These losses naturally account for the increasing steepness of the X-ray spectrum in the ``Diffuse'' and ``w2'' regions.

Alternatively, an acceleration mechanism like magnetic reconnection or stochastic acceleration could be in play. Such mechanisms could yield an exceedingly hard (Maxwellian-type) acceleration spectrum \citep{2014ApJ...783L..21S, 2015ASSL..407..311L}, which, after synchrotron cooling, converges to $E^{-2}$ (irrespective of the acceleration spectrum). It is noteworthy that the column density attains its maximum in the ``Head'' region (as discussed in Sections~\ref{sec:Analysis-SRS}~and~\ref{sec:Analysis-XSRSvC}). This may potentially signify jet entrainment, introducing an element of mixing and turbulence that could influence the shock acceleration process, potentially through internal shocks \citep{2013A&A...558A..19W}.

To delve further into the described scenario of the hard-spectrum acceleration and transport picture in this region, it is important to explore the synchrotron emission characteristics of the ``Head'' region, which signifies the innermost portion within the western lobe, characterized by its remarkably hard photon index of 1.55. Assuming that there is an equal distribution of energy between the  magnetic field and relativistic particles (both electrons and protons, which carry approximately 100 times more energy than electrons), and a volume of approximately $6.34\times10^{56}$ cm$^3$, the observed X-ray luminosity of $5.28\times10^{33}$~erg~s$^{-1}$ (within the 0.5--30~keV range) leads to an estimated equipartition magnetic field value $B$ of around $15~\mu$G in the context of synchrotron emission interpretation (e.g., \citealt{Vink_2020, Safi-Harb_2022, 2022PASJ...74.1143K, 1999acfp.book.....L}). It is worth noting that this estimate should be taken as a lower boundary, as the volume may be smaller given the knotty morphology of the emission region as observed by \chandra\ \citep{2025ApJ...993L..24T, 2022PASJ...74.1143K, Safi-Harb_2022}.

In this context, the energy essential to accelerate the electrons to X-ray energies within the ``Head'' region is approximately $5.9\times10^{43}$~erg. This, in turn, results in a radiative loss timescale of roughly 350~yr for 0.5--30~keV photons within a magnetic field of $15~\mu$G. This timescale is much shorter than the estimated age of W50/SS~433 \citep{1997ApJ...483..868S, 1998AJ....116.1842D, 2011MNRAS.414.2838G}, as expected. A more relevant comparison is with the transport time across the ``Head'' itself. In order for the advection time across the region of approximately 4~arcminutes (6.4~pc) to remain under the loss timescale, flow speeds of at least 5--6\% the speed of light are required, placing a local transport constraint on the X-ray-emitting electron population. 

At this point, it is important to notice the high degree of consistency between the eastern and western lobes of W50. Beginning with the initial emission $\sim$26--29~pc from SS~433, the structure of each jet appears consistent between the two, with a bright ``Head'' followed by diminishing intensity interior regions, and another bright interaction, or ``Lenticular'' region. Both the column densities, and photon indices follow the same pattern of spectral evolution between lobes as well. While the smaller ``Head'' extraction area leads to a lower luminosity than the eastern lobe's $1.1\times10^{34}$~erg~s$^{-1}$, the similar power-law properties, particularly the hard spectral index, lead to a magnetic field consistent with the 12~$\upmu$G seen in the eastern counterpart \citep{Safi-Harb_2022}.

It is widely accepted that the influence of SS~433's jets extends to the W50 nebula, reaching as far as the radio ears. However, a substantial portion of the jet's energy appears to be directed towards kinetic motion or an unobserved outflow. Notably, the X-ray luminosity within the western lobe, approximately $L_X\sim3\times10^{34}$~erg~s$^{-1}$, constitutes a minute fraction of the total power of the SS~433 jets. This jet power has been estimated to range from $3.2\times10^{38}$~erg~s$^{-1}$ based on \chandra\ HETG data \citep{Marshall2002} to $5\times10^{39}$~erg~s$^{-1}$ from \xmm\ data \citep{2005A&A...431..575B} to roughly $10^{40}$~erg~s$^{-1}$ based on ASCA data \citep{2006ApJ...637..486K}.

This study reveals that the onset of hard X-ray emission occurs at a distance of approximately 26.5 pc from SS~433, with a noticeable gap in X-ray emission between SS~433 and the ``Head'' region. Interestingly, compact X-ray knots have been observed within the arcsecond-scale X-ray jets emanating from SS~433 
(note that \(1''\) corresponds to \(2.7\times10^{-2}\)~pc at the distance of SS~433). These knots display rapid variations, with changes in intensity by a factor of 1.5--2.0 occurring on timescales ranging from days to as brief as hours. These observations suggest either an influence from SS~433 itself (``pumping") or a sequence of shocks, hinting at the presence of a swift, concealed outflow within the jet \citep{2005MNRAS.358..860M}. This SS~433 outflow might serve as the source of the initial particles that are subsequently accelerated upon reaching the ``Head'' region of the western lobe.

\begin{figure}[htb]
    \centering
    \begin{subfigure}{\columnwidth}
        \centering
        \includegraphics[width=\columnwidth]{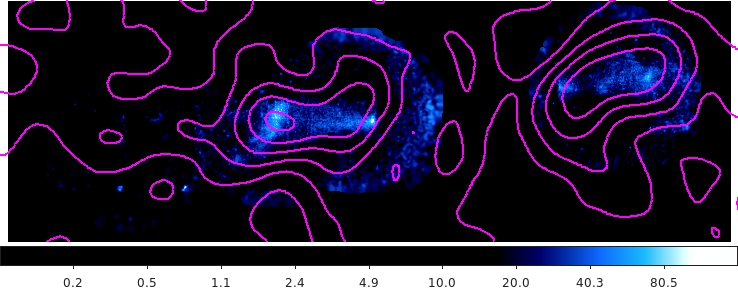}
        \caption{Broadband. 
        } \label{fig:hessBroad}
    \end{subfigure}
    \newline 
    \begin{subfigure}{\columnwidth}
        \centering
        \includegraphics[width=\columnwidth]{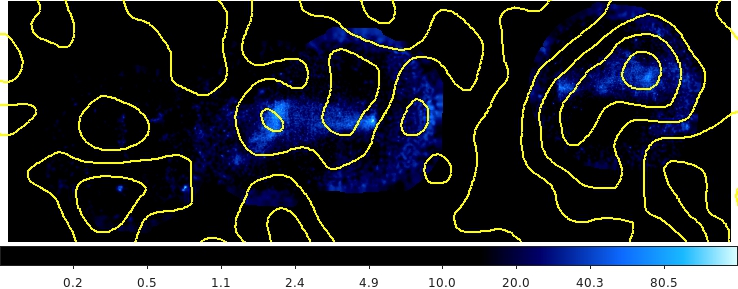}
        \caption{Low: $0.8-2.5$~TeV.} \label{fig:hess0.8-2.5}
    \end{subfigure}
    \newline 
    \begin{subfigure}{\columnwidth}
        \centering
        \includegraphics[width=\columnwidth]{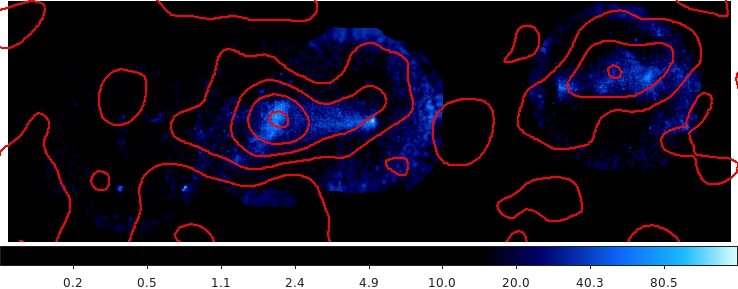}
        \caption{Mid: $2.5-10$~TeV.} \label{fig:hess2.5-10}
    \end{subfigure}
    \newline 
    \begin{subfigure}{\columnwidth}
        \centering
        \includegraphics[width=\columnwidth]{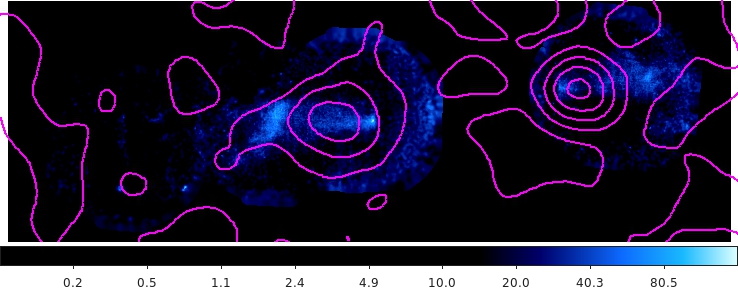}
        \caption{High: $>10$~TeV. } \label{fig:hess10+}
    \end{subfigure}
    \caption{Contour maps of H.E.S.S. fluxes \citep{hess_2024} overlaid on \xmm\ mosaic of the eastern and western lobes, with the image parameters chosen to highlight the keV-TeV comparison. This image shows that the X-ray emission resolves the sites of particle acceleration within the TeV emission, with the highest energy TeV photons peaking closest to the ``Head'' acceleration region.}
    \label{fig:hess}
\end{figure}

Recent analysis from H.E.S.S. \citep{hess_2024} supports this interpretation. Figure~\ref{fig:hess} illustrates flux contour maps of the H.E.S.S. observation split into 0.8--2.5~TeV, 2.5--10~TeV, $>10$~TeV, and Broadband, overlaid on the \xmm\ observations of W50. The symmetry of the eastern and western lobes is further maintained through these maps, as the emission can be seen to move inward towards the respective ``Head'' regions as the energy increases towards $>10$~TeV. That is to say, over the entire H.E.S.S. band, the TeV emission gets averaged out across the entirety of the X-ray jets; however, as the focus shifts to the highest energy emission, so too does this emission shift to focus on the beginning ``Head'' regions of the X-ray jets.

This phenomenon aligns with observations in extra-galactic jets \citep{2003ASPC..300..151H} where the hard spectrum component of synchrotron emission is anticipated to be visible only at the highest frequencies at the acceleration site, and in PWNe where the onset of X-ray emission corresponds to the pulsar wind termination shock \citep[e.g.,][]{2008AIPC..983..171K}, and where a hard spectral index is observed to become steeper as one moves away from the pulsar engine.

The observation of hard X-ray emission reaching energies of around 30~keV implies the presence of electrons with energies, E$_e$, following the relationship $\left(\text{E}_e/10~\text{TeV}\right)\sim0.5\left(\text{B}/1~\text{mG}\right)^{-1/2}\left(\text{E}_\gamma/1~\text{keV}\right)^{1/2}$ \citep[e.g.,][]{1995Natur.378..255K}. For the estimated magnetic field strength of 15~$\upmu$G, this corresponds to electron energies of approximately 180~TeV. This observation lends further support to the notion that this unique microquasar serves as an efficient site for particle acceleration, extending up to energies in the hundreds of TeV range, making it a promising PeVatron as further suggested by high-energy TeV observations \citep{2018arXiv181001892H, hess_2024, 2025NSRev..12af496L}. 

\subsection{Broadband Emission and Jet Modeling} \label{sec:bejm}

The semi-analytic models and SED calculations presented in \ref{sec:Mod} demonstrate that the observed hard X-ray spectrum and extended morphology of the western jet are not consistent with the considered 
toy models. The models that adopt a constant jet speed fail to reproduce an extended region bright in X-rays. Jet plasma dilution and weakening of the magnetic field cause a rapid decrease of the X-ray surface brightness. Furthermore, such jet models are not consistent with the hydrodynamic picture that places a shock at the ``Head'' region.

Models that adopt  a constant density outflow also face certain difficulties: models with the poloidal magnetic field, which decays fast with distance, result in X-ray surface brightness that is localized close to the head region, and relatively fast steepening of the X-ray spectrum along the jet due to decreasing strength of the magnetic field. The models that include only toroidal magnetic field lead to extended X-ray morphology with reasonably hard X-ray spectrum extending beyond \(60\)~pc from SS~433, however, the ``Head'' region appears to be very faint compared to the observation.

This mismatch between the model prediction and observations can be a result of oversimplified adopted models, and a better agreement can be achieved by introducing spatial variation of the magnetic field or in-situ particle reacceleration. However, there is another, in fact a very natural, possibility that can be discussed within the framework of the simplest approach: there should be both toroidal and poloidal components of the magnetic field at the acceleration site.

\begin{figure*}[htb]
    \begin{subfigure}[t]{0.475\textwidth}
        \includegraphics[width=\textwidth]{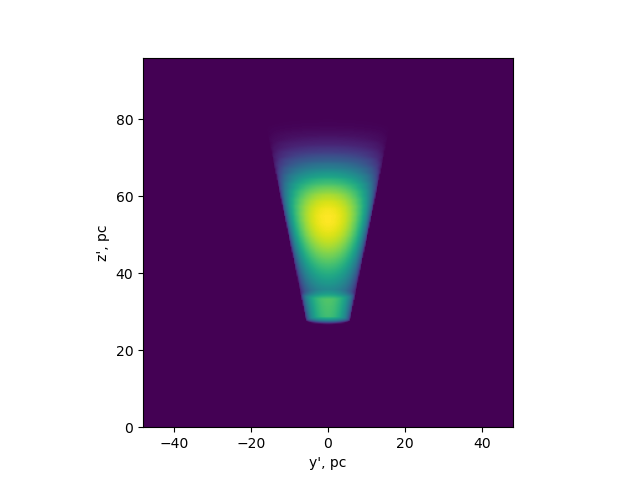}
        \caption{Synthetic synchrotron surface brightness at \(3\)~keV.} \label{fig:hybridSB}
    \end{subfigure}
    \hfill
    \begin{subfigure}[t]{0.475\textwidth}
        \includegraphics[width=\textwidth]{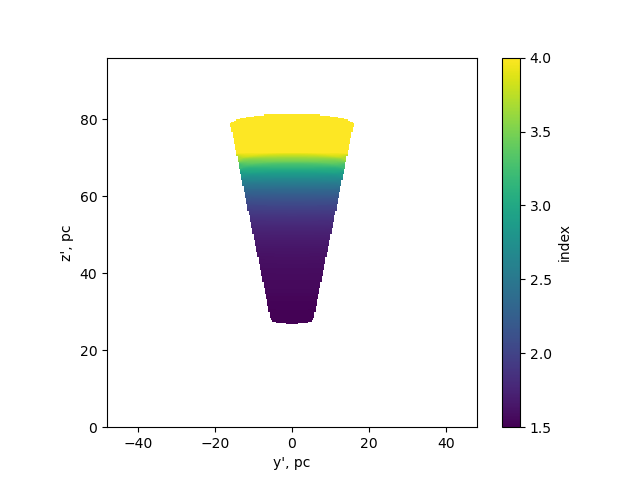}
        \caption{Surface slope of synthetic synchrotron emission obtained in a narrow band of energies around \(3\)~keV.} \label{fig:hybridSS}
    \end{subfigure}
    \caption{Surface brightness and slope of the X-ray emission computed for a toy model that adopts both poloidal (localized close to the innermost regions) and toroidal magnetic fields.} \label{fig:hybrid}
\end{figure*}

In large scale outflows the toroidal component of the magnetic field should dominate because, in the conical geometry considered here, the poloidal component decreases rapidly with distance (\(B_z\propto Z^{-2}\)). On the other hand, the IXPE observations show that in the innermost part of the X-ray lobe, the poloidal component dominates. Presence of this component also seems to be a necessary element for efficient operation of DSA. This then should be a locally generated field that can decay very quickly downstream. In contrast, the initial toroidal field should survive and in fact get amplified in a constant density flow. Under these assumptions, it is possible to better reproduce the surface brightness and extended regions demonstrating hard X-ray emission, which is demonstrated in Fig.~\ref{fig:hybrid}.

\subsection{Implications for Cosmic-ray Production}

Another important result revealed with the simulation presented in \ref{sec:Mod} is the small fraction of the jet kinetic luminosity transferred to relativistic electrons, \(\kappa_e\sim10^{-3}\). If the overall efficiency of the acceleration process is about \(10\%\), then relativistic protons should receive approximately \(100\) times more energy, however their inefficient radiation cooling in the jet prevents any noticeable contribution to the detected emission. These protons are transported down the jet with minor energy losses both due to the radiative cooling and adiabatic (in constant density jet), making SS~433 a persistent source of cosmic-ray protons with luminosity at the level of \(10^{38}\unit{erg\,s^{-1}}\).

In the case of W50, the non-thermal X-ray emissions are believed to be attributed to synchrotron emission. This inference is supported by the estimated magnetic field strength in the western lobe (as discussed previously in this section) and the observed hard X-ray photon index, both of which favor synchrotron emission over inverse Compton processes.
Additionally, the \textit{IXPE} study of the eastern jet, which displays almost identical properties to the western jet, was shown to have a synchrotron origin \citep{2024ApJ...961L..12K}.

\section{Conclusions} \label{sec:Conc}
This work has introduced the first combined observation of the western lobe utilizing both \nustar and \xmm\ instruments. Hard X-ray emission was successfully detected, resolved, and characterized originating from W50's western lobe. The western jet appears as a sharp boundary starting at approximately 17$\arcmin$ ($\sim$26.5~pc at a distance of 5.5~kpc)
from SS~433, notably recognized as the ``Head," featuring a knotty structure with a scale of approximately 4$\arcmin$ (6.4~pc). The ``Head'' region's emission exhibits a distinctive hard non-thermal X-ray spectrum, marked by a photon index of 1.55, a result corroborated by \nustar and \xmm\ joint observations (0.5--30 keV). Further away from SS~433, the photon index gradually steepens, reaching $\sim$1.7 in the ``Diffuse'' region, and approximately 2 in the ``w2'' region. These hard X-ray knots serve as indicators of the locations where particle acceleration takes place within the jet. 

Using a semi-analytic treatment of particle transport in a conical jet, combined with the X-ray and TeV SEDs and synthetic X-ray brightness maps, this study constrains the magnetic field in the western jet to around 10--20~$\mu$G and the fraction of jet power channelled into non-thermal electrons to roughly (0.3--0.6)\%. In this regime, a single electron population can simultaneously reproduce the observed hard X-ray synchrotron emission and the TeV $\gamma$-rays, while proton-proton interactions are far too inefficient to explain the $\gamma$-ray luminosity of the jet. This firmly favors a leptonic origin for both the X-ray and TeV emission and identifies W50 as an efficient Galactic accelerator of electrons well over 100 TeV. A modest fraction of the jet energy obtained by relativistic electrons in the ``Head'' region allows a very efficient acceleration of protons with power of \(10^{38}\ergs\), which remain invisible in $\gamma$-rays due to inefficient radiation in the jet. It is noted that LHAASO detected an extended component at energies above 100 TeV coinciding with a nearby cloud and suggesting a hadronic component \citep{2025NSRev..12af496L}. Future neutrino observations may reveal this source as a hadronic PeVatron.

Future work will extend these results by combining deeper, high-resolution X-ray observations of both lobes with more comprehensive multi-wavelength SED and dynamical modeling, including the surrounding radio shell and the large-scale environment. Furthermore, upcoming and approved \textit{IXPE} observations of the ``w1'' and "e2" regions will shed light on the magnetic field geometry in these acceleration zones.
These endeavours aim to further comprehension of this extraordinary system, holding significant relevance for probing the outflows of various galactic and extra-galactic sources, including other microquasars, pulsar wind nebulae, active galactic nuclei, and ultraluminous X-ray sources. Upcoming high-resolution and broadband capabilities such as with \textit{AXIS}, \textit{NewATHENA}, \textit{CTA}, \textit{ngVLA} or \textit{SKA} will provide an unprecedented and sensitive view of W50 and other such systems.

\vspace{0.25cm}

\begin{acknowledgments}
We are grateful to Yoshiyuki Inoue and Tatsuki Fujiwara for helpful discussions and Kazuho Kayama for support in preparing the X-ray proposals.
We also thank the referee for a thorough reading of the manuscript and for suggestions that helped with its clarity.
This research made use of NASA's Astrophysics Data System (ADS) and of data and software provided by the High Energy Astrophysics Science Archive Research Center (HEASARC), which is a service of the Astrophysics Science Division at NASA/GSFC.
This work made use of data from the \nustar mission, a project led by the California Institute of Technology, managed by the Jet Propulsion Laboratory, and funded by NASA.
This work also made use of data obtained with XMM--Newton, a European Space Agency science mission with instruments and contributions directly funded by ESA Member States and NASA. 
S.S.H. acknowledges support from the Natural Sciences and Engineering Research Council of Canada (NSERC) through the Discovery Grants and Canada Research Chairs programs, and from the Canadian Space Agency. B.M.I. is supported by an NSERC PGS-D. K.M. and S.Z. acknowledge support from NASA grant NNH20ZDA001N-NUSTAR.
N.T. acknowledges support by the Japan Society for the Promotion of Science (JSPS) KAKENHI grant Nos. JP22K14064, JP24H01819, and JP26H00820.

\facilities{NuSTAR, XMM, CXO}
\software{
HEASoft (v6.28; \url{https://heasarc.gsfc.nasa.gov/docs/software/lheasoft/}),
XMM-SAS (v19.1.0; \url{https://www.cosmos.esa.int/web/xmm-newton/sas}),
CIAO (v4.12; \citealt{2006SPIE.6270E..1VF}),
NuSTARDAS (v2.0.0; \url{https://heasarc.gsfc.nasa.gov/docs/nustar/analysis/}), 
SAOImage DS9 (\url{https://sites.google.com/cfa.harvard.edu/saoimageds9}),
XSPEC (v12.11.1; \citealt{1996ASPC..101...17A})
}
\end{acknowledgments}

\appendix
\restartappendixnumbering

\section{Appendices:  Additional Material}

\subsection{Background analysis -- NuSTAR background simulations} \label{sec:nustarbkg}

The \nustar telescope is susceptible to four types of background contamination: stray-light background photons (also called zero-bounce photons as they do not get reflected off the optics) leaking through aperture stops from sources $1.5^{\circ}$ to $5^{\circ}$ off-axis, ghost-ray background consisting of one-bounce photons from sources $3\arcmin$ to $40\arcmin$ off-axis, focused two-bounce photons from diffuse background emission in the FoV, and internal detector background which usually dominates above $\sim30$ keV. 

\begin{figure}[htb]
    \centering
    \begin{subfigure}[t]{0.3\textwidth}
        \centering
        \includegraphics[width=\textwidth]{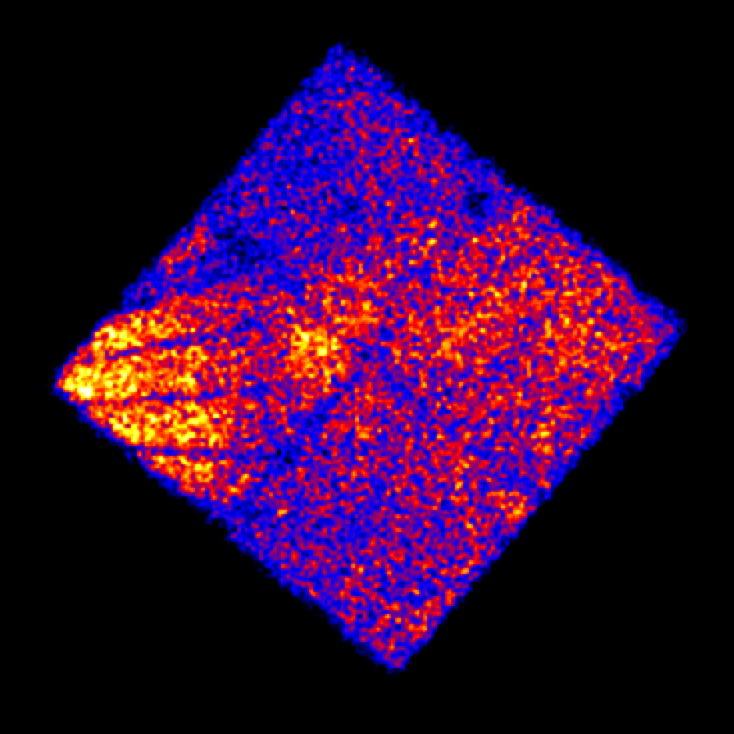}
        \caption{Original \nustar FPMA image} 
    \end{subfigure}
    \hfill
    \begin{subfigure}[t]{0.3\textwidth}
        \centering
        \includegraphics[width=\textwidth]{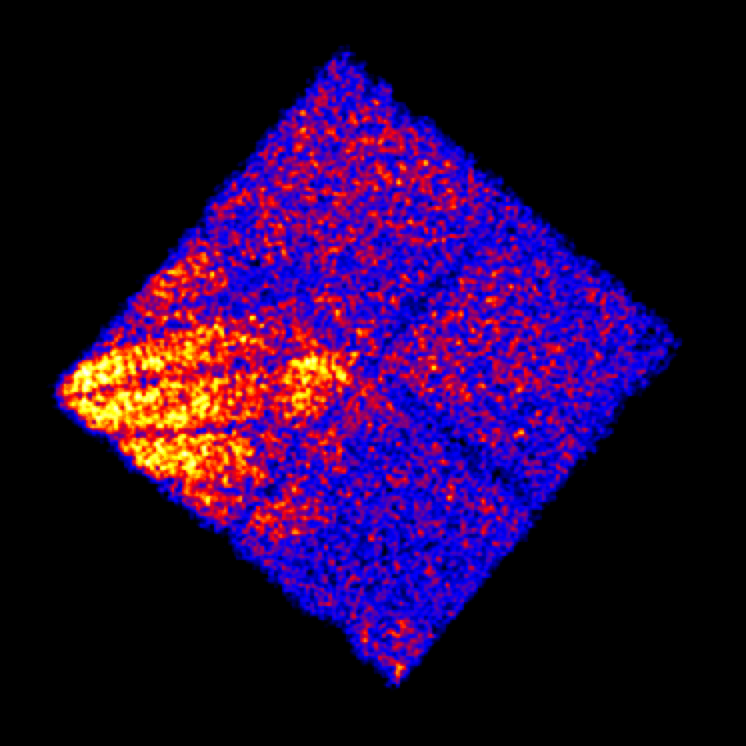}
        \caption{Original \nustar FPMB image} 
    \end{subfigure}
    \hfill
    \begin{subfigure}[t]{0.3\textwidth}
        \centering
        \includegraphics[width=\textwidth]{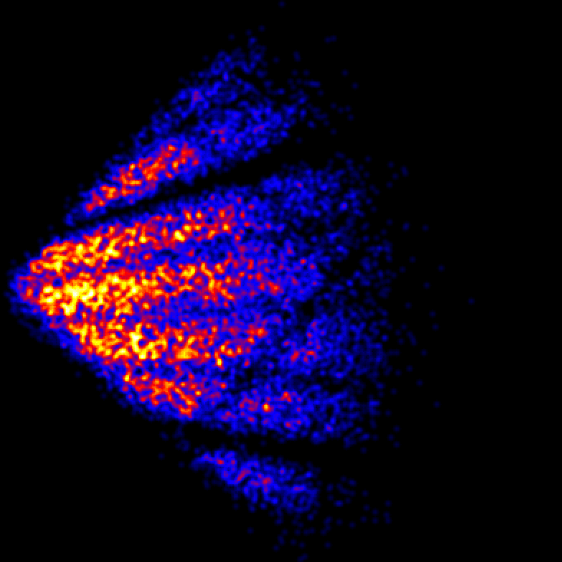}
        \caption{Ghost ray background (2 bounce photons) extracted from NuSIM simulation} 
    \end{subfigure}
    \caption{Comparison of ghost ray features in \nustar observation and NuSIM simulation} \label{fig:GRbgd}
\end{figure}

The \nustar images from both FPMA and FPMB evidently exhibit the ghost-ray background patterns radially distributed from the direction of SS~433 located on the left-hand side. Thus, the {\tt NuSIM} software was first utilized to assess the ghost-ray background contamination from SS~433. {\tt NuSIM} is a ray-tracing Monte Carlo program that simulates photon events on the detector planes. In each {\tt NuSIM} simulation, the position and approximate X-ray spectrum of SS~433 were input (an absorbed power-law model with $\Gamma = 2$) as well as the pointing vector and exposure that match with those of the actual \nustar observation on W50 western lobe. For each photon event, including both the source and background emission, NuSIM registers whether it underwent zero (stray light), one (ghost-ray), or two (focused) bounces. The one-bounce photons only mostly produced by SS~433 (which is the brightest X-ray source in the neighborhood by far) were extracted, and confirmed that the simulation produced the ghost-ray pattern as seen in the actual \nustar observation. The comparison between the raw \nustar FPMA, FPMB, and simulated ghost-ray images is presented in Figure~\ref{fig:GRbgd}.

\begin{figure}[htb]
    \centering
    \includegraphics[width=0.4 \textwidth]{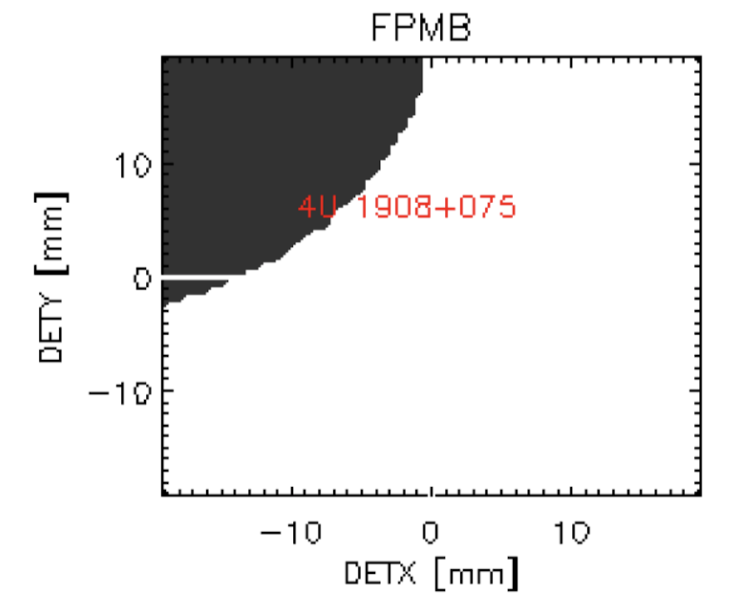}
    \caption{Stray light background affecting FPMB according to simulation from {\tt nustar\_stray\_light}. 
    Note that this image is shown in detector coordinates.
    }
    \label{fig:nustraylight}
\end{figure}

Then, {\tt nustar\_stray\_light} was used to examine potential contamination from stray-light background caused by nearby X-ray point sources. For a given pointing direction,  {\tt nustar\_stray\_light} searches for possible sources that could produce stray light background and simulates the pattern of stray light background on the detector plane. For this observation, no bright stray-light background source that affects FPMA was found, while 4U 1909+07 produces stray-light background in a part of the FPMB detector, as shown in Figure~\ref{fig:nustraylight}. Source-free regions were selected by avoiding these regions heavily contaminated by the ghost-ray and stray light background in subsequent stray-light diffuse background analysis using {\tt Nuskybgd} \citep{Wik2014}.

\begin{figure}[htb]
    \centering
    \begin{subfigure}[t]{0.45\textwidth}
        \centering
        \includegraphics[width=\textwidth]{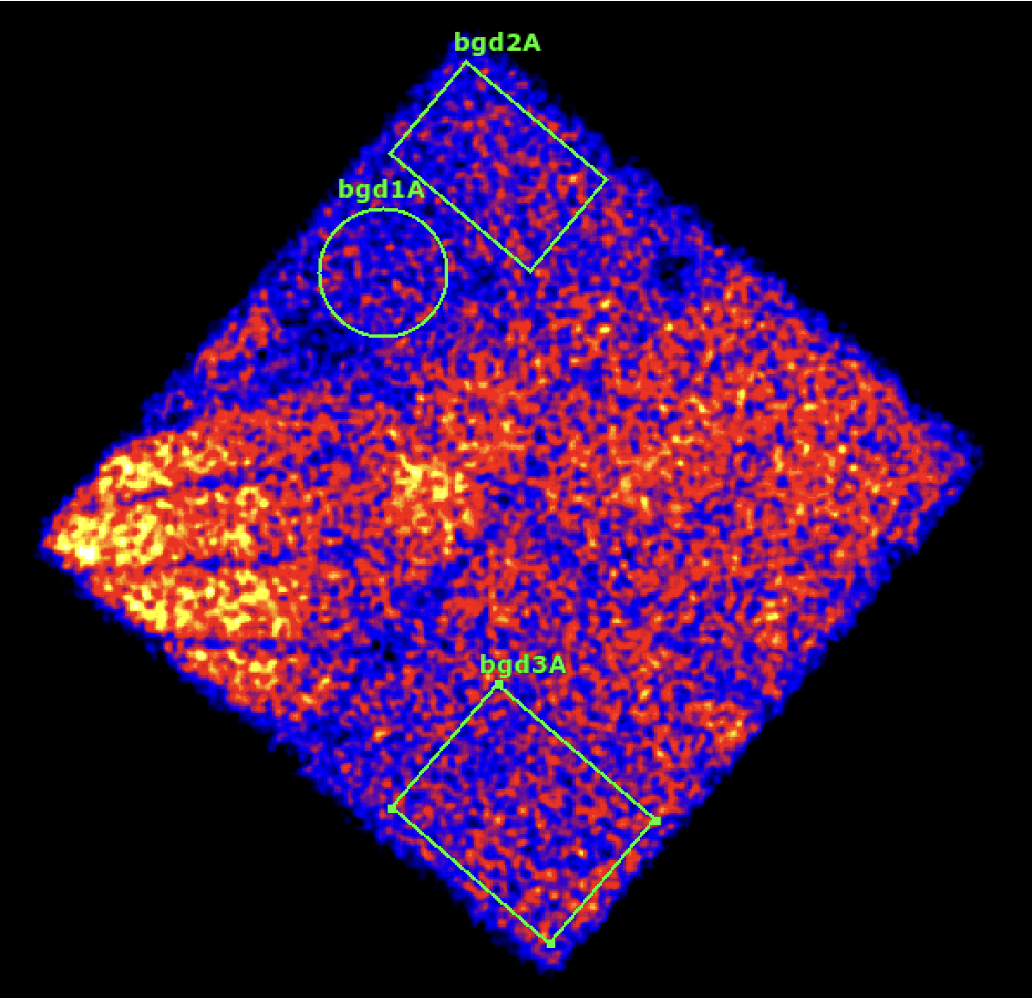}
        \caption{Source-free regions on FPMA selected for {\tt Nuskybgd} model fitting} 
    \end{subfigure}
    \hfill
    \begin{subfigure}[t]{0.45\textwidth}
        \centering
        \includegraphics[width=\textwidth]{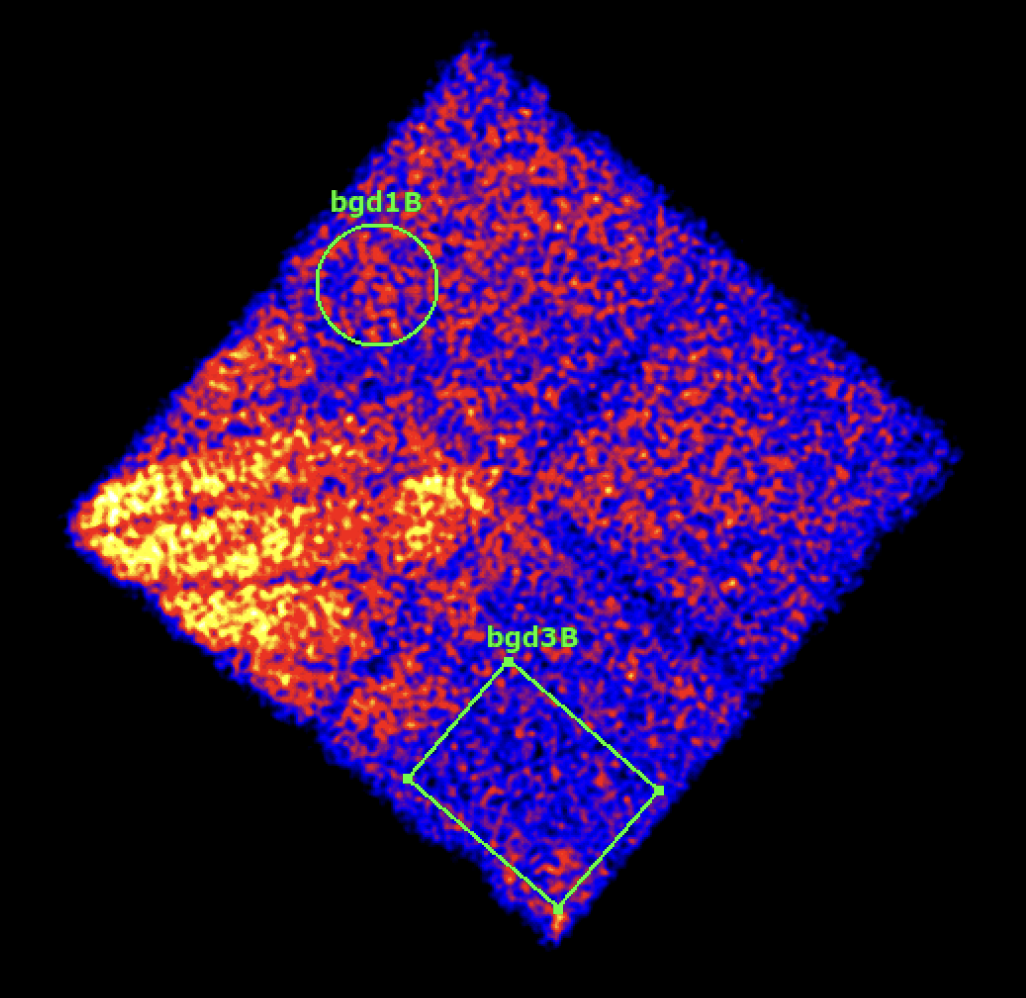}
        \caption{Source-free regions on FPMB selected for {\tt Nuskybgd} model fitting} 
    \end{subfigure}
    \caption{Source-free regions selected for {\tt Nuskybgd} model fitting.} \label{fig:nuskybgdregion}
\end{figure}

Besides the ghost-ray and stray-light background components discussed above, stray-light background contamination from the CXB must be considered. The diffuse background emission produces a non-uniform stray-light background pattern on the detector plane. Hence, a conventional method of background extraction from a nearby region might not be accurate for \nustar spectral and imaging analysis. Thus, background models were constructed by fitting multiple spectra extracted away from the source (i.e. ``w1'') and regions contaminated by the ghost-ray and stray-light background from SS~433 and 4U 1908+075, respectively, with {\tt Nuskybgd}. More specifically, source-free regions were first selected on the FPMA and FPMB, as shown in Figure~\ref{fig:nuskybgdregion}. The spectra were fitted to the model in {\tt Nuskybgd}, which takes into account different background components, including the focused background and instrument background. Then, for a given energy band, source region, and detector module, {\tt Nuskybgd} allows the production of model background spectra and images. Using {\tt Nuskybgd}, simulated background spectra were produced for the ``Head'', ``Diffuse'', and ``Full'' regions and background-subtracted images in different energy bands. Background-subtracted images produced from {\tt Nuskybgd} were used for the subsequent imaging analysis. For \nustar spectral analysis of the three source regions, fitting results in the 3--10 keV band were compared with those from \xmm\ data using both the simulated and locally-extracted background spectra. It was found that the simulated background spectra from {\tt Nuskybgd} did not produce consistent fitting results with the \xmm\ spectral analysis in the overlapping 3--10 keV band. The \nustar photon counts in 3--4 keV were higher when the simulated background spectra were used, making the best-fit photon indices softer than those from the \xmm\ spectral fitting. This is possibly due to another local diffuse soft X-ray emission that appears more prominently below 4 keV but is not included in the {\tt Nuskybgd}, which models the CXB stray-light and internal background components. In the \nustar spectral analysis, the background spectra extracted locally around the source were adopted since they provided more consistent results with the \xmm\ analysis. 

\subsection{Modeling of the non-thermal emission from the jet}  \label{sec:Mod}

Spectral and morphology modeling presented in the manuscript are based on a simplified semi-analytic approach, which is introduced below. It relies on a phenomenological description of the flow and fundamental relations between key physical parameters of the jet. 
A specific jet geometry is first assumed. The simplest choice is a conical jet, i.e., one with a constant half-opening angle \(\theta_j\) \citep[the jet precession angle of about \(10^\circ\) apparently determines the opening angle of the large-scale jet, although linear regression by][revealed a much more narrow, \(\approx2.6^\circ\pm0.3^\circ\), region bright in X-ray band]{2022PASJ...74.1143K}. In that case the jet radius increases linearly with distance along the jet axis:
\begin{equation}
  R(z) = z \sin\theta_j\,.
\end{equation}

While X-ray data favor this configuration \citep[see, e.g.,][]{2022PASJ...74.1143K}, other geometries~--~such as a parabolic jet, \(R \propto z^{1/2}\), or a cylindrical jet, \(R = \text{const}\)~--~cannot be excluded \textit{a priori}. Even if a conical geometry is preferred observationally, the discussion is kept sufficiently general to accommodate a wide range of jet shapes.

For stationary jets, the mass flux through any cross-section remains constant (here and below relativistic corrections to jet parameters are neglected):
\begin{equation}
  R^2(z)\, v(z)\, \rho(z) = \text{const}\,.
\end{equation}

Thus, for a given jet geometry, the bulk speed \(v(z)\) and plasma density \(\rho(z)\) are tightly coupled.

To estimate the magnetic field, a simple treatment is adopted based on the ideal MHD frozen-in condition. Under this assumption, the toroidal and poloidal components of the magnetic field scale with density and jet radius as:
\begin{equation}
  B_\varphi(z) = B_{\varphi 0} \frac{\rho(z) R(z)}{\rho_0 R_0} = B_{\varphi 0} \frac{v_0 R_0}{v(z) R(z)}\,,
\end{equation}
and
\begin{equation}
  B_z(z) = B_{z0} \frac{R_0^2}{R^2(z)}\,.
\end{equation}

These relations highlight a well-known feature: the toroidal component tends to dominate in large-scale outflows (especially since typically \(v < v_0\)). On the other hand, Imaging X-ray Polarimetry Explorer (IXPE) observations indicate that the magnetic field near the eastern ``Head'' knot is likely aligned with the jet axis \citep{2024ApJ...961L..12K}. Therefore, both configurations are considered in the phenomenological treatment.

This simplified approach allows the determination of the (M)HD structure of the jet by specifying a single function ---the jet speed--- provided that the jet geometry is constrained observationally.

Synchrotron surface brightness depends on the transport of relativistic particles and the distribution of the magnetic field. To compute synthetic brightness maps, one must first calculate the distribution of relativistic electrons along the jet, taking into account variations in the advection speed and various energy loss processes. This energy distribution is then convolved with the single-particle synchrotron spectrum to yield the monochromatic emissivity. Finally, if the bulk velocity is significant, Doppler boosting must be included, and the resulting emissivity should be integrated along the line of sight to produce synthetic synchrotron maps.

The evolution of non-thermal particles is governed by the continuity equation, which, for a mildly relativistic jet, can be written as:
\begin{equation}
  \pdv{}{z}\qty(v n) + \pdv{}{\gamma}\qty(\dot{\gamma} n) = q(\gamma, z)\,.
\end{equation}
Here, \(\mathrm{d}N = n\,\mathrm{d}z\,\mathrm{d}\gamma\) is the number of relativistic particles with Lorentz factor \(\gamma\) at a distance \(z\) from the jet origin (it is assumed the jet is narrow); \(\mathrm{d}N = q\,\mathrm{d}z\,\mathrm{d}\gamma\,\mathrm{d}t\) is the number of injected relativistic particles; and \(\dot{\gamma}(\gamma, z)\) describes the energy losses of the particles.

While the energy losses for protons in the jet are dominated by the adiabatic cooling, there are several important channels, that influence the distribution of relativistic electrons. The total energy loss rate for relativistic electrons includes synchrotron radiation, inverse Compton (IC) scattering, and adiabatic cooling:
\begin{equation}
  \dot{\gamma} = \dot{\gamma}_{\mathrm{syn}} + \dot{\gamma}_{\mathrm{IC}} + \dot{\gamma}_{\mathrm{ad}}\,.
\end{equation}

The synchrotron cooling rate is determined by the local magnetic field strength and the particle energy:
\begin{equation}
  \dot{\gamma}_{\mathrm{syn}} = -\frac{4}{9} \frac{e^4 B^2 \gamma^2}{m^2 c^3}\,.
\end{equation}

The energy loss due to IC scattering is computed assuming the target photon field is a superposition of several gray-body components:
\begin{equation}
  \dot{\gamma}_{\mathrm{IC}} = \dot{\gamma}_{\mathrm{CMBR}} + \dot{\gamma}_{\mathrm{FIR}} + \dot{\gamma}_{\mathrm{NIR}}\,,
\end{equation}
where these contributions include:
\begin{itemize}
  \item Cosmic microwave background (CMBR): \(T_{\mathrm{CMBR}} = 2.7~\mathrm{K}\), \(w_{\mathrm{CMBR}} = 0.25~\mathrm{eV\,cm^{-3}}\);
  \item Far-infrared background (FIR): \(T_{\mathrm{FIR}} = 80~\mathrm{K}\), \(w_{\mathrm{FIR}} = 0.8~\mathrm{eV\,cm^{-3}}\);
  \item Near-infrared background (NIR): \(T_{\mathrm{NIR}} = 3000~\mathrm{K}\), \(w_{\mathrm{NIR}} = 1.0~\mathrm{eV\,cm^{-3}}\);
\end{itemize}

The rate of adiabatic energy losses is determined by the structure of the flow in the jet:
\begin{equation}
  \dot{\gamma}_{\mathrm{ad}} = \frac{\gamma}{3} \dv{\ln \rho}{t} = \frac{\gamma v}{3 \rho} \dv{\rho}{z}\,,
\end{equation}
which is equivalent to the more practical form:
\begin{equation}
  \dot{\gamma}_{\mathrm{ad}} = -\frac{\gamma v}{3} \left( \frac{1}{v} \dv{v}{z} + \frac{2}{R} \dv{R}{z} \right)\,.
\end{equation}

Depending on the flow structure, adiabatic compression may also lead to heating of the plasma, which also affects the non-thermal component of the plasma. In addition to possible adiabatic heating, other (re)acceleration processes may take place in the jet. These processes extract a fraction of the internal energy and transfer it to the non-thermal particle population.
Within the framework adopted here, such processes determine the particle injection spectrum \(q\).

Motivated by the jet morphology observed in the X-ray and TeV energy bands \citep[see, e.g.,][]{hess_2024}, non-thermal particles are assumed to be predominantly injected at the innermost X-ray knot, with possible additional injection sites along the jet (at \(z_1, \dots, z_k\)):
\begin{equation}
  q(\gamma, z) = q_0(\gamma) \delta(z - z_0) + \sum\limits_{i=1}^{k} q_i(\gamma) \delta(z - z_i)\,.
\end{equation}

Each injection component \(q_i\) is modeled as a power-law spectrum with a super-exponential cutoff:
\begin{equation}
  q_i(\gamma) = A_i \gamma^{-\alpha_i} \exp\left[-\left( \frac{\gamma}{\gamma_{\mathrm{max},i}} \right)^{\beta_i} \right] \theta(\gamma - \gamma_{\mathrm{min},i})\,.
\end{equation}
Here, \(\alpha_i \approx 2\) is the spectral index, \(\beta_i \sim 1\) is the cutoff index \citep[note that the analytic solution by][yields \(\beta_e=2\)]{2007A&A...465..695Z}, \(\gamma_{\mathrm{max},i}\) is the maximum Lorentz factor, \(A_i\) is a normalization factor, and \(\gamma_{\mathrm{min},i}\) is the minimum energy.

Several arguments allow the estimation of the parameters of the injection spectrum. In particular, the maximum energy for electrons is commonly obtained by equating the acceleration time to the total energy loss time. The acceleration time is typically estimated as:
\begin{equation}
  t_{\mathrm{acc}} \approx \frac{\eta\, m_e c^2 \gamma}{e B c}\,,
\end{equation}
which yields numerically:
\begin{equation}
  t_{\mathrm{acc}} \approx 6 \times 10^7 \left( \frac{\eta}{10^3} \right) \left( \frac{\gamma}{10^7} \right) \left( \frac{B}{10~\upmu\mathrm{G}} \right)^{-1}~\mathrm{s}\,.
\end{equation}

Here, \(\eta \gg 1\) is a dimensionless acceleration efficiency parameter. In mildly relativistic jets, assuming highly efficient acceleration, the theoretical minimum is \(\eta_{\mathrm{min}} = 2\pi c^2 / v^2 \sim 10^3\).

If jets accelerate very high-energy (\(E > 100\)~GeV) or even ultra-high-energy (\(E > 100\)~TeV) electrons, then synchrotron losses typically dominate:
\begin{equation}
  t_{\mathrm{syn}} \approx 8 \times 10^{11} \left( \frac{\gamma}{10^7} \right)^{-1} \left( \frac{B}{10~\upmu\mathrm{G}} \right)^{-2}~\mathrm{s}\,.
\end{equation}

Balancing the acceleration and synchrotron loss timescales gives:
\begin{equation}
  \gamma_{\mathrm{max}} \approx 10^9 \left( \frac{\eta}{10^3} \right)^{-1/2} \left( \frac{B}{10~\upmu\mathrm{G}} \right)^{-1/2}\,,
\end{equation}
implying that electrons can be accelerated into the ultra-high-energy regime.

Another constraint is related to the power of the acceleration mechanism. The total energy transferred to non-thermal particles is given by:
\begin{equation}
  L_{\mathrm{acc}} = A m_e c^2 \int\limits_{\gamma_{\mathrm{min}}}^{\infty} \dd{\gamma}\, \gamma^{1 - \alpha} \exp\left[-\left( \frac{\gamma}{\gamma_{\mathrm{max}}} \right)^\beta \right]\,,
\end{equation}
and must remain a fraction of the total jet power, i.e., \(L_{\mathrm{acc}} = \kappa_{\mathrm{acc}} L_{\mathrm{j}}\), where \(\kappa_{\mathrm{acc}} < 0.1\).

For an injection spectrum close to the canonical value, \(\alpha \approx 2\), the integral simplifies to approximately
\[
L_{\mathrm{acc}} \approx A m_e c^2 \ln\left( \frac{\gamma_{\mathrm{max}}}{\gamma_{\mathrm{min}}} \right)\,,
\]
which has only a weak dependence on \(\gamma_{\mathrm{min/max}}\). Therefore, this constraint primarily determines the normalization factor \(A\).

For a typical jet magnetic field strength of \(B \sim 10~\upmu\mathrm{G}\), electrons with Lorentz factor \(\gamma \sim 10^5\) emit synchrotron photons predominantly at energies of \(\sim 10^{-3}~\mathrm{eV}\), i.e., in the infrared band. Moreover, electrons at this energy do not cool efficiently on the characteristic dynamical timescale of the jet,
\[
t_{\mathrm{dyn}} \sim \frac{z_0}{v} \sim 10^{10}~\mathrm{s}
\]
(for \(z_0 \sim 10~\mathrm{pc}\) and \(v \sim 0.1c\)).

Thus for modeling aimed at fitting the X-ray energy band, one can set \(\gamma_{\mathrm{min}} = 10^4\), and increase this parameter if the resulting radio flux exceeds the level observed.

To compute the jet morphology seen in the X-ray band, it is necessary to integrate the emissivity along paths parallel to the line of sight:
\begin{equation}
  S_{\nu'}\qty(y',z') = \int\limits_{-\infty}^{\infty}\dd{x'} j'_{\nu'}\qty(x',y',z')\,.
\end{equation}
Here, \((x',y',z')\) is a Cartesian coordinate system whose origin coincides with the jet origin; the \((y',z')\) plane is the “plane of the sky”; and the jet axis lies in the \((x',z')\) plane. If the jet is inclined by an angle \(\theta_o\) with respect to the plane of the sky, then the coordinate transformation between the jet frame and the integration frame is simply a rotation about the \(y\)-axis (which coincides with the \(y'\)-axis) by the angle \(\theta_o\).

The emission in the jet is assumed to be homogeneous, and the emissivity can be approximated as
\begin{equation}
  \begin{split}
    j'_{\nu'}\qty(x',y',z') &= j_\nu\qty(z) \qty(\frac{\nu'}{\nu})^2 \times \\
                            &\theta\qty(R(z) - \sqrt{x^2 + y^2})\,,
  \end{split}
\end{equation}
where \(x, y, z\) are coordinates in the jet frame, where the jet axis is aligned with the \(z\)-axis, and \(\nu'\) and \(\nu\) are related by the Lorentz transformation between the plasma co-moving frame and the observer's frame.

Even if the jet is only mildly relativistic and relativistic effects on jet dynamics are negligible, Doppler boosting can still significantly affect the observed surface brightness. Under a simplified treatment of the (M)HD structure of the jet, the jet speed is modelled as a function of distance along the jet, \(v(z)\). At each point in the jet cross section, the local velocity direction is determined by the jet shape and position:
\begin{equation}
  \vec{v}_{\mathrm{loc}} = v(z) \frac{(x \kappa_v, y \kappa_v, 1)}{\sqrt{(x \kappa_v)^2 + (y \kappa_v)^2 + 1}}\,,
\end{equation}
where \(\kappa_v = (\dv*{R}{z})/R\) determines the local opening angle of the jet. The line-of-sight direction in the \((x,y,z)\) coordinate system is \(\vec{n}_o = (\cos\theta_o, 0, \sin\theta_o)\), so the cosine of the angle between the local velocity and the line of sight is
\begin{equation}
  \mu = \frac{\vec{v}_{\mathrm{loc}} \cdot \vec{n}_o}{v(z)} = \frac{x \kappa_v \cos\theta_o + \sin\theta_o}{\sqrt{(x \kappa_v)^2 + (y \kappa_v)^2 + 1}}\,.
\end{equation}
The Doppler factor defines the relation between emitted and observed frequencies, \(\nu' = \Doppler \nu\), where
\begin{equation}
  \Doppler = \frac{1}{\Gamma \qty(1 - \frac{v}{c} \mu)}\,, \quad \Gamma = \frac{1}{\sqrt{1 - \qty(\frac{v}{c})^2}}\,.
\end{equation}

Combining the above expressions, the surface brightness becomes:
\begin{equation}
  \begin{split}
    S_{\nu'}\qty(y',z') &= \int\limits_{-\infty}^{\infty} \dd{x'}\, j_{\nu'/\Doppler}(z)\, \Doppler^2 \times\\
                        &\theta\qty(R(z) - \sqrt{x^2 + y^2})\,.
  \end{split}
\end{equation}
Here, the synchrotron emissivity \(j_\nu(z)\) is computed in the jet co-moving frame using standard prescriptions. It depends on the local magnetic field strength and the electron energy distribution, both assumed to vary only with the distance \(z\) along the jet.

Accounting for the Doppler boosting of the IC emission requires a different treatment. The reason for that is the external origin of the target fields. Following the approach used above for the synchrotron emission would require first transforming the photon fields to the jet co-moving frame, computing the emissivity in this frame (which is not isotropic in either frame), and then transforming the resulting emissivity back to the laboratory frame. The same result, however, can be achieved with a simpler reasoning if the calculation is performed instead directly in the laboratory frame using the Lorentz-invariant distribution function \citep[see, e.g.,][]{2018MNRAS.481.1455K}:
\begin{equation}
  j'_{\nu'} = c \int \overline{\dv{\sigma}{\nu'}} p'^2 f'\qty(\vec{r}', p' \vec{n}_o)\, \dd{p'}\, \dd{n}_{ph}\,,
\end{equation}
where \(f'\) is the distribution function in phase space and \(\vec{n}_o\) is the direction of the line of sight. Here it is assumed that the photon field \(\dd{n}_{ph}\) is isotropic in the laboratory frame. The differential cross section averaged over scattering angles is used, \(\overline{\dv{\sigma}{\nu'}}\). The distribution function \(f'\) is Lorentz-invariant and can be obtained from the energy distribution in the jet co-moving frame as
\begin{equation}
  \begin{split}
    f'\qty(\vec{r}', p' \vec{n}_o) &= f\qty(\vec{r}, \vec{p}) \\
                                   &= \frac{1}{4\pi m_e^2 c^2 \gamma p} \frac{n}{\pi R^2(z)}\,.
  \end{split}
\end{equation}
Here, Lorentz transformations to the co-moving frame yield:
\begin{equation}
  \gamma = \gamma' \Gamma \qty(1 - \frac{v p'}{m_e c^2 \gamma'} \mu)\,.
\end{equation}
This relation is derived from the Lorentz-invariant scalar product of the particle four-momentum and the fluid four-velocity. For ultra-relativistic particles, \(p' = m_e c \gamma\), one obtains \(\gamma' = \gamma \Doppler\), thus
\begin{equation}
  j'_{\nu'} = \frac{1}{4\pi^2 R^2(z)} \int \overline{\dv{\sigma}{\nu'}} \Doppler^2 n\qty(\gamma' / \Doppler)\, \dd{\gamma'}\, \dd{n}_{ph}\,.
\end{equation}
Although this expression resembles the standard one, it is important to note that the calculation is performed assuming an isotropic photon field in the laboratory frame. Since the photon fields are gray-bodies, the integration over the photon distribution can be replaced with analytic approximations \citep[see in][]{2014ApJ...783..100K}:
\begin{equation}
  j'_{\nu'} = \frac{1}{4\pi^2 R^2(z)} \int \eval{\overline{\dv{\sigma}{\nu'}}}_{bb} \Doppler^2 n\qty(\gamma' / \Doppler)\, \dd{\gamma'}\,.
\end{equation}
To obtain the IC brightness distributions, one needs to integrate this expression along the line of sight.

It is also important to estimate the possible contribution to the jet emission from hadronic processes. At the multi
parsec distance from the companion star, the only relevant process is proton-proton (pp) interactions. Protons propagating
down the jet lose a fraction of their energy via the pp channel:
\begin{equation}
  \Delta E = - \frac{\rho E}{\tau_{pp}m_p}\frac{\Delta z}{v}\,,
\end{equation}
where \(\tau_{pp}=1/(c\kappa \sigma_{pp})\approx 10^{15}\unit{s\,cm^{-3}}\).
Integrating this relation along the jet from the acceleration site gives,
\begin{equation}
  \frac{E_0}{E} = \exp[\int\limits_{Z\mathrm{min}}^{Z\mathrm{max}} \dd{z}\frac{\rho }{\tau_{pp} m_pv}]\,.
\end{equation}
The density of the jet plasma and its speed are related by \(\pi R^2 \rho v= \dot{M}\), where
\(\dot{M} \sim 3\times10^{19}\unit{g\,s^{-1}}\) for the case of SS~433. Thus, one obtains
\begin{equation}
  \frac{E_0}{E} = \exp[\int\limits_{Z\mathrm{min}}^{Z\mathrm{max}} \dd{z}\frac{\pi\rho^2 R^2 }{\tau_{pp} m_p\dot{M}}]\,,
\end{equation}
which for a conic jet with constant density, \(\rho_0\), yields
\begin{equation}
  \frac{E_0}{E} = \exp[\frac{\pi\rho_0^2 Z_{\mathrm{max}}^3\tan^2\theta_j }{3\tau_{pp} m_p\dot{M}}]\,.
\end{equation}
Finally, after some simplifications
\begin{equation}
  \frac{E_0}{E} = \exp[\frac{\rho_0 Z_{\mathrm{max}}^3}{3\tau_{pp} m_pZ^2_{\mathrm{min}}v_0}]\,.
\end{equation}
The coefficient under the exponent determines the efficiency of the pp process in the jet. For the fiducial values of the jet parameters, this coefficient is \(\approx 10^{-8}\). For a jet with a constant speed it should be even further decreased by a factor of \(3 Z_{\mathrm{min}}^2/Z_{\mathrm{max}}^2\approx 3\times 10^{-2}\). Thus proton-proton emission from the jet should be below \(10^{30}\unit{erg\,s^{-1}}\), thus it cannot be responsible even for the jet $\gamma$-ray emission, which exceeds \(10^{32}\unit{erg\,s^{-1}}\).

As explained above, for each jet geometry there are four basic models: constant speed with toroidal magnetic field (hereafter ``Model~A''); constant speed with poloidal magnetic field (``Model~B''); constant density with toroidal magnetic field (``Model~C''); and constant density with poloidal magnetic field (``Model~D''). These models are tested focusing on the western jet. The following data are used: the spectral energy distribution in X-rays (``XMM Full'' in Table~\ref{tab:data} and Fig.~\ref{fig:coord+regs}) and TeV $\gamma$-rays. For illustrative purposes, the \textit{Chandra} X-ray spectrum is also used, scaled to match the TeV spectrum extraction region \citep{hess_2024}. Even under the simplest assumptions, the number of model parameters describing the emission is significant; therefore, a detailed exploration of the parameter space is beyond the scope of this paper. For simplicity, a conical jet geometry with \(\theta_{j} = 12^\circ\) is adopted. The initial flow speed, magnetization, injection index, and the fraction of jet power transferred to relativistic electrons at the innermost part of the considered region (at \(Z\mysub{min} = 26.5~\unit{pc}\)) are adjusted to match the fluxes in the X-ray and TeV bands. The particle distribution is computed in the jet segment between \(Z\mysub{min}\) and \(Z\mysub{max} = 80~\unit{pc}\).

The initial magnetic field in the jet is chosen to match the ratio of the synchrotron (X-ray) and inverse Compton (TeV)
components, and the injection power is defined by the flux level. TeV emission detected with H.E.S.S. from the western
jet was approximately \(4\times 10^{-13}\unit{erg\,cm^{-2}\,s^{-1}}\) and the \xmm\ flux is at the level of
\(2\times10^{-11}\unit{erg\,cm^{-2}\,s^{-1}}\), thus the measured ratio of these components is \(\sim50\). The jet initial speed and injection index control the position of the cooling break and the spectrum hardness, respectively. 
The \xmm\ spectrum appears to be hard at keV energies, requiring either a cooling break at \(\hbar\omega\mysub{br}\gtrsim 3\)~keV or a very
hard injection spectrum. These spectral properties allow one to define the required model parameters for each scenario.

\begin{figure*}[htb]
    \begin{subfigure}[t]{0.475\textwidth}
        \includegraphics[width=\textwidth]{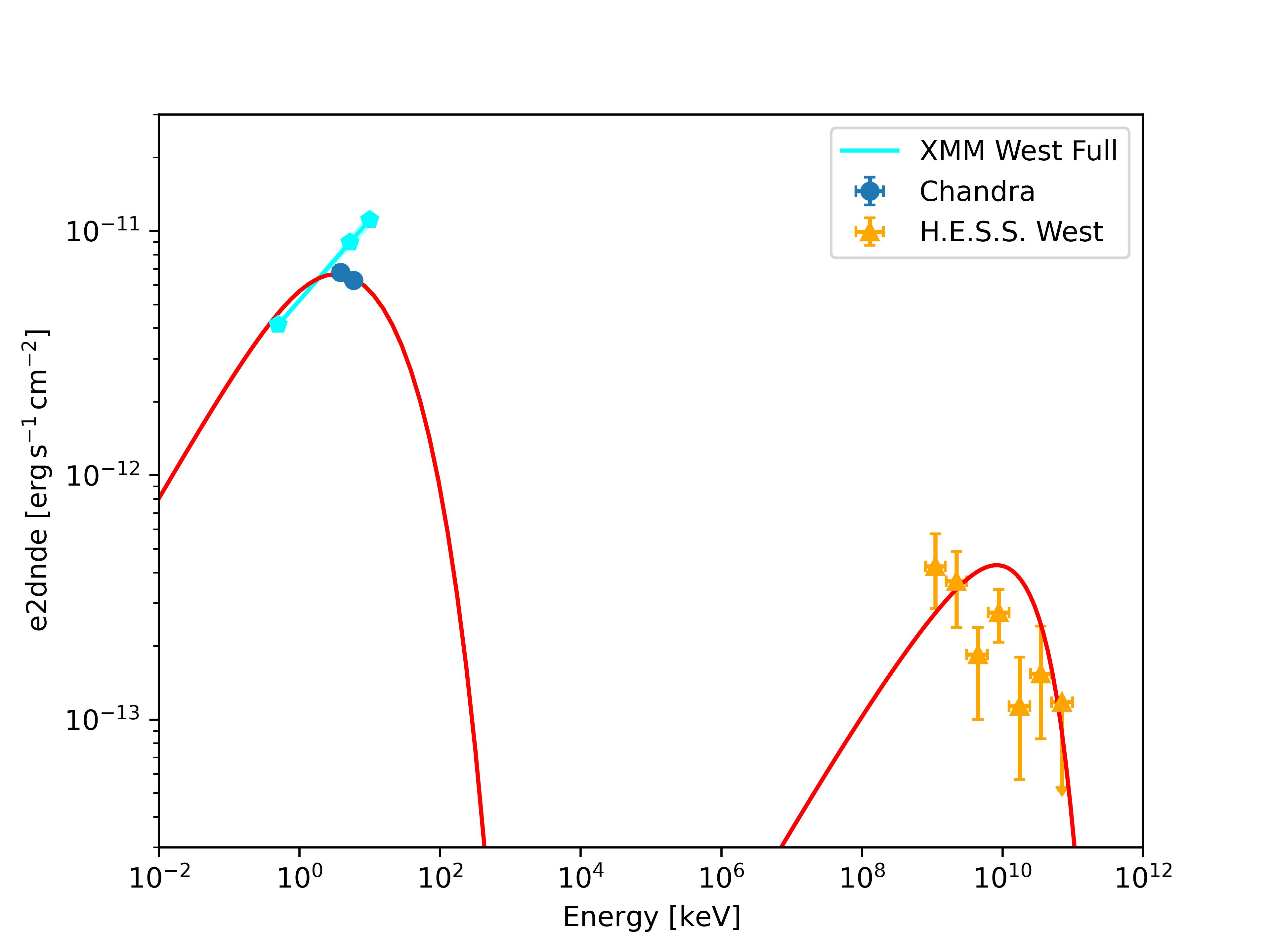}
        \caption{Model (A).} 
    \end{subfigure}
    \hfill
    \begin{subfigure}[t]{0.475\textwidth}
        \includegraphics[width=\textwidth]{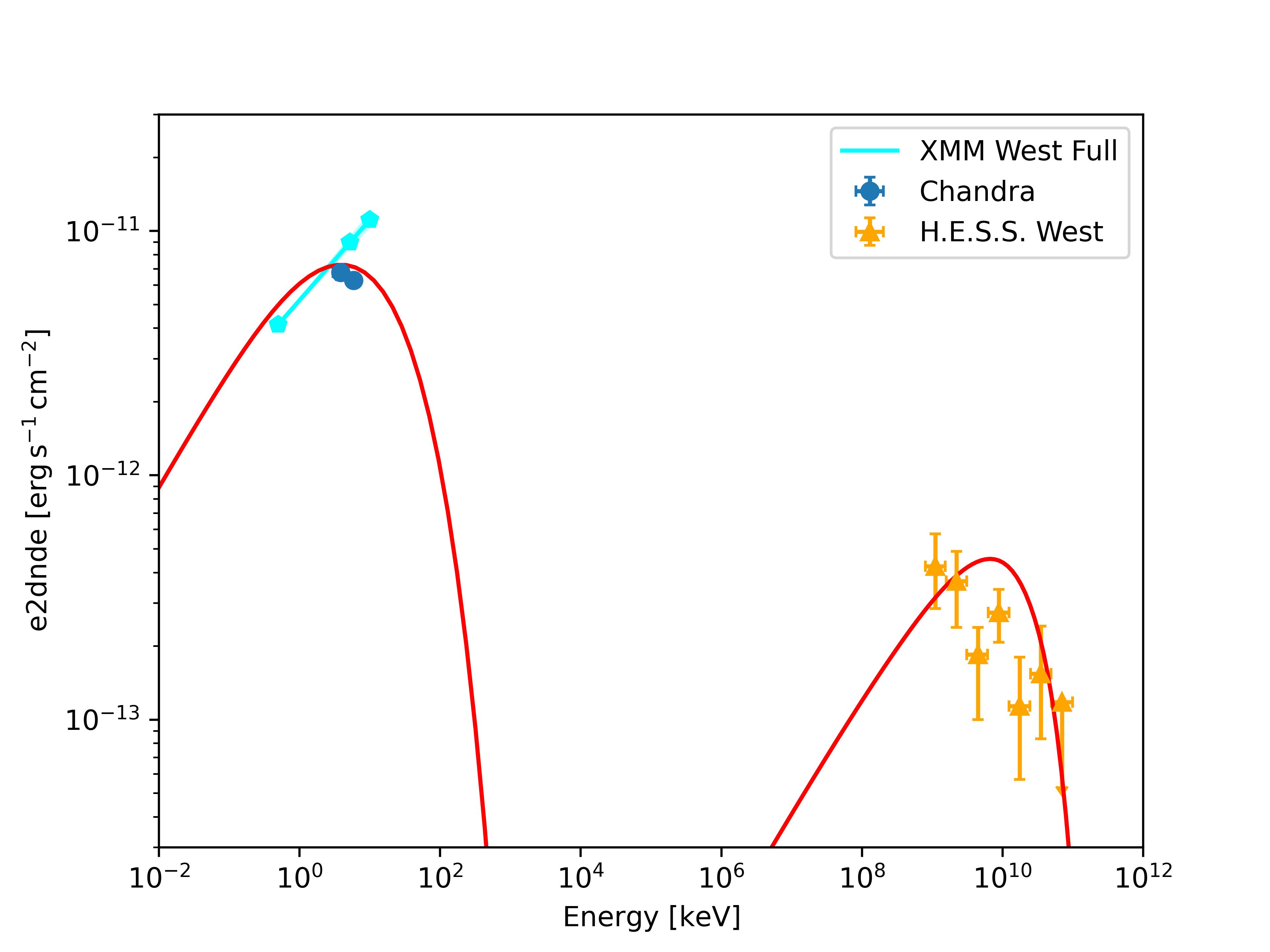}
        \caption{Model (B).} 
    \end{subfigure}
    \begin{subfigure}[t]{0.475\textwidth}
        \includegraphics[width=\textwidth]{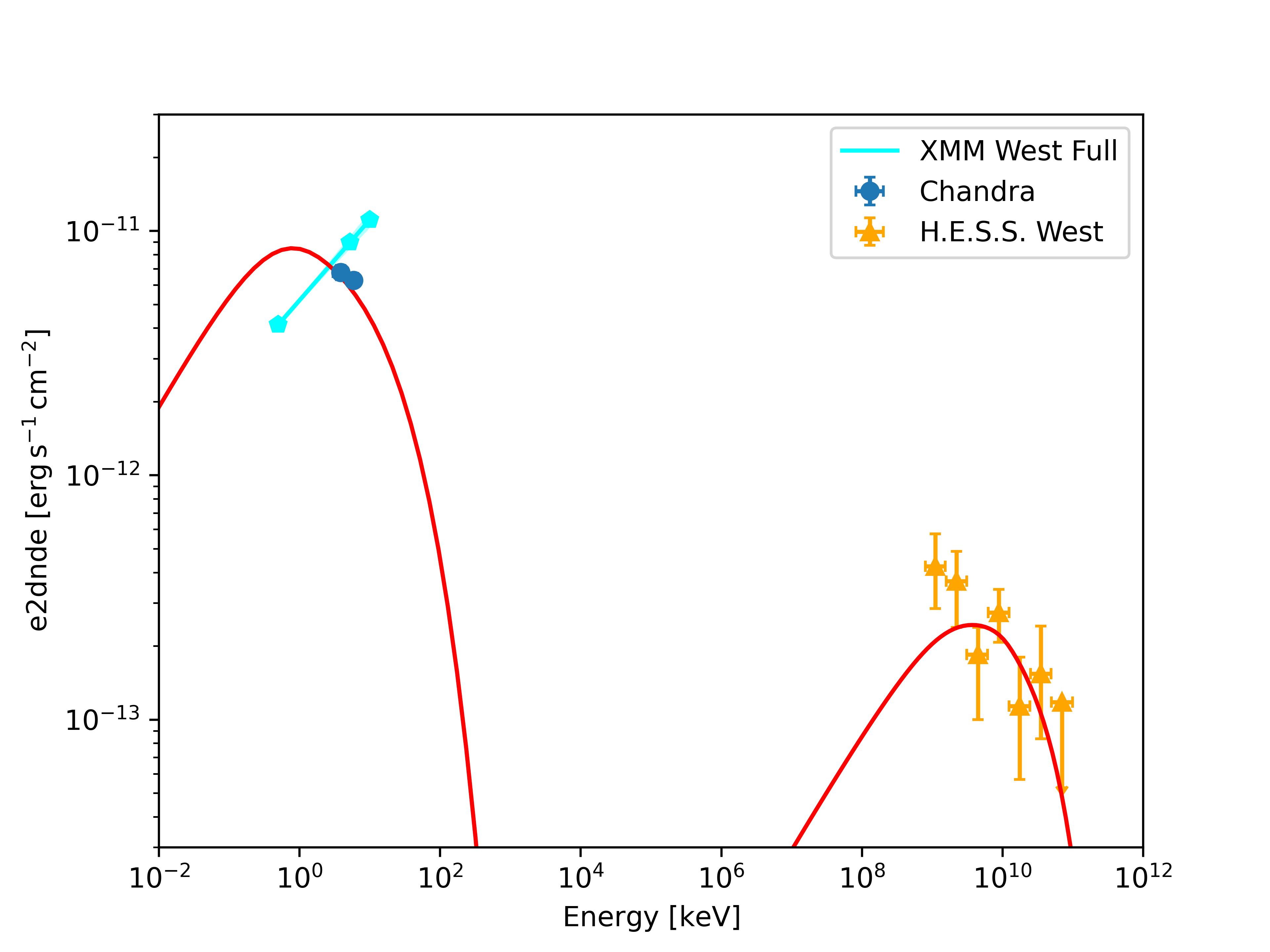}
        \caption{Model (C).} 
    \end{subfigure}
    \hfill
    \begin{subfigure}[t]{0.475\textwidth}
        \includegraphics[width=\textwidth]{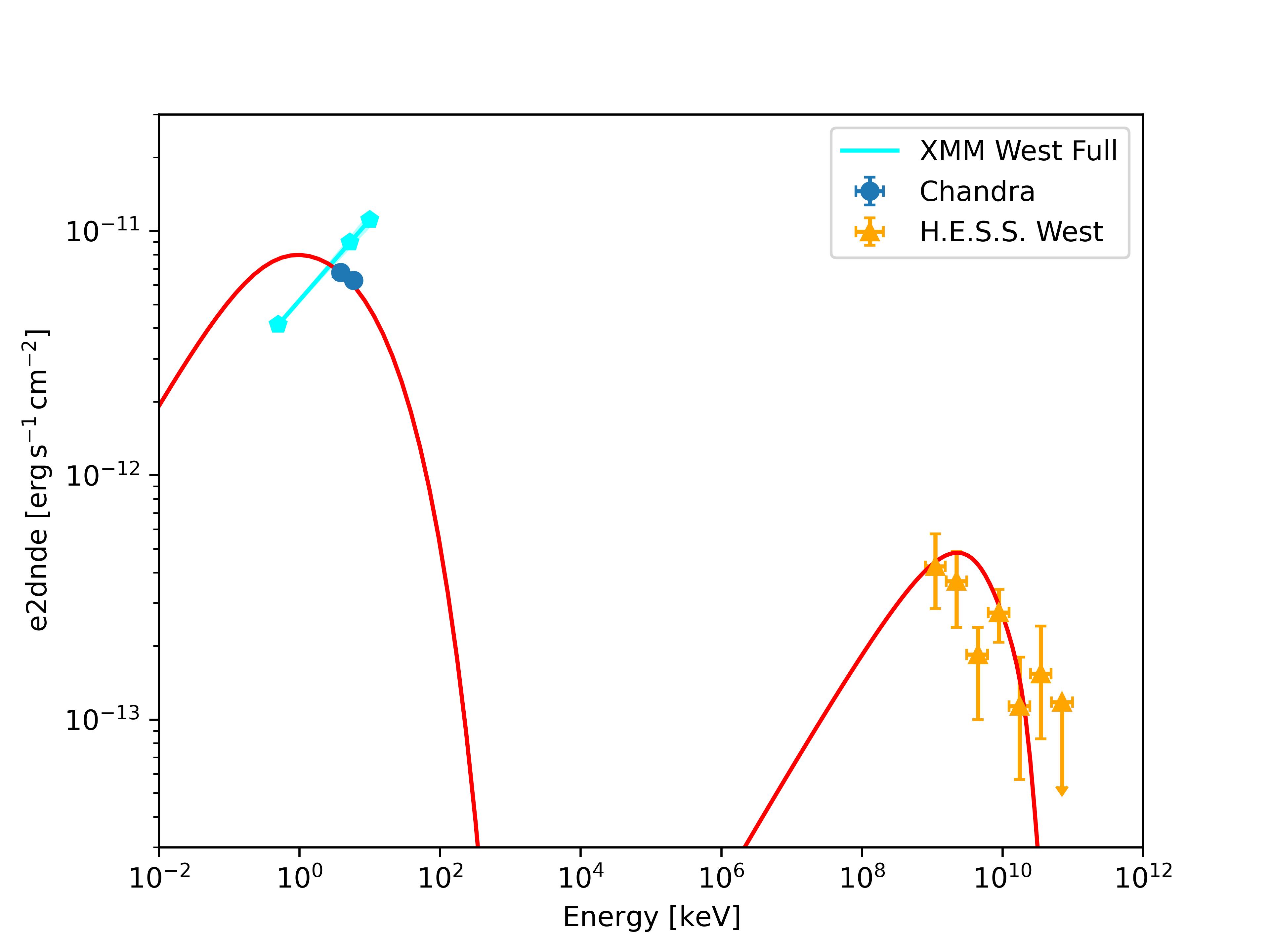}
        \caption{Model (D).} 
    \end{subfigure}
    \begin{subfigure}[t]{0.475\textwidth}
        \includegraphics[width=\textwidth]{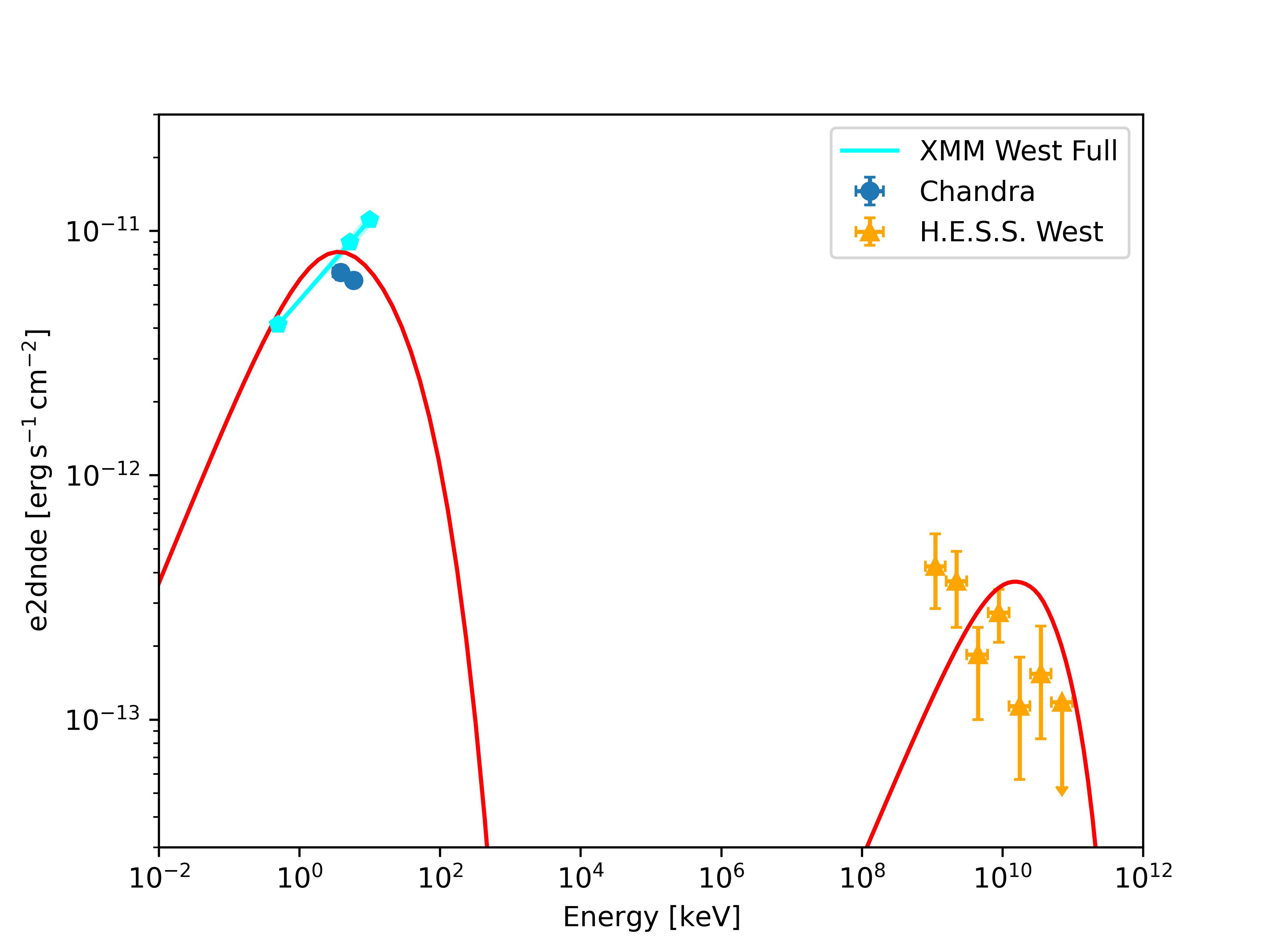}
        \caption{Model (C hard).} 
    \end{subfigure}
    \hfill
    \begin{subfigure}[t]{0.475\textwidth}
        \includegraphics[width=\textwidth]{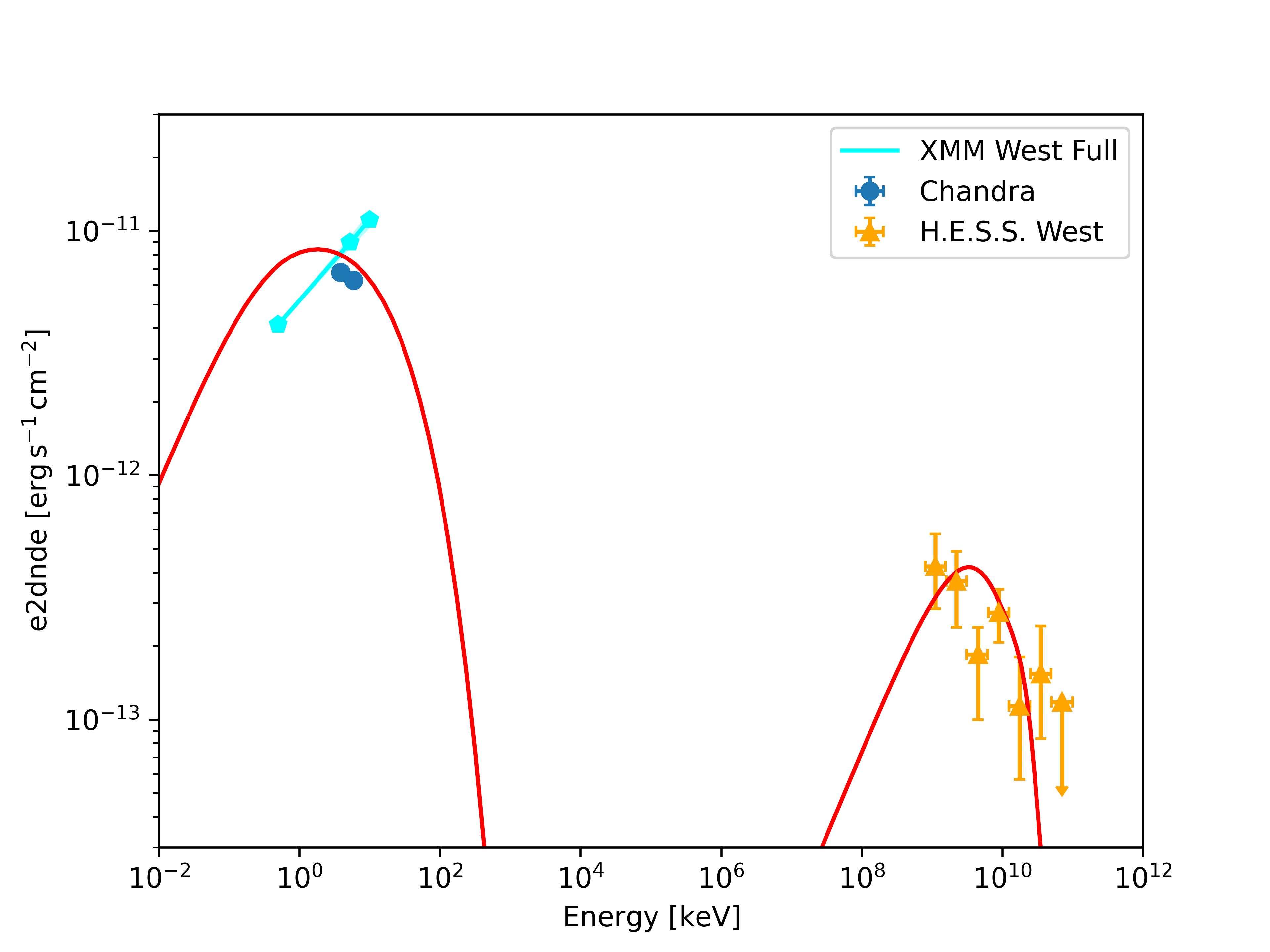}
        \caption{Model (D hard).} 
    \end{subfigure}
    \caption{Synchrotron and inverse Compton spectral energy distribution.} \label{fig:sed_west}
\end{figure*}

\emph{Model A:} In a constant speed jet to achieve the flux ratio one needs a magnetic field at the level of \(20\unit{\upmu G}\) and the position of the cooling break requires \(v_0\gg 0.1 c\). In Fig.~\ref{fig:sed_west} an example calculation is shown for \(B_{\varphi,0}=20\unit{\upmu G}\), \(v_0=0.5c\), and \(\kappa_e=1.5\%\).  \emph{Model B:} poloidal magnetic field decreases faster, thus similar SED requires a slightly higher strength of the field compared to Model A, \(B_{z,0}=25\unit{\upmu G}\) and lower power of the electron accelerator, \(\kappa_e=1.2\%\). A calculation for these parameters is shown in Fig.~\ref{fig:sed_west}.

Models B and D share the geometry of the magnetic field, but absence of adiabatic losses in constant-density case makes IC scattering efficient
in a larger part of the jet. Thus, Model D requires (as shown above) stronger magnetic field, and lower power of the
electron accelerator. In Fig.~\ref{fig:sed_west} the calculation for the Model D is shown with
\(B_{z,0}=50\unit{\upmu G}\). This strong magnetic field causes a cooling break moving down to
\(\hbar\omega\mysub{br}\approx1\)~keV, which hardens matching the X-ray slope revealed with \xmm. To achieve a
better spectral agreement, one can adopt a hard injection spectrum. For example, in Fig.~\ref{fig:sed_west} a
calculation with \(\alpha_e=1.6\) is shown.

Model C, which adopts a constant density and toroidal magnetic field implies even more efficient cooling, thus matching
the hard X-ray slope is even more challenging than in Model D. In Fig.~\ref{fig:sed_west} the SED is shown adopting a
quite weak initial field of \(B_{z,0}=10\unit{\upmu G}\). Adopting a harder injection spectrum, and even weaker magnetic
field, one can achieve a better agreement with the broadband spectra, e.g., in Fig.~\ref{fig:sed_west} also demonstrated is
a case with \(\alpha_e=1.6\).

\begin{table*}[htb]
    \caption{Other model parameters have fixed values for all cases: \(v_0=0.1c\), \(\beta_e=2\), \(\eta=10^3(c/v_0)^2\), \(Z\mysub{min}=26.5\)~pc, \(Z\mysub{max}=80\)~pc, \(\gamma\mysub{e,min}=10^4\), \(\theta_{o}=10^\circ\), \(\theta\mysub{j}=0.2\unit{rad}\approx11.5^\circ\). } \label{tab:models} \vspace{-0.1cm}
    \centering
    \begin{tabular}{l|c|c|c|c|c|c}
        \\
        \hline
          Parameters&\multicolumn{6}{c}{Models} \\ \hline
          & Model A & Model B & Model C & Model D & Model Ch &  Model Dh\\\hline
          \(B_{\varphi,0}\unit{\upmu G}\)&\(18\)&0&\(10\)&0&\(6\)&0\\
          \(B_{z,0}\unit{\upmu G}\)&0&\(25\)&0&\(50\)&0&\(50\)\\
          jet model&\multicolumn{2}{c|}{constant speed}&\multicolumn{4}{c}{constant density}\\
          \(\kappa_e\)&\(7\times10^{-3}\)&\(8\times10^{-3}\)&\(1\times10^{-3}\)&\(2\times10^{-3}\)&\(4\times10^{-4}\)&\(7\times10^{-4}\)\\
          \(\alpha_e\)&\multicolumn{4}{c|}{\(2\)}&\multicolumn{2}{c}{\(1.6\)}\\
          \hline
    \end{tabular}
\end{table*}

\clearpage


\bibliography{W50west}

@ARTICLE{1992ApJ...392..458C,
       author = {{Cioffi}, Denis F. and {Blondin}, John M.},
        title = "{The Evolution of Cocoons Surrounding Light, Extragalactic Jets}",
      journal = {\apj},
         year = 1992,
        month = jun,
       volume = {392},
        pages = {458},
          doi = {10.1086/171445},
       adsurl = {https://ui.adsabs.harvard.edu/abs/1992ApJ...392..458C}
}

@ARTICLE{2018ApJ...863..103S,
       author = {{Su}, Yang and {Zhou}, Xin and {Yang}, Ji and {Chen}, Yang and {Chen}, Xuepeng and {Zhang}, Shaobo},
        title = "{The Large-scale Interstellar Medium of SS 433/W50 Revisited}",
      journal = {\apj},
         year = 2018,
        month = aug,
       volume = {863},
       number = {1},
          eid = {103},
        pages = {103},
          doi = {10.3847/1538-4357/aad04e},
archivePrefix = {arXiv},
       eprint = {1807.03737},
 primaryClass = {astro-ph.GA},
       adsurl = {https://ui.adsabs.harvard.edu/abs/2018ApJ...863..103S}
}

@ARTICLE{2023PASJ...75..338S,
       author = {{Sakemi}, Haruka and {Machida}, Mami and {Yamamoto}, Hiroaki and {Tachihara}, Kengo},
        title = "{Molecular clouds at the eastern edge of radio nebula W 50}",
      journal = {\pasj},
         year = 2023,
        month = apr,
       volume = {75},
       number = {2},
        pages = {338-350},
          doi = {10.1093/pasj/psad001},
archivePrefix = {arXiv},
       eprint = {2301.13333},
 primaryClass = {astro-ph.GA},
       adsurl = {https://ui.adsabs.harvard.edu/abs/2023PASJ...75..338S}
}

@BOOK{1999acfp.book.....L,
    author = {{Lang}, K.~R.},
    title = "{Astrophysical formulae}",
    year = 1999,
    adsurl = {https://ui.adsabs.harvard.edu/abs/1999acfp.book.....L},
    publisher = {New York : Springer}
}

@ARTICLE{2024ApJ...975L..28C,
    author = {{Chi}, Yi-Heng and {Huang}, Jiahui and {Zhou}, Ping and {Feng}, Hua and {Li}, Xiang-Dong and {Markoff}, Sera B. and {Safi-Harb}, Samar and {Olivera-Nieto}, Laura},
    title = "{An X-Ray Shell Reveals the Supernova Explosion for Galactic Microquasar SS 433}",
    journal = {\apjl},
    year = 2024,
    month = nov,
    volume = {975},
    number = {2},
    eid = {L28},
    pages = {L28},
    doi = {10.3847/2041-8213/ad84ed},
    archivePrefix = {arXiv},
    eprint = {2410.06510},
    primaryClass = {astro-ph.HE},
    adsurl = {https://ui.adsabs.harvard.edu/abs/2024ApJ...975L..28C}
}

@ARTICLE{2025ApJ...993L..24T,
       author = {{Tsuji}, Naomi and {Inoue}, Yoshiyuki and {Khangulyan}, Dmitry and {Mori}, Kaya and {Safi-Harb}, Samar and {Tanaka}, Takaaki and {Olivera-Nieto}, Laura and {Mac Intyre}, Brydyn and {Kayama}, Kazuho and {Tsuru}, Takeshi Go and {Uchida}, Hiroyuki and {Fujiwara}, Tatsuki and {Aharonian}, Felix},
        title = "{The First Proper Motion Measurement of the Acceleration Regions in the Large-scale Jets of SS 433 Powering the W50 Nebula}",
      journal = {\apjl},
         year = 2025,
        month = nov,
       volume = {993},
       number = {1},
          eid = {L24},
        pages = {L24},
          doi = {10.3847/2041-8213/ae0e73},
archivePrefix = {arXiv},
       eprint = {2510.06431},
 primaryClass = {astro-ph.HE},
       adsurl = {https://ui.adsabs.harvard.edu/abs/2025ApJ...993L..24T}
}

@ARTICLE{2026A&A...707A.278S,
       author = {{Sunyaev}, Rashid and {Khabibullin}, Ildar and {Churazov}, Eugene and {Gilfanov}, Marat and {Medvedev}, Pavel and {Sazonov}, Sergey},
        title = "{X-ray panorama of the SS 433/W50 complex by SRG/eROSITA}",
      journal = {\aap},
         year = 2026,
        month = mar,
       volume = {707},
          eid = {A278},
        pages = {A278},
          doi = {10.1051/0004-6361/202557726},
archivePrefix = {arXiv},
       eprint = {2510.14938},
 primaryClass = {astro-ph.HE},
       adsurl = {https://ui.adsabs.harvard.edu/abs/2026A&A...707A.278S}
}

@ARTICLE{2025NSRev..12af496L,
       author = {{LHAASO Collaboration} and {Cao}, Zhen and {Aharonian}, Felix and {Bai}, Yun-Xiang and {Bao}, Yi-Wei and {Bastieri}, Denis and {Bi}, Xiao-Jun and {Bi}, Yu-Jiang and {Bian}, Wen-Yi and {Bukevich}, Anatoly V. and {Cai}, Chengmiao and {Cao}, Wen-Yu and {Cao}, Zhe and {Chang}, Jin and {Chang}, Jin-Fan and {Chen}, Aming and {Chen}, En-Sheng and {Chen}, Guohai and {Chen}, Hua-Xi and {Chen}, Liang and {Chen}, Long and {Chen}, Ming-Jun and {Chen}, Ma-Li and {Chen}, Qi-Hui and {Chen}, Shi and {Chen}, Su-Hong and {Chen}, Song-Zhan and {Chen}, Tian-Lu and {Chen}, Xiao-Bin and {Chen}, Xuejian and {Chen}, Yang and {Cheng}, Ning and {Cheng}, Yao-Dong and {Chung Chu}, Ming and {Cui}, Ming-Yang and {Cui}, Shu-Wang and {Cui}, Xiao-Hong and {Cui}, Yi-Dong and {Dai}, Ben-Zhong and {Dai}, Hong-Liang and {Dai}, Zigao and {Luobu}, Danzeng and {Diao}, Yang-Xuan and {Dong}, Xu-Qiang and {Duan}, Kai-Kai and {Fan}, Jun-Hui and {Fan}, Yi-Zhong and {Fang}, Jun and {Fang}, Jian-Hua and {Fang}, Kun and {Feng}, Cun-Feng and {Feng}, Hua and {Feng}, Li and {Feng}, Shaohui and {Feng}, Xiao-Ting and {Feng}, Yi and {Feng}, You-Liang and {Gabici}, Stefano and {Gao}, Bo and {Gao}, Chuan-Dong and {Gao}, Qi and {Gao}, Wei and {Gao}, Wei-Kang and {Ge}, Maomao and {Ge}, Ting-Ting and {Geng}, Lisi and {Giacinti}, Gwenael and {Gong}, Guanghua and {Gou}, Quanbu and {Gu}, Min-Hao and {Guo}, Fu-Lai and {Guo}, Jing and {Guo}, Xiao-Lei and {Guo}, Yi-Qing and {Guo}, Ying-Ying and {Han}, Yi-Ang and {Hannuksela}, Otto A. and {Hasan}, Mariam and {He}, Hui-Hai and {He}, Hao-Ning and {He}, Jia-Yin and {He}, Xinyu and {He}, Yu and {Hern{\'a}ndez-Cadena}, Sergio and {Hou}, Bo-Wen and {Hou}, Chao and {Hou}, Xian and {Hu}, Hong-Bo and {Hu}, Shi-Cong and {Huang}, Chen and {Huang}, Dai-Hui and {Huang}, Jiajun and {Huang}, Tian-Qi and {Huang}, Wen-Jun and {Huang}, Xing-Tao and {Huang}, Xiao-Yuan and {Huang}, Yong and {Huang}, Yi-Yun and {Ji}, Xiao-Lu and {Jia}, Huan-Yu and {Jia}, Kang and {Jiang}, Hou-Bing and {Jiang}, Kun and {Jiang}, Xiao-Wei and {Jiang}, Ze-Jun and {Jin}, Min and {Kaci}, Samy and {Kang}, Ming-Ming and {Karpikov}, Ivan and {Khangulyan}, Dmitry and {Kuleshov}, Denis and {Kurinov}, Kirill and {Li}, Bing-Bing and {Li}, Cheng and {Li}, Cong and {Li}, Dan and {Li}, Fei and {Li}, Haibo and {Li}, Huicai and {Li}, Jian and {Li}, Jie and {Li}, Kai and {Li}, Long and {Li}, Rong-Lan and {Li}, Si-Da and {Li}, Tian-Yang and {Li}, Wen-Lian and {Li}, Xiu-Rong and {Li}, Xin and {Li}, Yuan and {Li}, Yizhuo and {Li}, Zhe and {Li}, Zhuo and {Liang}, En-Wei and {Liang}, Yun-Feng and {Lin}, Su-Jie and {Liu}, Bing and {Liu}, Cheng and {Liu}, Dong and {Liu}, Dang-Bo and {Liu}, Hu and {Liu}, Hai-Dong and {Liu}, Jia and {Liu}, Jia-Li and {Liu}, Ji-Ren and {Liu}, Mao-Yuan and {Liu}, Ruo-Yu and {Liu}, Si-Ming and {Liu}, Wei and {Liu}, X. and {Liu}, Yi and {Liu}, Yu and {Liu}, Yi-Nong and {Lou}, Yu-Qing and {Luo}, Qing and {Luo}, Yu and {Lv}, Hong-Kui and {Ma}, Bo-Qiang and {Ma}, Ling-Ling and {Ma}, Xin-Hua and {Mao}, Ji-Rong and {Min}, Zhen and {Mitthumsiri}, Warit and {Mou}, Guo-Bin and {Mu}, Hui-Jun and {Neronov}, Andrii and {Ng}, Kenny Chun Yu and {Ni}, Ming-Yang and {Nie}, Lin and {Ou}, Le-Jian and {Pattarakijwanich}, Petchara and {Pei}, Zhi-Yuan and {Qi}, Jin-Can and {Qi}, Meng-Yao and {Qin}, Jia-Jun and {Raza}, Ali and {Ren}, Chong-Yang and {Ruffolo}, David and {S{\'a}iz}, Alejandro and {Semikoz}, Dmitri and {Shao}, Lang and {Shchegolev}, Oleg and {Shen}, Yun-Zhi and {Sheng}, Xiang-Dong and {Shi}, Zhaodong and {Shu}, Fu-Wen and {Song}, Hui-Chao and {Stenkin}, Yuri V. and {Stepanov}, Vladimir and {Su}, Yang and {Sun}, Dongxu and {Sun}, Hao and {Sun}, Qinning and {Sun}, Xiaona and {Sun}, Zhibin and {Hussain Tabasam}, Nabeel and {Takata}, Jumpei and {Tam}, Pak Hin Thomas and {Tan}, Hong-Bin and {Tang}, Qingwen},
        title = "{Ultrahigh-Energy Gamma-ray Emission Associated with Black Hole-Jet Systems}",
      journal = {National Science Review},
         year = 2025,
        month = dec,
       volume = {12},
       number = {12},
          eid = {nwaf496},
        pages = {nwaf496},
          doi = {10.1093/nsr/nwaf496},
archivePrefix = {arXiv},
       eprint = {2410.08988},
 primaryClass = {astro-ph.HE},
       adsurl = {https://ui.adsabs.harvard.edu/abs/2025NSRev..12af496L}
}

@ARTICLE{Marshall2002,
       author = {{Marshall}, Herman L. and {Canizares}, Claude R. and {Schulz}, Norbert S.},
        title = "{The High-Resolution X-Ray Spectrum of SS 433 Using the Chandra HETGS}",
      journal = {\apj},
         year = 2002,
        month = jan,
       volume = {564},
       number = {2},
        pages = {941-952},
          doi = {10.1086/324398},
archivePrefix = {arXiv},
       eprint = {astro-ph/0108206},
 primaryClass = {astro-ph},
       adsurl = {https://ui.adsabs.harvard.edu/abs/2002ApJ...564..941M}
}

@ARTICLE{Wik2014,
       author = {{Wik}, Daniel R. and {Hornstrup}, A. and {Molendi}, S. and {Madejski}, G. and {Harrison}, F.~A. and {Zoglauer}, A. and {Grefenstette}, B.~W. and {Gastaldello}, F. and {Madsen}, K.~K. and {Westergaard}, N.~J. and {Ferreira}, D.~D.~M. and {Kitaguchi}, T. and {Pedersen}, K. and {Boggs}, S.~E. and {Christensen}, F.~E. and {Craig}, W.~W. and {Hailey}, C.~J. and {Stern}, D. and {Zhang}, W.~W.},
        title = "{NuSTAR Observations of the Bullet Cluster: Constraints on Inverse Compton Emission}",
      journal = {\apj},
         year = 2014,
        month = sep,
       volume = {792},
       number = {1},
          eid = {48},
        pages = {48},
          doi = {10.1088/0004-637X/792/1/48},
archivePrefix = {arXiv},
       eprint = {1403.2722},
 primaryClass = {astro-ph.HE},
       adsurl = {https://ui.adsabs.harvard.edu/abs/2014ApJ...792...48W}
}

@ARTICLE{1999ApJ...512..784S,
       author = {{Safi-Harb}, Samar and {Petre}, Robert},
        title = "{Rossi X-Ray Timing Explorer Observations of the Eastern Lobe of W50 Associated with SS 433}",
      journal = {\apj},
         year = 1999,
        month = feb,
       volume = {512},
       number = {2},
        pages = {784-792},
          doi = {10.1086/306803},
       adsurl = {https://ui.adsabs.harvard.edu/abs/1999ApJ...512..784S}
}

@ARTICLE{1997ApJ...483..868S,
       author = {{Safi-Harb}, Samar and {{\"O}gelman}, Hakki},
        title = "{ROSAT and ASCA Observations of W50 Associated with the Peculiar Source SS 433}",
      journal = {\apj},
         year = 1997,
        month = jul,
       volume = {483},
       number = {2},
        pages = {868-881},
          doi = {10.1086/304274},
       adsurl = {https://ui.adsabs.harvard.edu/abs/1997ApJ...483..868S}
}

@ARTICLE{1994PASJ...46L.109Y,
       author = {{Yamauchi}, Shigeo and {Kawai}, Nobuyuki and {Aoki}, Takashi},
        title = "{A Non-Thermal X-Ray Spectrum from the Supernova Remnant W 50}",
      journal = {\pasj},
         year = 1994,
        month = jun,
       volume = {46},
        pages = {L109-L113},
       adsurl = {https://ui.adsabs.harvard.edu/abs/1994PASJ...46L.109Y}
}

@ARTICLE{1998AJ....116.1842D,
       author = {{Dubner}, G.~M. and {Holdaway}, M. and {Goss}, W.~M. and {Mirabel}, I.~F.},
        title = "{A High-Resolution Radio Study of the W50-SS 433 System and the Surrounding Medium}",
      journal = {\aj},
         year = 1998,
        month = oct,
       volume = {116},
       number = {4},
        pages = {1842-1855},
          doi = {10.1086/300537},
       adsurl = {https://ui.adsabs.harvard.edu/abs/1998AJ....116.1842D}
}

@ARTICLE{2007A&A...463..611B,
       author = {{Brinkmann}, W. and {Pratt}, G.~W. and {Rohr}, S. and {Kawai}, N. and {Burwitz}, V.},
        title = "{XMM-Newton observations of the eastern jet of SS 433}",
      journal = {\aap},
         year = 2007,
        month = feb,
       volume = {463},
       number = {2},
        pages = {611-619},
          doi = {10.1051/0004-6361:20065570},
archivePrefix = {arXiv},
       eprint = {astro-ph/0610781},
 primaryClass = {astro-ph},
       adsurl = {https://ui.adsabs.harvard.edu/abs/2007A&A...463..611B}
}

@ARTICLE{1984ARA&A..22..507M,
       author = {{Margon}, Bruce},
        title = "{Observations of SS 433}",
      journal = {\araa},
         year = 1984,
        month = jan,
       volume = {22},
        pages = {507-536},
          doi = {10.1146/annurev.aa.22.090184.002451},
       adsurl = {https://ui.adsabs.harvard.edu/abs/1984ARA&A..22..507M}
}

@ARTICLE{2011MNRAS.414.2838G,
       author = {{Goodall}, Paul T. and {Alouani-Bibi}, Fathallah and {Blundell}, Katherine M.},
        title = "{When microquasar jets and supernova collide: hydrodynamically simulating the SS 433-W 50 interaction}",
      journal = {\mnras},
         year = 2011,
        month = jul,
       volume = {414},
       number = {4},
        pages = {2838-2859},
          doi = {10.1111/j.1365-2966.2011.18388.x},
archivePrefix = {arXiv},
       eprint = {1101.3486},
 primaryClass = {astro-ph.SR},
       adsurl = {https://ui.adsabs.harvard.edu/abs/2011MNRAS.414.2838G}
}

@ARTICLE{2004ApJ...616L.159B,
       author = {{Blundell}, Katherine M. and {Bowler}, Michael G.},
        title = "{Symmetry in the Changing Jets of SS 433 and Its True Distance from Us}",
      journal = {\apjl},
         year = 2004,
        month = dec,
       volume = {616},
       number = {2},
        pages = {L159-L162},
          doi = {10.1086/426542},
archivePrefix = {arXiv},
       eprint = {astro-ph/0410456},
 primaryClass = {astro-ph},
       adsurl = {https://ui.adsabs.harvard.edu/abs/2004ApJ...616L.159B}
}

@ARTICLE{farnes2017,
       author = {{Farnes}, J.~S. and {Gaensler}, B.~M. and {Purcell}, C. and {Sun}, X.~H. and {Haverkorn}, M. and {Lenc}, E. and {O'Sullivan}, S.~P. and {Akahori}, T.},
        title = "{Interacting large-scale magnetic fields and ionized gas in the W50/SS433 system}",
      journal = {\mnras},
         year = 2017,
        month = jun,
       volume = {467},
       number = {4},
        pages = {4777-4801},
          doi = {10.1093/mnras/stx338},
archivePrefix = {arXiv},
       eprint = {1604.06552},
 primaryClass = {astro-ph.GA},
       adsurl = {https://ui.adsabs.harvard.edu/abs/2017MNRAS.467.4777F}
}

@ARTICLE{2005AdSpR..35.1062M,
       author = {{Moldowan}, A. and {Safi-Harb}, S. and {Fuchs}, Y. and {Dubner}, G.},
        title = "{A multi-wavelength study of the western lobe of W50 powered by the galactic microquasar SS 433}",
      journal = {Advances in Space Research},
         year = 2005,
        month = jan,
       volume = {35},
       number = {6},
        pages = {1062-1065},
          doi = {10.1016/j.asr.2005.01.086},
archivePrefix = {arXiv},
       eprint = {astro-ph/0501361},
 primaryClass = {astro-ph},
       adsurl = {https://ui.adsabs.harvard.edu/abs/2005AdSpR..35.1062M}
}

@ARTICLE{2005A&A...439..635A,
       author = {{Aharonian}, F. and {Akhperjanian}, A. and {Beilicke}, M. and {Bernl{\"o}hr}, K. and {B{\"o}rst}, H. -G. and {Bojahr}, H. and {Bolz}, O. and {Coarasa}, T. and {Contreras}, J. and {Cortina}, J. and {Denninghoff}, S. and {Fonseca}, V. and {Girma}, M. and {G{\"o}tting}, N. and {Heinzelmann}, G. and {Hermann}, G. and {Heusler}, A. and {Hofmann}, W. and {Horns}, D. and {Jung}, I. and {Kankanyan}, R. and {Kestel}, M. and {Kohnle}, A. and {Konopelko}, A. and {Kranich}, D. and {Lampeitl}, H. and {Lopez}, M. and {Lorenz}, E. and {Lucarelli}, F. and {Mang}, O. and {Mazin}, D. and {Meyer}, H. and {Mirzoyan}, R. and {Moralejo}, A. and {O{\~n}a-Wilhelmi}, E. and {Panter}, M. and {Plyasheshnikov}, A. and {P{\"u}hlhofer}, G. and {de Los Reyes}, R. and {Rhode}, W. and {Ripken}, J. and {Rowell}, G.~P. and {Sahakian}, V. and {Samorski}, M. and {Schilling}, M. and {Siems}, M. and {Sobzynska}, D. and {Stamm}, W. and {Tluczykont}, M. and {Vitale}, V. and {V{\"o}lk}, H.~J. and {Wiedner}, C.~A. and {Wittek}, W.},
        title = "{TeV gamma-ray observations of SS-433 and a survey of the surrounding field with the HEGRA IACT-System}",
      journal = {\aap},
         year = 2005,
        month = aug,
       volume = {439},
       number = {2},
        pages = {635-643},
          doi = {10.1051/0004-6361:20042248},
       adsurl = {https://ui.adsabs.harvard.edu/abs/2005A&A...439..635A}
}

@ARTICLE{2018Natur.562...82A,
       author = {{Abeysekara}, A.~U. and {Albert}, A. and {Alfaro}, R. and {Alvarez}, C. and {{\'A}lvarez}, J.~D. and {Arceo}, R. and {Arteaga-Vel{\'a}zquez}, J.~C. and {Avila Rojas}, D. and {Ayala Solares}, H.~A. and {Belmont-Moreno}, E. and {BenZvi}, S.~Y. and {Brisbois}, C. and {Caballero-Mora}, K.~S. and {Capistr{\'a}n}, T. and {Carrami{\~n}ana}, A. and {Casanova}, S. and {Castillo}, M. and {Cotti}, U. and {Cotzomi}, J. and {Couti{\~n}o de Le{\'o}n}, S. and {De Le{\'o}n}, C. and {De la Fuente}, E. and {D{\'\i}az-V{\'e}lez}, J.~C. and {Dichiara}, S. and {Dingus}, B.~L. and {DuVernois}, M.~A. and {Ellsworth}, R.~W. and {Engel}, K. and {Espinoza}, C. and {Fang}, K. and {Fleischhack}, H. and {Fraija}, N. and {Galv{\'a}n-G{\'a}mez}, A. and {Garc{\'\i}a-Gonz{\'a}lez}, J.~A. and {Garfias}, F. and {Gonz{\'a}lez-Mu{\~n}oz}, A. and {Gonz{\'a}lez}, M.~M. and {Goodman}, J.~A. and {Hampel-Arias}, Z. and {Harding}, J.~P. and {Hernandez}, S. and {Hinton}, J. and {Hona}, B. and {Hueyotl-Zahuantitla}, F. and {Hui}, C.~M. and {H{\"u}ntemeyer}, P. and {Iriarte}, A. and {Jardin-Blicq}, A. and {Joshi}, V. and {Kaufmann}, S. and {Kar}, P. and {Kunde}, G.~J. and {Lauer}, R.~J. and {Lee}, W.~H. and {Le{\'o}n Vargas}, H. and {Li}, H. and {Linnemann}, J.~T. and {Longinotti}, A.~L. and {Luis-Raya}, G. and {L{\'o}pez-Coto}, R. and {Malone}, K. and {Marinelli}, S.~S. and {Martinez}, O. and {Martinez-Castellanos}, I. and {Mart{\'\i}nez-Castro}, J. and {Matthews}, J.~A. and {Miranda-Romagnoli}, P. and {Moreno}, E. and {Mostaf{\'a}}, M. and {Nayerhoda}, A. and {Nellen}, L. and {Newbold}, M. and {Nisa}, M.~U. and {Noriega-Papaqui}, R. and {Pretz}, J. and {P{\'e}rez-P{\'e}rez}, E.~G. and {Ren}, Z. and {Rho}, C.~D. and {Rivi{\`e}re}, C. and {Rosa-Gonz{\'a}lez}, D. and {Rosenberg}, M. and {Ruiz-Velasco}, E. and {Salesa Greus}, F. and {Sandoval}, A. and {Schneider}, M. and {Schoorlemmer}, H. and {Seglar Arroyo}, M. and {Sinnis}, G. and {Smith}, A.~J. and {Springer}, R.~W. and {Surajbali}, P. and {Taboada}, I. and {Tibolla}, O. and {Tollefson}, K. and {Torres}, I. and {Vianello}, G. and {Villase{\~n}or}, L. and {Weisgarber}, T. and {Werner}, F. and {Westerhoff}, S. and {Wood}, J. and {Yapici}, T. and {Yodh}, G. and {Zepeda}, A. and {Zhang}, H. and {Zhou}, H.},
        title = "{Very-high-energy particle acceleration powered by the jets of the microquasar SS 433}",
      journal = {\nat},
         year = 2018,
        month = oct,
       volume = {562},
       number = {7725},
        pages = {82-85},
          doi = {10.1038/s41586-018-0565-5},
       adsurl = {https://ui.adsabs.harvard.edu/abs/2018Natur.562...82A}
}

@ARTICLE{2018arXiv181001892H,
       author = {{HAWC Collaboration} and {Abeysekara}, A.~U. and {Albert}, A. and {Alfaro}, R. and {Alvarez}, C. and {{\'A}lvarez}, J.~D. and {Arceo}, R. and {Arteaga-Vel{\'a}zquez}, J.~C. and {Avila Rojas}, D. and {Ayala Solares}, H.~A. and {Belmont-Moreno}, E. and {BenZvi}, S.~Y. and {Brisbois}, C. and {Caballero-Mora}, K.~S. and {Capistr{\'a}n}, T. and {Carrami{\~n}ana}, A. and {Casanova}, S. and {Castillo}, M. and {Cotti}, U. and {Cotzomi}, J. and {Couti{\~n}o de Le{\'o}n}, S. and {De Le{\'o}n}, C. and {De la Fuente}, E. and {D{\'\i}az-V{\'e}lez}, J.~C. and {Dichiara}, S. and {Dingus}, B.~L. and {DuVernois}, M.~A. and {Ellsworth}, R.~W. and {Engel}, K. and {Espinoza}, C. and {Fang}, K. and {Fleischhack}, H. and {Fraija}, N. and {Galv{\'a}n-G{\'a}mez}, A. and {Garc{\'\i}a-Gonz{\'a}lez}, J.~A. and {Garfias}, F. and {Gonz{\'a}lez Mu{\~n}oz}, A. and {Gonz{\'a}lez}, M.~M. and {Goodman}, J.~A. and {Hampel-Arias}, Z. and {Harding}, J.~P. and {Hernandez}, S. and {Hinton}, J. and {Hona}, B. and {Hueyotl-Zahuantitla}, F. and {Hui}, C.~M. and {H{\"u}ntemeyer}, P. and {Iriarte}, A. and {Jardin-Blicq}, A. and {Joshi}, V. and {Kaufmann}, S. and {Kar}, P. and {Kunde}, G.~J. and {Lauer}, R.~J. and {Lee}, W.~H. and {Le{\'o}n Vargas}, H. and {Li}, H. and {Linnemann}, J.~T. and {Longinotti}, A.~L. and {Luis-Raya}, G. and {L{\'o}pez-Coto}, R. and {Malone}, K. and {Marinelli}, S.~S. and {Martinez}, O. and {Martinez-Castellanos}, I. and {Mart{\'\i}nez-Castro}, J. and {Matthews}, J.~A. and {Miranda-Romagnoli}, P. and {Moreno}, E. and {Mostaf{\'a}}, M. and {Nayerhoda}, A. and {Nellen}, L. and {Newbold}, M. and {Nisa}, M.~U. and {Noriega-Papaqui}, R. and {P{\'e}rez-P{\'e}rez}, E.~G. and {Pretz}, J. and {Ren}, Z. and {Rho}, C.~D. and {Rivi{\`e}re}, C. and {Rosa-Gonz{\'a}lez}, D. and {Rosenberg}, M. and {Ruiz-Velasco}, E. and {Salesa Greus}, F. and {Sandoval}, A. and {Schneider}, M. and {Schoorlemmer}, H. and {Seglar Arroyo}, M. and {Sinnis}, G. and {Smith}, A.~J. and {Springer}, R.~W. and {Surajbali}, P. and {Taboada}, I. and {Tibolla}, O. and {Tollefson}, K. and {Torres}, I. and {Vianello}, G. and {Villase{\~n}or}, L. and {Weisgarber}, T. and {Werner}, F. and {Westerhoff}, S. and {Wood}, J. and {Yapici}, T. and {Yodh}, G. and {Zepeda}, A. and {Zhang}, H. and {Zhou}, H.},
        title = "{Very high energy particle acceleration powered by the jets of the microquasar SS 433}",
      journal = {arXiv e-prints},
         year = 2018,
        month = oct,
          eid = {arXiv:1810.01892},
        pages = {arXiv:1810.01892},
archivePrefix = {arXiv},
       eprint = {1810.01892},
 primaryClass = {astro-ph.HE},
       adsurl = {https://ui.adsabs.harvard.edu/abs/2018arXiv181001892H}
}

@ARTICLE{2001A&A...365L..27T,
       author = {{Turner}, M.~J.~L. and {Abbey}, A. and {Arnaud}, M. and {Balasini}, M. and {Barbera}, M. and {Belsole}, E. and {Bennie}, P.~J. and {Bernard}, J.~P. and {Bignami}, G.~F. and {Boer}, M. and {Briel}, U. and {Butler}, I. and {Cara}, C. and {Chabaud}, C. and {Cole}, R. and {Collura}, A. and {Conte}, M. and {Cros}, A. and {Denby}, M. and {Dhez}, P. and {Di Coco}, G. and {Dowson}, J. and {Ferrando}, P. and {Ghizzardi}, S. and {Gianotti}, F. and {Goodall}, C.~V. and {Gretton}, L. and {Griffiths}, R.~G. and {Hainaut}, O. and {Hochedez}, J.~F. and {Holland}, A.~D. and {Jourdain}, E. and {Kendziorra}, E. and {Lagostina}, A. and {Laine}, R. and {La Palombara}, N. and {Lortholary}, M. and {Lumb}, D. and {Marty}, P. and {Molendi}, S. and {Pigot}, C. and {Poindron}, E. and {Pounds}, K.~A. and {Reeves}, J.~N. and {Reppin}, C. and {Rothenflug}, R. and {Salvetat}, P. and {Sauvageot}, J.~L. and {Schmitt}, D. and {Sembay}, S. and {Short}, A.~D.~T. and {Spragg}, J. and {Stephen}, J. and {Str{\"u}der}, L. and {Tiengo}, A. and {Trifoglio}, M. and {Tr{\"u}mper}, J. and {Vercellone}, S. and {Vigroux}, L. and {Villa}, G. and {Ward}, M.~J. and {Whitehead}, S. and {Zonca}, E.},
        title = "{The European Photon Imaging Camera on XMM-Newton: The MOS cameras}",
      journal = {\aap},
         year = 2001,
        month = jan,
       volume = {365},
        pages = {L27-L35},
          doi = {10.1051/0004-6361:20000087},
archivePrefix = {arXiv},
       eprint = {astro-ph/0011498},
 primaryClass = {astro-ph},
       adsurl = {https://ui.adsabs.harvard.edu/abs/2001A&A...365L..27T}
}

@ARTICLE{2001A&A...365L..18S,
       author = {{Str{\"u}der}, L. and {Briel}, U. and {Dennerl}, K. and {Hartmann}, R. and {Kendziorra}, E. and {Meidinger}, N. and {Pfeffermann}, E. and {Reppin}, C. and {Aschenbach}, B. and {Bornemann}, W. and {Br{\"a}uninger}, H. and {Burkert}, W. and {Elender}, M. and {Freyberg}, M. and {Haberl}, F. and {Hartner}, G. and {Heuschmann}, F. and {Hippmann}, H. and {Kastelic}, E. and {Kemmer}, S. and {Kettenring}, G. and {Kink}, W. and {Krause}, N. and {M{\"u}ller}, S. and {Oppitz}, A. and {Pietsch}, W. and {Popp}, M. and {Predehl}, P. and {Read}, A. and {Stephan}, K.~H. and {St{\"o}tter}, D. and {Tr{\"u}mper}, J. and {Holl}, P. and {Kemmer}, J. and {Soltau}, H. and {St{\"o}tter}, R. and {Weber}, U. and {Weichert}, U. and {von Zanthier}, C. and {Carathanassis}, D. and {Lutz}, G. and {Richter}, R.~H. and {Solc}, P. and {B{\"o}ttcher}, H. and {Kuster}, M. and {Staubert}, R. and {Abbey}, A. and {Holland}, A. and {Turner}, M. and {Balasini}, M. and {Bignami}, G.~F. and {La Palombara}, N. and {Villa}, G. and {Buttler}, W. and {Gianini}, F. and {Lain{\'e}}, R. and {Lumb}, D. and {Dhez}, P.},
        title = "{The European Photon Imaging Camera on XMM-Newton: The pn-CCD camera}",
      journal = {\aap},
         year = 2001,
        month = jan,
       volume = {365},
        pages = {L18-L26},
          doi = {10.1051/0004-6361:20000066},
       adsurl = {https://ui.adsabs.harvard.edu/abs/2001A&A...365L..18S}
}

@INPROCEEDINGS{2006SPIE.6270E..1VF,
       author = {{Fruscione}, Antonella and {McDowell}, Jonathan C. and {Allen}, Glenn E. and {Brickhouse}, Nancy S. and {Burke}, Douglas J. and {Davis}, John E. and {Durham}, Nick and {Elvis}, Martin and {Galle}, Elizabeth C. and {Harris}, Daniel E. and {Huenemoerder}, David P. and {Houck}, John C. and {Ishibashi}, Bish and {Karovska}, Margarita and {Nicastro}, Fabrizio and {Noble}, Michael S. and {Nowak}, Michael A. and {Primini}, Frank A. and {Siemiginowska}, Aneta and {Smith}, Randall K. and {Wise}, Michael},
        title = "{CIAO: Chandra's data analysis system}",
    booktitle = {Society of Photo-Optical Instrumentation Engineers (SPIE) Conference Series},
         year = 2006,
       editor = {{Silva}, David R. and {Doxsey}, Rodger E.},
       series = {Society of Photo-Optical Instrumentation Engineers (SPIE) Conference Series},
       volume = {6270},
        month = jun,
          eid = {62701V},
        pages = {62701V},
          doi = {10.1117/12.671760},
       adsurl = {https://ui.adsabs.harvard.edu/abs/2006SPIE.6270E..1VF}
}

@ARTICLE{2000ApJ...542..914W,
       author = {{Wilms}, J. and {Allen}, A. and {McCray}, R.},
        title = "{On the Absorption of X-Rays in the Interstellar Medium}",
      journal = {\apj},
         year = 2000,
        month = oct,
       volume = {542},
       number = {2},
        pages = {914-924},
          doi = {10.1086/317016},
archivePrefix = {arXiv},
       eprint = {astro-ph/0008425},
 primaryClass = {astro-ph},
       adsurl = {https://ui.adsabs.harvard.edu/abs/2000ApJ...542..914W}
}

@INPROCEEDINGS{1996ASPC..101...17A,
       author = {{Arnaud}, K.~A.},
        title = "{XSPEC: The First Ten Years}",
    booktitle = {Astronomical Data Analysis Software and Systems V},
         year = 1996,
       editor = {{Jacoby}, George H. and {Barnes}, Jeannette},
       series = {Astronomical Society of the Pacific Conference Series},
       volume = {101},
        month = jan,
        pages = {17},
       adsurl = {https://ui.adsabs.harvard.edu/abs/1996ASPC..101...17A}
}

@ARTICLE{2018A&A...612A..14M,
       author = {{MAGIC Collaboration} and {Ahnen}, M.~L. and {Ansoldi}, S. and {Antonelli}, L.~A. and {Arcaro}, C. and {Babi{\'c}}, A. and {Banerjee}, B. and {Bangale}, P. and {Barres de Almeida}, U. and {Barrio}, J.~A. and {Becerra Gonz{\'a}lez}, J. and {Bednarek}, W. and {Bernardini}, E. and {Berti}, A. and {Biasuzzi}, B. and {Biland}, A. and {Blanch}, O. and {Bonnefoy}, S. and {Bonnoli}, G. and {Borracci}, F. and {Carosi}, R. and {Carosi}, A. and {Chatterjee}, A. and {Colin}, P. and {Colombo}, E. and {Contreras}, J.~L. and {Cortina}, J. and {Covino}, S. and {Cumani}, P. and {da Vela}, P. and {Dazzi}, F. and {de Angelis}, A. and {de Lotto}, B. and {de O{\~n}a Wilhelmi}, E. and {di Pierro}, F. and {Doert}, M. and {Dom{\'\i}nguez}, A. and {Dominis Prester}, D. and {Dorner}, D. and {Doro}, M. and {Einecke}, S. and {Eisenacher Glawion}, D. and {Elsaesser}, D. and {Engelkemeier}, M. and {Fallah Ramazani}, V. and {Fern{\'a}ndez-Barral}, A. and {Fidalgo}, D. and {Fonseca}, M.~V. and {Font}, L. and {Fruck}, C. and {Galindo}, D. and {Garc{\'\i}a L{\'o}pez}, R.~J. and {Garczarczyk}, M. and {Gaug}, M. and {Giammaria}, P. and {Godinovi{\'c}}, N. and {Gora}, D. and {Griffiths}, S. and {Guberman}, D. and {Hadasch}, D. and {Hahn}, A. and {Hassan}, T. and {Hayashida}, M. and {Herrera}, J. and {Hose}, J. and {Hrupec}, D. and {Hughes}, G. and {Ishio}, K. and {Konno}, Y. and {Kubo}, H. and {Kushida}, J. and {Kuve{\v{z}}di{\'c}}, D. and {Lelas}, D. and {Lindfors}, E. and {Lombardi}, S. and {Longo}, F. and {L{\'o}pez}, M. and {L{\'o}pez-Oramas}, A. and {Majumdar}, P. and {Makariev}, M. and {Maneva}, G. and {Manganaro}, M. and {Mannheim}, K. and {Maraschi}, L. and {Mariotti}, M. and {Mart{\'\i}nez}, M. and {Mazin}, D. and {Menzel}, U. and {Minev}, M. and {Mirzoyan}, R. and {Moralejo}, A. and {Moreno}, V. and {Moretti}, E. and {Munar-Adrover}, P. and {Neustroev}, V. and {Niedzwiecki}, A. and {Nievas Rosillo}, M. and {Nilsson}, K. and {Nishijima}, K. and {Noda}, K. and {Nogu{\'e}s}, L. and {Paiano}, S. and {Palacio}, J. and {Paneque}, D. and {Paoletti}, R. and {Paredes}, J.~M. and {Paredes-Fortuny}, X. and {Pedaletti}, G. and {Peresano}, M. and {Perri}, L. and {Persic}, M. and {Prada Moroni}, P.~G. and {Prandini}, E. and {Puljak}, I. and {Garcia}, J.~R. and {Reichardt}, I. and {Rhode}, W. and {Rib{\'o}}, M. and {Rico}, J. and {Saito}, T. and {Satalecka}, K. and {Schroeder}, S. and {Schweizer}, T. and {Shore}, S.~N. and {Sillanp{\"a}{\"a}}, A. and {Sitarek}, J. and {{\v{S}}nidari{\'c}}, I. and {Sobczynska}, D. and {Stamerra}, A. and {Strzys}, M. and {Suri{\'c}}, T. and {Takalo}, L. and {Tavecchio}, F. and {Temnikov}, P. and {Terzi{\'c}}, T. and {Tescaro}, D. and {Teshima}, M. and {Torres}, D.~F. and {Torres-Alb{\`a}}, N. and {Treves}, A. and {Vanzo}, G. and {Vazquez Acosta}, M. and {Vovk}, I. and {Ward}, J.~E. and {Will}, M. and {Wu}, M.~H. and {Zari{\'c}}, D. and {H.~E.~S.~S. Collaboration} and {Abdalla}, H. and {Abramowski}, A. and {Aharonian}, F. and {Ait Benkhali}, F. and {Akhperjanian}, A.~G. and {Andersson}, T. and {Ang{\"u}ner}, E.~O. and {Arakawa}, M. and {Arrieta}, M. and {Aubert}, P. and {Backes}, M. and {Balzer}, A. and {Barnard}, M. and {Becherini}, Y. and {Becker Tjus}, J. and {Berge}, D. and {Bernhard}, S. and {Bernl{\"o}hr}, K. and {Blackwell}, R. and {B{\"o}ttcher}, M. and {Boisson}, C. and {Bolmont}, J. and {Bordas}, P. and {Bregeon}, J. and {Brun}, F. and {Brun}, P. and {Bryan}, M. and {B{\"u}chele}, M. and {Bulik}, T. and {Capasso}, M. and {Carr}, J. and {Casanova}, S. and {Cerruti}, M. and {Chakraborty}, N. and {Chalme-Calvet}, R. and {Chaves}, R.~C.~G. and {Chen}, A. and {Chevalier}, J. and {Chr{\'e}tien}, M. and {Coffaro}, M. and {Colafrancesco}, S. and {Cologna}, G. and {Condon}, B. and {Conrad}, J. and {Cui}, Y. and {Davids}, I.~D. and {Decock}, J. and {Degrange}, B. and {Deil}, C. and {Devin}, J. and {Dewilt}, P. and {Dirson}, L. and {Djannati-Ata{\"\i}}, A. and {Domainko}, W. and {Donath}, A. and {Drury}, L.~O. 'c. and {Dutson}, K. and {Dyks}, J. and {Edwards}, T. and {Egberts}, K. and {Eger}, P. and {Ernenwein}, J. -P. and {Eschbach}, S. and {Farnier}, C. and {Fegan}, S. and {Fernandes}, M.~V. and {Fiasson}, A. and {Fontaine}, G. and {F{\"o}rster}, A. and {Funk}, S. and {F{\"u}{\ss}ling}, M. and {Gabici}, S. and {Gajdus}, M. and {Gallant}, Y.~A. and {Garrigoux}, T. and {Giavitto}, G. and {Giebels}, B. and {Glicenstein}, J.~F. and {Gottschall}, D. and {Goyal}, A. and {Grondin}, M. -H. and {Hahn}, J. and {Haupt}, M. and {Hawkes}, J. and {Heinzelmann}, G. and {Henri}, G. and {Hermann}, G. and {Hervet}, O. and {Hinton}, J.~A. and {Hofmann}, W. and {Hoischen}, C. and {Holler}, M. and {Horns}, D. and {Ivascenko}, A. and {Iwasaki}, H. and {Jacholkowska}, A. and {Jamrozy}, M. and {Janiak}, M. and {Jankowsky}, D. and {Jankowsky}, F. and {Jingo}, M. and {Jogler}, T. and {Jouvin}, L. and {Jung-Richardt}, I. and {Kastendieck}, M.~A. and {Katarzy{\'n}ski}, K. and {Katsuragawa}, M. and {Katz}, U. and {Kerszberg}, D. and {Khangulyan}, D. and {Kh{\'e}lifi}, B. and {Kieffer}, M. and {King}, J. and {Klepser}, S. and {Klochkov}, D. and {Klu{\'z}niak}, W. and {Kolitzus}, D. and {Komin}, Nu. and {Kosack}, K. and {Krakau}, S. and {Kraus}, M. and {Kr{\"u}ger}, P.~P. and {Laffon}, H. and {Lamanna}, G. and {Lau}, J. and {Lees}, J. -P. and {Lefaucheur}, J. and {Lefranc}, V. and {Lemi{\`e}re}, A. and {Lemoine-Goumard}, M. and {Lenain}, J. -P. and {Leser}, E. and {Lohse}, T. and {Lorentz}, M. and {Liu}, R. and {L{\'o}pez-Coto}, R. and {Lypova}, I. and {Marandon}, V. and {Marcowith}, A. and {Mariaud}, C. and {Marx}, R. and {Maurin}, G. and {Maxted}, N. and {Mayer}, M. and {Meintjes}, P.~J. and {Meyer}, M. and {Mitchell}, A.~M.~W. and {Moderski}, R. and {Mohamed}, M. and {Mohrmann}, L. and {Mor{\r{a}}}, K. and {Moulin}, E. and {Murach}, T. and {Nakashima}, S. and {de Naurois}, M. and {Niederwanger}, F. and {Niemiec}, J. and {Oakes}, L. and {O'Brien}, P. and {Odaka}, H. and {{\"O}ttl}, S. and {Ohm}, S. and {Ostrowski}, M. and {Oya}, I. and {Padovani}, M. and {Panter}, M. and {Parsons}, R.~D. and {Pekeur}, N.~W. and {Pelletier}, G. and {Perennes}, C. and {Petrucci}, P. -O. and {Peyaud}, B. and {Piel}, Q. and {Pita}, S. and {Poon}, H. and {Prokhorov}, D. and {Prokoph}, H. and {P{\"u}hlhofer}, G. and {Punch}, M. and {Quirrenbach}, A. and {Raab}, S. and {Reimer}, A. and {Reimer}, O. and {Renaud}, M. and {de Los Reyes}, R. and {Richter}, S. and {Rieger}, F. and {Romoli}, C. and {Rowell}, G. and {Rudak}, B. and {Rulten}, C.~B. and {Safi-Harb}, S. and {Sahakian}, V. and {Saito}, S. and {Salek}, D. and {Sanchez}, D.~A. and {Santangelo}, A. and {Sasaki}, M. and {Schlickeiser}, R. and {Sch{\"u}ssler}, F. and {Schulz}, A. and {Schwanke}, U. and {Schwemmer}, S. and {Seglar-Arroyo}, M. and {Settimo}, M. and {Seyffert}, A.~S. and {Shafi}, N. and {Shilon}, I. and {Simoni}, R. and {Sol}, H. and {Spanier}, F. and {Spengler}, G. and {Spies}, F. and {Stawarz}, {\L}. and {Steenkamp}, R. and {Stegmann}, C. and {Stycz}, K. and {Sushch}, I. and {Takahashi}, T. and {Tavernet}, J. -P. and {Tavernier}, T. and {Taylor}, A.~M. and {Terrier}, R. and {Tibaldo}, L. and {Tiziani}, D. and {Tluczykont}, M. and {Trichard}, C. and {Tsuji}, N. and {Tuffs}, R. and {Uchiyama}, Y. and {van der Walt}, D.~J. and {van Eldik}, C. and {van Rensburg}, C. and {van Soelen}, B. and {Vasileiadis}, G. and {Veh}, J. and {Venter}, C. and {Viana}, A. and {Vincent}, P. and {Vink}, J. and {Voisin}, F. and {V{\"o}lk}, H.~J. and {Vuillaume}, T. and {Wadiasingh}, Z. and {Wagner}, S.~J. and {Wagner}, P. and {Wagner}, R.~M. and {White}, R. and {Wierzcholska}, A. and {Willmann}, P. and {W{\"o}rnlein}, A. and {Wouters}, D. and {Yang}, R. and {Zabalza}, V. and {Zaborov}, D. and {Zacharias}, M. and {Zanin}, R. and {Zdziarski}, A.~A. and {Zech}, A. and {Zefi}, F. and {Ziegler}, A. and {Zywucka}, N.},
        title = "{Constraints on particle acceleration in SS433/W50 from MAGIC and H.E.S.S. observations}",
      journal = {\aap},
         year = 2018,
        month = apr,
       volume = {612},
          eid = {A14},
        pages = {A14},
          doi = {10.1051/0004-6361/201731169},
archivePrefix = {arXiv},
       eprint = {1707.03658},
 primaryClass = {astro-ph.HE},
       adsurl = {https://ui.adsabs.harvard.edu/abs/2018A&A...612A..14M}
}

@INPROCEEDINGS{2003ASPC..300..151H,
       author = {{Harris}, D.~E.},
        title = "{What Can We Learn about Extragalactic Radio Jets from X-Ray Data?}",
    booktitle = {Radio Astronomy at the Fringe},
         year = 2003,
       editor = {{Zensus}, J. Anton and {Cohen}, Marshall H. and {Ros}, Eduardo},
       series = {Astronomical Society of the Pacific Conference Series},
       volume = {300},
        month = jan,
        pages = {151},
archivePrefix = {arXiv},
       eprint = {astro-ph/0302097},
 primaryClass = {astro-ph},
       adsurl = {https://ui.adsabs.harvard.edu/abs/2003ASPC..300..151H}
}

@ARTICLE{2007MNRAS.381..308B,
       author = {{Boumis}, P. and {Meaburn}, J. and {Alikakos}, J. and {Redman}, M.~P. and {Akras}, S. and {Mavromatakis}, F. and {L{\'o}pez}, J.~A. and {Caulet}, A. and {Goudis}, C.~D.},
        title = "{Deep optical observations of the interaction of the SS 433 microquasar jet with the W50 radio continuum shell}",
      journal = {\mnras},
         year = 2007,
        month = oct,
       volume = {381},
       number = {1},
        pages = {308-318},
          doi = {10.1111/j.1365-2966.2007.12276.x},
archivePrefix = {arXiv},
       eprint = {0707.4243},
 primaryClass = {astro-ph},
       adsurl = {https://ui.adsabs.harvard.edu/abs/2007MNRAS.381..308B}
}

@ARTICLE{2020ApJ...889L...5F,
       author = {{Fang}, Ke and {Charles}, Eric and {Blandford}, Roger D.},
        title = "{GeV-TeV Counterparts of SS 433/W50 from Fermi-LAT and HAWC Observations}",
      journal = {\apjl},
         year = 2020,
        month = jan,
       volume = {889},
       number = {1},
          eid = {L5},
        pages = {L5},
          doi = {10.3847/2041-8213/ab62b8},
archivePrefix = {arXiv},
       eprint = {2001.03599},
 primaryClass = {astro-ph.HE},
       adsurl = {https://ui.adsabs.harvard.edu/abs/2020ApJ...889L...5F}
}

@INPROCEEDINGS{2017ICRC...35..713K,
       author = {{Kar}, P. and {VERITAS Collaboration}},
        title = "{VERITAS Observations of High-Mass X-Ray Binary SS 433}",
    booktitle = {35th International Cosmic Ray Conference (ICRC2017)},
         year = 2017,
       series = {International Cosmic Ray Conference},
       volume = {301},
        month = jan,
          eid = {713},
        pages = {713},
archivePrefix = {arXiv},
       eprint = {1708.04967},
 primaryClass = {astro-ph.HE},
       adsurl = {https://ui.adsabs.harvard.edu/abs/2017ICRC...35..713K}
}

@ARTICLE{2005MNRAS.358..860M,
       author = {{Migliari}, S. and {Fender}, R.~P. and {Blundell}, K.~M. and {M{\'e}ndez}, M. and {van der Klis}, M.},
        title = "{Rapid variability of the arcsec-scale X-ray jets of SS 433}",
      journal = {\mnras},
         year = 2005,
        month = apr,
       volume = {358},
       number = {3},
        pages = {860-868},
          doi = {10.1111/j.1365-2966.2005.08791.x},
archivePrefix = {arXiv},
       eprint = {astro-ph/0501097},
 primaryClass = {astro-ph},
       adsurl = {https://ui.adsabs.harvard.edu/abs/2005MNRAS.358..860M}
}

@ARTICLE{2014ApJ...789...72N,
       author = {{Nynka}, Melania and {Hailey}, Charles J. and {Reynolds}, Stephen P. and {An}, Hongjun and {Baganoff}, Frederick K. and {Boggs}, Steven E. and {Christensen}, Finn E. and {Craig}, William W. and {Gotthelf}, Eric V. and {Grefenstette}, Brian W. and {Harrison}, Fiona A. and {Krivonos}, Roman and {Madsen}, Kristin K. and {Mori}, Kaya and {Perez}, Kerstin and {Stern}, Daniel and {Wik}, Daniel R. and {Zhang}, William W. and {Zoglauer}, Andreas},
        title = "{NuSTAR Study of Hard X-Ray Morphology and Spectroscopy of PWN G21.5-0.9}",
      journal = {\apj},
         year = 2014,
        month = jul,
       volume = {789},
       number = {1},
          eid = {72},
        pages = {72},
          doi = {10.1088/0004-637X/789/1/72},
archivePrefix = {arXiv},
       eprint = {1405.3239},
 primaryClass = {astro-ph.HE},
       adsurl = {https://ui.adsabs.harvard.edu/abs/2014ApJ...789...72N}
}

@ARTICLE{2011ApJ...740...64L,
       author = {{Lefa}, E. and {Rieger}, F.~M. and {Aharonian}, F.},
        title = "{Formation of Very Hard Gamma-Ray Spectra of Blazars in Leptonic Models}",
      journal = {\apj},
         year = 2011,
        month = oct,
       volume = {740},
       number = {2},
          eid = {64},
        pages = {64},
          doi = {10.1088/0004-637X/740/2/64},
archivePrefix = {arXiv},
       eprint = {1106.4201},
 primaryClass = {astro-ph.HE},
       adsurl = {https://ui.adsabs.harvard.edu/abs/2011ApJ...740...64L}
}

@ARTICLE{2014ApJ...783L..21S,
       author = {{Sironi}, Lorenzo and {Spitkovsky}, Anatoly},
        title = "{Relativistic Reconnection: An Efficient Source of Non-thermal Particles}",
      journal = {\apjl},
         year = 2014,
        month = mar,
       volume = {783},
       number = {1},
          eid = {L21},
        pages = {L21},
          doi = {10.1088/2041-8205/783/1/L21},
archivePrefix = {arXiv},
       eprint = {1401.5471},
 primaryClass = {astro-ph.HE},
       adsurl = {https://ui.adsabs.harvard.edu/abs/2014ApJ...783L..21S}
}

@INPROCEEDINGS{2015ASSL..407..311L,
       author = {{Lazarian}, Alex and {Eyink}, Gregory L. and {Vishniac}, Ethan T. and {Kowal}, Grzegorz},
        title = "{Magnetic Reconnection in Astrophysical Environments}",
    booktitle = {Magnetic Fields in Diffuse Media},
         year = 2015,
       editor = {{Lazarian}, Alexander and {de Gouveia Dal Pino}, Elisabete M. and {Melioli}, Claudio},
       series = {Astrophysics and Space Science Library},
       volume = {407},
        month = jan,
        pages = {311},
          doi = {10.1007/978-3-662-44625-6\_12},
archivePrefix = {arXiv},
       eprint = {1407.6356},
 primaryClass = {astro-ph.SR},
       adsurl = {https://ui.adsabs.harvard.edu/abs/2015ASSL..407..311L}
}

@ARTICLE{2013A&A...558A..19W,
       author = {{Wykes}, Sarka and {Croston}, Judith H. and {Hardcastle}, Martin J. and {Eilek}, Jean A. and {Biermann}, Peter L. and {Achterberg}, Abraham and {Bray}, Justin D. and {Lazarian}, Alex and {Haverkorn}, Marijke and {Protheroe}, Ray J. and {Bromberg}, Omer},
        title = "{Mass entrainment and turbulence-driven acceleration of ultra-high energy cosmic rays in Centaurus A}",
      journal = {\aap},
         year = 2013,
        month = oct,
       volume = {558},
          eid = {A19},
        pages = {A19},
          doi = {10.1051/0004-6361/201321622},
archivePrefix = {arXiv},
       eprint = {1305.2761},
 primaryClass = {astro-ph.HE},
       adsurl = {https://ui.adsabs.harvard.edu/abs/2013A&A...558A..19W}
}

@ARTICLE{1987AJ.....94.1633E,
       author = {{Elston}, R. and {Baum}, S.},
        title = "{High-Resolution Radio Observations of W50, the Remnant Associated with SS 433}",
      journal = {\aj},
         year = 1987,
        month = dec,
       volume = {94},
        pages = {1633},
          doi = {10.1086/114594},
       adsurl = {https://ui.adsabs.harvard.edu/abs/1987AJ.....94.1633E}
}

@ARTICLE{2003ApJ...591..361G,
       author = {{Gotthelf}, E.~V.},
        title = "{X-Ray Spectra of Young Pulsars and Their Wind Nebulae: Dependence on Spin-Down Energy Loss Rate}",
      journal = {\apj},
         year = 2003,
        month = jul,
       volume = {591},
       number = {1},
        pages = {361-365},
          doi = {10.1086/375124},
archivePrefix = {arXiv},
       eprint = {astro-ph/0303155},
 primaryClass = {astro-ph},
       adsurl = {https://ui.adsabs.harvard.edu/abs/2003ApJ...591..361G}
}

@INPROCEEDINGS{2008AIPC..983..171K,
       author = {{Kargaltsev}, O. and {Pavlov}, G.~G.},
        title = "{Pulsar Wind Nebulae in the Chandra Era}",
    booktitle = {40 Years of Pulsars: Millisecond Pulsars, Magnetars and More},
         year = 2008,
       editor = {{Bassa}, C. and {Wang}, Z. and {Cumming}, A. and {Kaspi}, V.~M.},
       series = {American Institute of Physics Conference Series},
       volume = {983},
        month = feb,
        pages = {171-185},
          doi = {10.1063/1.2900138},
archivePrefix = {arXiv},
       eprint = {0801.2602},
 primaryClass = {astro-ph},
       adsurl = {https://ui.adsabs.harvard.edu/abs/2008AIPC..983..171K}
}

@ARTICLE{2004ASPRv..12....1F,
       author = {{Fabrika}, S.},
        title = "{The jets and supercritical accretion disk in SS433}",
      journal = {\apspr},
         year = 2004,
        month = jan,
       volume = {12},
        pages = {1-152},
archivePrefix = {arXiv},
       eprint = {astro-ph/0603390},
 primaryClass = {astro-ph},
       adsurl = {https://ui.adsabs.harvard.edu/abs/2004ASPRv..12....1F}
}

@ARTICLE{1974A&A....32..375V,
       author = {{Velusamy}, T. and {Kundu}, M.~R.},
        title = "{Observations of intensity and linear polarization in supernova remnants at 11 cm wavelength.}",
      journal = {\aap},
         year = 1974,
        month = jun,
       volume = {32},
        pages = {375-390},
       adsurl = {https://ui.adsabs.harvard.edu/abs/1974A&A....32..375V}
}

@ARTICLE{1980ApJ...236L..23V,
       author = {{van den Bergh}, S.},
        title = "{The optical remnant of W50 (SS 433).}",
      journal = {\apjl},
         year = 1980,
        month = feb,
       volume = {236},
        pages = {L23},
          doi = {10.1086/183190},
       adsurl = {https://ui.adsabs.harvard.edu/abs/1980ApJ...236L..23V}
}

@ARTICLE{1995Natur.378..255K,
       author = {{Koyama}, K. and {Petre}, R. and {Gotthelf}, E.~V. and {Hwang}, U. and {Matsuura}, M. and {Ozaki}, M. and {Holt}, S.~S.},
        title = "{Evidence for shock acceleration of high-energy electrons in the supernova remnant SN1006}",
      journal = {\nat},
         year = 1995,
        month = nov,
       volume = {378},
       number = {6554},
        pages = {255-258},
          doi = {10.1038/378255a0},
       adsurl = {https://ui.adsabs.harvard.edu/abs/1995Natur.378..255K}
}

@ARTICLE{2004ApJ...608..698S,
       author = {{Sambruna}, Rita M. and {Gambill}, Jessica K. and {Maraschi}, L. and {Tavecchio}, F. and {Cerutti}, R. and {Cheung}, C.~C. and {Urry}, C. Megan and {Chartas}, G.},
        title = "{A Survey of Extended Radio Jets with Chandra and the Hubble Space Telescope}",
      journal = {\apj},
         year = 2004,
        month = jun,
       volume = {608},
       number = {2},
        pages = {698-720},
          doi = {10.1086/383124},
archivePrefix = {arXiv},
       eprint = {astro-ph/0401475},
 primaryClass = {astro-ph},
       adsurl = {https://ui.adsabs.harvard.edu/abs/2004ApJ...608..698S}
}

@ARTICLE{2006ApJ...637..486K,
       author = {{Kotani}, T. and {Trushkin}, S.~A. and {Valiullin}, R. and {Kinugasa}, K. and {Safi-Harb}, S. and {Kawai}, N. and {Namiki}, M.},
        title = "{A Massive Jet Ejection Event from the Microquasar SS 433 Accompanying Rapid X-Ray Variability}",
      journal = {\apj},
         year = 2006,
        month = jan,
       volume = {637},
       number = {1},
        pages = {486-493},
          doi = {10.1086/498387},
archivePrefix = {arXiv},
       eprint = {astro-ph/0509411},
 primaryClass = {astro-ph},
       adsurl = {https://ui.adsabs.harvard.edu/abs/2006ApJ...637..486K}
}

@ARTICLE{2005A&A...431..575B,
       author = {{Brinkmann}, W. and {Kotani}, T. and {Kawai}, N.},
        title = "{XMM-Newton observations of SS 433  I. EPIC spectral analysis}",
      journal = {\aap},
         year = 2005,
        month = feb,
       volume = {431},
        pages = {575-586},
          doi = {10.1051/0004-6361:20041768},
       adsurl = {https://ui.adsabs.harvard.edu/abs/2005A&A...431..575B}
}

@ARTICLE{Safi-Harb_2022,
    doi = {10.3847/1538-4357/ac7c05},
    url = {https://dx.doi.org/10.3847/1538-4357/ac7c05},
    year = {2022},
    month = {aug},
    publisher = {The American Astronomical Society},
    volume = {935},
    number = {2},
    pages = {163},
    author = {Samar Safi-Harb and Brydyn {Mac Intyre} and Shuo Zhang and Isaac Pope and Shuhan Zhang and Nathan Saffold and Kaya Mori and Eric V. Gotthelf and Felix Aharonian and Matthew Band and Chelsea Braun and Ke Fang and Charles Hailey and Melania Nynka and Chang D. Rho},
    title = {Hard X-Ray Emission from the Eastern Jet of SS 433 Powering the W50 “Manatee” Nebula: Evidence for Particle Reacceleration},
    journal = {The Astrophysical Journal}
}

@ARTICLE{2006MNRAS.371..829S,
    author = {{Sanders}, J.~S.},
    title = "{Contour binning: a new technique for spatially resolved X-ray spectroscopy applied to Cassiopeia A}",
    journal = {\mnras},
    year = 2006,
    month = sep,
    volume = {371},
    number = {2},
    pages = {829-842},
    doi = {10.1111/j.1365-2966.2006.10716.x},
    archivePrefix = {arXiv},
    eprint = {astro-ph/0606528},
    primaryClass = {astro-ph},
    adsurl = {https://ui.adsabs.harvard.edu/abs/2006MNRAS.371..829S}
}

@inbook{Vink_2020, 
    place={Cham, Switzerland}, 
    title={12.1.2 The Minimum Energy Requirement and the Van der Laan Mechanism}, 
    booktitle={Physics and evolution of Supernova Remnants}, 
    publisher={Springer}, 
    author={Vink, Jacco},
    pages={327--330},
    year={2020}
}

@ARTICLE{2007A&A...465..695Z,
       author = {{Zirakashvili}, V.~N. and {Aharonian}, F.},
        title = "{Analytical solutions for energy spectra of electrons accelerated by nonrelativistic shock-waves in shell type supernova remnants}",
      journal = {\aap},
         year = 2007,
        month = apr,
       volume = {465},
       number = {3},
        pages = {695-702},
          doi = {10.1051/0004-6361:20066494},
archivePrefix = {arXiv},
       eprint = {astro-ph/0612717},
 primaryClass = {astro-ph},
       adsurl = {https://ui.adsabs.harvard.edu/abs/2007A&A...465..695Z}
}

@article{hess_2024,
   title={Acceleration and transport of relativistic electrons in the jets of the microquasar SS 433},
   volume={383},
   ISSN={1095-9203},
   url={http://dx.doi.org/10.1126/science.adi2048},
   DOI={10.1126/science.adi2048},
   number={6681},
   journal={Science},
   publisher={American Association for the Advancement of Science (AAAS)},
   author={Aharonian, F. and Benkhali, F. Ait and Aschersleben, J. and Ashkar, H. and Backes, M. and Martins, V. Barbosa and Batzofin, R. and Becherini, Y. and Berge, D. and Bernlöhr, K. and Bi, B. and Böttcher, M. and Boisson, C. and Bolmont, J. and de Lavergne, M. de Bony and Borowska, J. and Bouyahiaoui, M. and Breuhaus, M. and Brose, R. and Brown, A. M. and Brun, F. and Bruno, B. and Bulik, T. and Burger-Scheidlin, C. and Caroff, S. and Casanova, S. and Cecil, R. and Celic, J. and Cerruti, M. and Chand, T. and Chandra, S. and Chen, A. and Chibueze, J. and Chibueze, O. and Cotter, G. and Dai, S. and Mbarubucyeye, J. Damascene and Djannati-Ataï, A. and Dmytriiev, A. and Doroshenko, V. and Egberts, K. and Einecke, S. and Ernenwein, J.-P. and Filipovic, M. and Fontaine, G. and Füßling, M. and Funk, S. and Gabici, S. and Ghafourizadeh, S. and Giavitto, G. and Glawion, D. and Glicenstein, J.-F. and Grolleron, G. and Haerer, L. and Hinton, J. A. and Hofmann, W. and Holch, T. L. and Holler, M. and Horns, D. and Jamrozy, M. and Jankowsky, F. and Jardin-Blicq, A. and Joshi, V. and Jung-Richardt, I. and Kasai, E. and Katarzyński, K. and Khatoon, R. and Khélifi, B. and Klepser, S. and Kluźniak, W. and Komin, Nu. and Kosack, K. and Kostunin, D. and Kundu, A. and Lang, R. G. and Le Stum, S. and Leitl, F. and Lemière, A. and Lenain, J.-P. and Leuschner, F. and Lohse, T. and Luashvili, A. and Lypova, I. and Mackey, J. and Malyshev, D. and Malyshev, D. and Marandon, V. and Marchegiani, P. and Marcowith, A. and Martí-Devesa, G. and Marx, R. and Mehta, A. and Mitchell, A. and Moderski, R. and Mohrmann, L. and Montanari, A. and Moulin, E. and Murach, T. and Nakashima, K. and de Naurois, M. and Niemiec, J. and Noel, A. Priyana and Ohm, S. and Olivera-Nieto, L. and de Ona Wilhelmi, E. and Ostrowski, M. and Panny, S. and Panter, M. and Parsons, R. D. and Peron, G. and Prokhorov, D. A. and Pühlhofer, G. and Punch, M. and Quirrenbach, A. and Reichherzer, P. and Reimer, A. and Reimer, O. and Ren, H. and Renaud, M. and Reville, B. and Rieger, F. and Rowell, G. and Rudak, B. and Ricarte, H. Rueda and Ruiz-Velasco, E. and Sahakian, V. and Salzmann, H. and Santangelo, A. and Sasaki, M. and Schäfer, J. and Schüssler, F. and Schwanke, U. and Shapopi, J. N. S. and Sol, H. and Specovius, A. and Spencer, S. and Stawarz, L. and Steenkamp, R. and Steinmassl, S. and Steppa, C. and Streil, K. and Sushch, I. and Suzuki, H. and Takahashi, T. and Tanaka, T. and Taylor, A. M. and Terrier, R. and Tsirou, M. and Tsuji, N. and Unbehaun, T. and van Eldik, C. and Vecchi, M. and Veh, J. and Venter, C. and Vink, J. and Wach, T. and Wagner, S. J. and Werner, F. and White, R. and Wierzcholska, A. and Wong, Yu Wun and Zacharias, M. and Zargaryan, D. and Zdziarski, A. A. and Zech, A. and Zouari, S. and Żywucka, N.},
   year={2024},
   month=jan, pages={402–406} 
}

@ARTICLE{2022PASJ...74.1143K,
       author = {{Kayama}, Kazuho and {Tanaka}, Takaaki and {Uchida}, Hiroyuki and {Tsuru}, Takeshi Go and {Sudoh}, Takahiro and {Inoue}, Yoshiyuki and {Khangulyan}, Dmitry and {Tsuji}, Naomi and {Yamamoto}, Hiroaki},
        title = "{Spatially resolved study of the SS 433/W 50 west region with Chandra: X-ray structure and spectral variation of non-thermal emission}",
      journal = {\pasj},
         year = 2022,
        month = oct,
       volume = {74},
       number = {5},
        pages = {1143-1156},
          doi = {10.1093/pasj/psac060},
archivePrefix = {arXiv},
       eprint = {2207.05924},
 primaryClass = {astro-ph.HE},
       adsurl = {https://ui.adsabs.harvard.edu/abs/2022PASJ...74.1143K}
}

@ARTICLE{2024ApJ...961L..12K,
       author = {{Kaaret}, Philip and {Ferrazzoli}, Riccardo and {Silvestri}, Stefano and {Negro}, Michela and {Manfreda}, Alberto and {Wu}, Kinwah and {Costa}, Enrico and {Soffitta}, Paolo and {Safi-Harb}, Samar and {Poutanen}, Juri and {Veledina}, Alexandra and {Di Marco}, Alessandro and {Slane}, Patrick and {Bianchi}, Stefano and {Ingram}, Adam and {Romani}, Roger W. and {Cibrario}, Nicol{\`o} and {Mac Intyre}, Brydyn and {Mikus̆incov{\'a}}, Romana and {Ratheesh}, Ajay and {Steiner}, James F. and {Svoboda}, Jiri and {Tugliani}, Stefano and {Agudo}, Iv{\'a}n and {Antonelli}, Lucio A. and {Bachetti}, Matteo and {Baldini}, Luca and {Baumgartner}, Wayne H. and {Bellazzini}, Ronaldo and {Bongiorno}, Stephen D. and {Bonino}, Raffaella and {Brez}, Alessandro and {Bucciantini}, Niccol{\`o} and {Capitanio}, Fiamma and {Castellano}, Simone and {Cavazzuti}, Elisabetta and {Chen}, Chien-Ting and {Ciprini}, Stefano and {De Rosa}, Alessandra and {Del Monte}, Ettore and {Di Gesu}, Laura and {Di Lalla}, Niccol{\`o} and {Donnarumma}, Immacolata and {Doroshenko}, Victor and {Dov{\v{c}}iak}, Michal and {Ehlert}, Steven R. and {Enoto}, Teruaki and {Evangelista}, Yuri and {Fabiani}, Sergio and {Garc{\'\i}a}, Javier A. and {Gunji}, Shuichi and {Hayashida}, Kiyoshi and {Heyl}, Jeremy and {Iwakiri}, Wataru and {Jorstad}, Svetlana G. and {Karas}, Vladimir and {Kislat}, Fabian and {Kitaguchi}, Takao and {Kolodziejczak}, Jeffery J. and {Krawczynski}, Henric and {La Monaca}, Fabio and {Latronico}, Luca and {Liodakis}, Ioannis and {Maldera}, Simone and {Marin}, Fr{\'e}d{\'e}ric and {Marinucci}, Andrea and {Marscher}, Alan P. and {Marshall}, Herman L. and {Massaro}, Francesco and {Matt}, Giorgio and {Mitsuishi}, Ikuyuki and {Mizuno}, Tsunefumi and {Muleri}, Fabio and {Ng}, Chi-Yung and {O'Dell}, Stephen L. and {Omodei}, Nicola and {Oppedisano}, Chiara and {Papitto}, Alessandro and {Pavlov}, George G. and {Peirson}, Abel L. and {Perri}, Matteo and {Pesce-Rollins}, Melissa and {Petrucci}, Pierre-Olivier and {Pilia}, Maura and {Possenti}, Andrea and {Puccetti}, Simonetta and {Ramsey}, Brian D. and {Rankin}, John and {Roberts}, Oliver J. and {Sgr{\`o}}, Carmelo and {Spandre}, Gloria and {Swartz}, Douglas A. and {Tamagawa}, Toru and {Tavecchio}, Fabrizio and {Taverna}, Roberto and {Tawara}, Yuzuru and {Tennant}, Allyn F. and {Thomas}, Nicholas E. and {Tombesi}, Francesco and {Trois}, Alessio and {Tsygankov}, Sergey S. and {Turolla}, Roberto and {Vink}, Jacco and {Weisskopf}, Martin C. and {Xie}, Fei and {Zane}, Silvia},
        title = "{X-Ray Polarization of the Eastern Lobe of SS 433}",
      journal = {\apjl},
         year = 2024,
        month = jan,
       volume = {961},
       number = {1},
          eid = {L12},
        pages = {L12},
          doi = {10.3847/2041-8213/ad103b},
archivePrefix = {arXiv},
       eprint = {2311.16313},
 primaryClass = {astro-ph.HE},
       adsurl = {https://ui.adsabs.harvard.edu/abs/2024ApJ...961L..12K}
}

@ARTICLE{2018MNRAS.481.1455K,
       author = {{Khangulyan}, Dmitry and {Bosch-Ramon}, Valent{\'\i} and {Uchiyama}, Yasunobu},
        title = "{Inverse Compton emission from relativistic jets in binary systems}",
      journal = {\mnras},
         year = 2018,
        month = dec,
       volume = {481},
       number = {2},
        pages = {1455-1468},
          doi = {10.1093/mnras/sty2356},
archivePrefix = {arXiv},
       eprint = {1808.09628},
 primaryClass = {astro-ph.HE},
       adsurl = {https://ui.adsabs.harvard.edu/abs/2018MNRAS.481.1455K}
}

@ARTICLE{2014ApJ...783..100K,
       author = {{Khangulyan}, D. and {Aharonian}, F.~A. and {Kelner}, S.~R.},
        title = "{Simple Analytical Approximations for Treatment of Inverse Compton Scattering of Relativistic Electrons in the Blackbody Radiation Field}",
      journal = {\apj},
         year = 2014,
        month = mar,
       volume = {783},
       number = {2},
          eid = {100},
        pages = {100},
          doi = {10.1088/0004-637X/783/2/100},
archivePrefix = {arXiv},
       eprint = {1310.7971},
 primaryClass = {astro-ph.HE},
       adsurl = {https://ui.adsabs.harvard.edu/abs/2014ApJ...783..100K}
}

@ARTICLE{1997MNRAS.291..162A,
       author = {{Aharonian}, F.~A. and {Atoyan}, A.~M. and {Kifune}, T.},
        title = "{Inverse Compton gamma radiation of faint synchrotron X-ray nebulae around pulsars}",
      journal = {\mnras},
         year = 1997,
        month = oct,
       volume = {291},
       number = {1},
        pages = {162-176},
          doi = {10.1093/mnras/291.1.162},
       adsurl = {https://ui.adsabs.harvard.edu/abs/1997MNRAS.291..162A}
}

@article{Sakai2026,
   title={Spectral and photometric variability of SS 433 observed with XRISM and simultaneous optical and near-infrared telescopes},
   volume={78},
   ISSN={2053-051X},
   url={http://dx.doi.org/10.1093/pasj/psaf152},
   DOI={10.1093/pasj/psaf152},
   number={2},
   journal={Publications of the Astronomical Society of Japan},
   publisher={Oxford University Press (OUP)},
   author={Sakai, Yusuke and Yamada, Shinya and Okada, Yuta and Takagi, Toshihiro and Usuki, Tomoya and Shidatsu, Megumi and Kobayashi, Shogo B and Petre, Robert and Ueda, Yoshihiro and Uchiyama, Hideki and Tan, Miho and Kotani, Taro and Igarashi, Taichi and Machida, Mami and Sakemi, Haruka and Kawai, Nobuyuki and Miura, Daiki and Yamaguchi, Hiroya and Fujiwara, Kanta and Hiramatsu, Daichi and Isogai, Keisuke and Kang, Chulsoo and Kimura, Mariko and Murata, Katsuhiro L and Nagayama, Takahiro and Nakamoto, Taichi and Namekata, Kosuke and Niida, Yuki and Niino, Yuu and Niwano, Masafumi and Oh, Kyuseok and Sako, Shigeyuki and Sasada, Mahito and Suzuki, Hiromasa and Taguchi, Kenta and Takahashi, Ichiro and Uenishi, Miyu and Yatsu, Yoichi and Yoshimoto, Marina},
   year={2026},
   month=Feb, pages={436–453} 
}

@ARTICLE{2009ApJ...703..662R,
       author = {{Reynolds}, Stephen P.},
        title = "{Synchrotron-Loss Spectral Breaks in Pulsar-Wind Nebulae and Extragalactic Jets}",
      journal = {\apj},
         year = 2009,
        month = sep,
       volume = {703},
       number = {1},
        pages = {662-670},
          doi = {10.1088/0004-637X/703/1/662},
archivePrefix = {arXiv},
       eprint = {0907.4756},
 primaryClass = {astro-ph.HE},
       adsurl = {https://ui.adsabs.harvard.edu/abs/2009ApJ...703..662R}
}

\end{document}